\documentclass[superscriptaddress]{revtex4-2}

\usepackage[utf8]{inputenc}
\usepackage[english]{babel}
\usepackage{graphicx}
\usepackage{dcolumn}
\usepackage{bm}
\usepackage{amsmath}    
\usepackage{verbatim}   
\usepackage{color}      
\usepackage{hyperref}   
\usepackage{mathrsfs}
\usepackage{upgreek}
\usepackage{braket}
\usepackage{physics}
\usepackage{amssymb}
\usepackage{amsmath}
\usepackage{subfig}
\usepackage{float}
\usepackage{booktabs}
\usepackage{cancel}
\usepackage{caption}
\usepackage{bigints}
\usepackage{trackchanges}
\newcommand{\ham}[0]{\mathcal{H}}

\begin{document}

\title{An Overview on the Nature of the Bounce in LQC and PQM}
\author{G. Barca}\email{Corresponding author. E-mail: gabriele.barca@uniroma1.it}
\affiliation{Department of Physics, “Sapienza” University of Rome, P.le Aldo Moro, 5 (00185) Roma, Italy}
\author{E. Giovannetti}
\affiliation{Department of Physics, “Sapienza” University of Rome, P.le Aldo Moro, 5 (00185) Roma, Italy}
\author{G. Montani}
\affiliation{Department of Physics, “Sapienza” University of Rome, P.le Aldo Moro, 5 (00185) Roma, Italy}\affiliation{ENEA, Fusion and Nuclear Safety Department, C.R. Frascati, Via E. Fermi, 45 (00044) Frascati (RM), Italy}

\begin{abstract}
We present a review on some of the basic aspects concerning quantum cosmology in the presence of cut-off physics as it has emerged in the literature during the last fifteen years. We first analyze how the Wheeler--DeWitt equation describes the quantum Universe dynamics, when a pure metric approach is concerned, showing how, in general, the primordial singularity is not removed by the quantum effects. We then analyze the main implications of applying the loop quantum gravity prescriptions to the minisuperspace model, i.e., we discuss the basic features of the so-called loop quantum cosmology. For the isotropic Universe dynamics, we compare the original approach, dubbed the $\mu_0$ scheme, and the most commonly accepted formulation for which the area gap is taken as physically scaled, i.e., the so-called $\bar{\mu}$ scheme. Furthermore, some fundamental results concerning the Bianchi Universes are discussed, especially with respect to the morphology of the Bianchi IX model. Finally, we consider some relevant criticisms developed over the last ten years about the real link existing between the full theory of loop quantum gravity and its minisuperspace implementation, especially with respect to the preservation of the internal $SU(2)$ symmetry. In the second part of the review, we consider the dynamics of the isotropic Universe and of the Bianchi models in the framework of polymer quantum mechanics. Throughout the paper, we focus on the effective semiclassical dynamics and study the full quantum theory only in some cases, such as the FLRW model and the Bianchi I model in the Ashtekar variables. We first address the polymerization in terms of the Ashtekar--Barbero--Immirzi connection and show how the resulting dynamics is isomorphic to the $\mu_0$ scheme of loop quantum cosmology with a critical energy density of the Universe that depends on the initial conditions of the dynamics. The following step is to analyze the polymerization of volume-like variables, both for the isotropic and Bianchi I models, and~we see that if the Universe volume (the cubed scale factor) is one of the configurational variables, then~the resulting dynamics is isomorphic to that one emerging in loop quantum cosmology for the $\bar{\mu}$ scheme, with the critical energy density value being fixed only by fundamental constants and the Immirzi parameter. Finally, we consider the polymer quantum dynamics of the homogeneous and inhomogeneous Mixmaster model by means of a metric approach. In particular, we compare the results obtained by using the volume variable, which leads to the emergence of a singularity- and chaos-free cosmology, to the use of the standard Misner variable. In the latter case, we deal with the surprising result of a cosmology that is still singular, and its chaotic properties depend on the ratio between the lattice steps for the isotropic and anisotropic variables. We conclude the review with some considerations of the problem of changing variables in the polymer representation of the minisuperspace dynamics. In particular, on a semiclassical level, we consider how the dynamics can be properly mapped in two different sets of variables (at the price of having to deal with a coordinate dependent lattice step), and we infer some possible implications on the equivalence of the $\mu_0$ and $\bar{\mu}$ scheme of loop quantum cosmology.
\end{abstract}

\maketitle

\section{Introduction}
Despite the fact that no self-consistent theory has been developed in quantum gravity (for the most interesting approaches, see \cite{deWitt1,deWitt2,deWitt3,Kuchar81,AshtekarVariables,Rovelli_Smolin_LQG,Rovelli_1995DiscretenessAreaVolume,Rovelli_1995SpinNetworks,CQGMontani,hartlehawking,ThiemannBook,RovelliVidotto,RovelliBook2004}), along the years, the arena of primordial cosmology has constituted a valuable test to estimate the predictivity of the proposed theories on the birth of the Universe and quantum evolution \cite{MontaniPrimordial}.

The most significant change in the point of view on how to approach the quantization of the gravitational degrees of freedom took place with the formulation of the so-called loop quantum gravity (LQG) \cite{ThiemannBook}, especially because this formulation was able to construct a kinematical Hilbert space and to justify spontaneously the emergence of discrete area and volume spectra. LQG relies on the possibility to reduce the gravitational phase space to that of a $SU(2)$ non-Abelian theory \cite{Rovelli_Smolin_LQG,Rovelli_1995DiscretenessAreaVolume,Rovelli_1995SpinNetworks,CQGMontani}, and then the quantization scheme is performed by using ``smeared'' (non-local) variables, such as the holonomy and flux variables, as suggested by the original Wilson loop formulation and by non-Abelian gauge theories on a lattice. Indeed, when adopting Astekar--Barbero--Immirzi (first order) variables, the~invariance of the gravitational action under the local rotation of the triad adapted to the spacetime foliation is expressed in the form of a Gauss constraint.

The implementation of this new approach to the cosmological setting leads to define, in a rather rigorous, mathematical way, the concept of a primordial Big Bounce, already hypothesized in the seventies. However, the cosmological implementation of LQG, commonly dubbed loop quantum cosmology (LQC), has the intrinsic limitation that the basic $SU(2)$-symmetry underlying the LQG formulation is unavoidably lost \cite{Cianfrani_Montani_Gaugefixing,Bojowald_2020BLAST} when the minisuperspace dynamics is addressed. This is due to the fact that the homogeneity constraint reduces the cosmological problem to a finite number of degrees of freedom; in~particular, it becomes impossible to perform the local rotation and preserve the structure constants of the Lie algebra associated to the specific isometry group. In this respect, we~could say that LQC requires a sort of gauge fixing of the full $SU(2)$ invariance; see~the analysis in \cite{Cianfrani_Montani_Gaugefixing2}, where this question is explicitly addressed. In addition, the problem of translating the quantum constraints from the full to the reduced level remains still open~\cite{ThiemannBook}.

Despite these limitations, LQC remains an interesting attempt to regularize the cosmological singularity, opening a new perspective on the origin and evolution of the Universe. Furthermore, the so-called ``effective formulation'' of LQC is isomorphic to the implementation of polymer quantum mechanics (PQM) \cite{Corichi_2007,CorichiContinuumLQC,AshtekarEffectiveLQC,Singh_2005EffectiveLQC} to the minisuperspace variables, typically the Universe scale factors. This correspondence allows to investigate some features of the LQC formulation by applying simplified formalisms to more complicated models, thus making them viable \cite{PolymerTAUBBattisti,battisti2009gup,PolymerMatterHossain_2010,PolymerTimeLawrie_2011,PolymerMatterKreienbuehl_2013,Lecian_2013,Cianfrani_2014,PolymerInflationHassan_2015,CyclicalBianchiIMoriconi_2016,SemiclassicalPolymerMatterHassan_2017,AnisotropyBigBounceMoriconi_2017,NaturalInflationAli_2017,CrinoPintaudi,CyclicalBianchiIMontani_2018,LiftinAmbiguitiesBen_Achour_2019,Mantero,stefano,DiAntonio,CascioliWKB,ProtectedSymmetryAchour_2019,Giovannetti_2019,EntropyCampolongo_2020,Federico,Silvia}.

Here, we provide a review of basic well-established results of LQC and more recent analyses in polymer quantum cosmology, with the purpose of better outlining the reliable achievements and the open questions in this sector of the quantum cosmological problem formulation. In order to better compare it with LQC, when presenting the cosmological implementation of PQM, we mainly focus on its effective dynamics, except for the less involved models, where a full quantum analysis is possible. In particular, two recent studies \cite{Federico,Silvia} applied the polymer framework to the formulation of the flat Friedmann--Lema\^itre-Robertson--Walker (FLRW) and Bianchi I models, respectively. The peculiarities of these two analyses lie in the comparison between the polymer quantum dynamics in terms of the Ashtekar--Barbero--Immirzi connections and in terms of volume-like configurational~variables.

In this review, we highlight how the evolution of the quantum Universe is sensitive to the considered set of configurational variables: when real connections are polymerized, the resulting picture resembles that which is commonly dubbed the $\mu_0$ scheme of LQC \cite{BojowaldOriginalLQC,Ashtekar0_2006,Ashtekar_2006}; on the other hand, the use of volume coordinates can be associated to the so-called $\bar{\mu}$ scenario \cite{Ashtekar2_2006,Ashtekar_2008,Ashtekar_2011Review}. In LQC, the difference in these two schemes is due to the cosmological implementation of the area element as a kinematical or a dynamical quantity (in the $\bar{\mu}$ scenario the area gap is rescaled for the momentum variable, i.e., the squared cosmic scale~factor). 

Actually, the $\mu_0$ and $\bar{\mu}$ schemes lead to very different pictures of the primordial quantum Universe: both correspond to a Big Bounce, but, while in the $\bar{\mu}$ scheme the critical energy density is fixed by fundamental constants and the Immirzi parameter only, in the $\mu_0$ dynamics it depends on the initial conditions for the wave packets. The fact that these two very different representations of the early Universe are associated with the polymerization of the two different sets of variables cited above offers an intriguing perspective to better interpret the real physics of the two scenarios and shows how PQM could shed light on the possible shortcomings of the LQC formalism.

We will also provide an interesting comparison of the cosmological implementation of the PQM in the metric representation. In particular, we will compare the analyses in~\cite{CrinoPintaudi,Giovannetti_2019} and \cite{stefano}, where the homogeneous and inhomogeneous Mixmaster dynamics is studied through the polymerization of the standard Misner isotropic variable $\alpha$ and of the Universe volume, respectively: the difference is only in using or not using a logarithm in the definition of the isotropic variable, but the implication is very deep since only the volume representation ensures a bouncing cosmology. Furthermore, questions concerning the chaotic or non-chaotic nature of the semiclassical Bianchi IX dynamics are addressed in some detail. 

Then, we will further present a coherent and detailed discussion of the relations existing between LQC and polymer quantum cosmology, also discussing some of the most relevant open questions, especially concerning the equivalence or non-equivalence of the resulting dynamics in different sets of configurational variables. 

We would like to stress that in this work, we will not consider the basic problem of the implementation of the Copenhagen School interpretation to the Universe quantum dynamics, common to all quantum cosmological formulations. We will briefly present this issue because we think that it is important to keep in mind such difficulties since they could perhaps drive the investigation toward a fully consistent theory of quantum gravity and, hence, quantum cosmology. However, our presentation will escape this puzzling basic interpretative question, as is often implicitly done in the literature. We must remark that there are many other approaches of a different nature that are able to replace the singularity with a Bounce, such as, for example, the Ekpyrotic scenario, massive gravity, and other modified gravity theories; for general reviews of these models, see \cite{OtherBounce1,OtherBounce2}.

The paper is organized as follows. In Section \ref{secgiovanni}, we introduce general features of the cosmological dynamics. We present the difficulties in implementing the Copenhagen interpretation to cosmology, then we describe the classical dynamics of homogeneous models (both isotropic and anisotropic) with some attention to the problem of time and to the definition of a cosmological clock. In Section \ref{secLQC}, we present LQC: first, we briefly summarize the features of LQG that are relevant for its cosmological implementation; then, we discuss in detail the two formulations of isotropic LQC that are the $\mu_0$ and $\bar{\mu}$ schemes, and also the implementation of the latter to the anisotropic sector; and finally, we conclude with a summary of critics and shortcomings that show the need for a different quantum mechanical approach to cosmology. In Section \ref{secPOL}, we present polymer cosmology: we first introduce the formalism of polymer quantum mechanics, and then we focus on its implementation to both the isotropic Universe and the anisotropic Bianchi models in different sets of variables (namely, the Ashtekar variables, the volume-like variables and the Misner-like ones). We conclude this section with a discussion on the results obtained with different sets and also present a possible way to recover an equivalence between them. Finally, in Section \ref{secConcl} we summarize the review and provide some final remarks. Throughout all the paper, we use the natural units $8\pi G=c=\hslash=1$.

\section{Cosmological Quantum Dynamics}
\label{secgiovanni}
The first attempt to implement the canonical quantum gravity approach developed in \cite{deWitt1,deWitt2,deWitt3} to the cosmological setting was due to the analysis proposed in \cite{MisnerQC69}, where the Bianchi IX Universe was studied within certain approximations, and the most relevant properties of its quantum dynamics were elucidated (for extensions of this approach to generic inhomogeneous models, see \cite{IMPONENTE_2004,Imponente_2006,Benini_2004frameindependence,Benini_2006InhomogeneousQuantumMixmaster,BENINI_2008CovariantInhomogeneousMixmaster,BENINI_2008ClassQuantInhomogeneousMixmaster,Kirillov93,Kirillov_2002}, and for a refined numerical study of the Bianchi IX quantization, see \cite{Graham_1995}). 

Before entering some technical aspects of canonical quantum cosmology in the metric formulation, it is mandatory to fix our attention to some intrinsic conceptual difficulties that we meet on the interpretative level, even if we could assume the construction of a Hilbert space and the determination of a suitable time variable to describe the quantum dynamics as solved. 

The standard interpretation of canonical quantum mechanics is due to the so-called Copenhagen school, which postulated some general prescriptions, validated by the analysis of atomic and molecular spectra and was never contradicted by experiments in modern relativistic particle physics. We briefly summarize here the Copenhagen school interpretation via its main statements (very difficult to implement in quantum cosmology). 
\begin{itemize}
    \item The concept of probability to find a physical system into a given state is the ``large numbers'' limit of the frequency by which that state is registered in repeated \mbox{experiments}.
    \item The ``measure'' operation on a given quantum system must be performed by a classical (or better, quasi-classical) observer, who induces a ``collapse'' of the wave function into a specific eigenstate by physically interacting with the quantum environment.
\end{itemize}

When referred to the cosmological setting, both the statements above have a very critical implementation. In fact, on one hand, we observe only one realization of the Universe, and no frequency approximating the probability can be determined; on the other hand, in a quantum Universe, it appears impossible (or at least ambiguous) to speak of a quasi-classical observer. The possibility to recover both the concepts postulated above would require that at least a portion of the Universe be in a quasi-classical state, so that the interaction of these degrees of freedom with the fully quantum ones offers an arena to recover the basic notion of the Copenhagen school interpretation (see, in this respect, the Universe wave function interpretation provided in \cite{Vilenkin}, where such a picture is~investigated).

We conclude by observing that the classical portion of the Universe mentioned above  cannot be identified with the present classical Universe thought of as an ``observer'' of the primordial quantum phases. This claim is supported by the following two considerations.
\begin{itemize}
    \item The information we receive from the quantum Universe in the Planck regime (mediated by the physics of the cosmic microwave background radiation (CMB)) is already a single classical determination of the quantum system, among all the possible ones.
    \item That information cannot be induced by a direct measurement on the primordial Universe, simply because it lives in our past light cone, and no physical interaction between our classical apparatus and the Planckian Universe can take place (even if we were able to detect photons directly emitted in the quantum phase).
\end{itemize}

Thus, in what follows, we will think of the Universe wave function as if it were associated to physical notions in principle, according to the Copenhagen school interpretation, without entering further into how it can be really demonstrated, or which alternative interpretation could be addressed.

\subsection{The Isotropic Universe}
The Robertson--Walker (RW) geometry describing the isotropic Universe is a very simple model, which has only one dynamical degree of freedom due to the high level of symmetry. If on a classical level its employment is well justified by a large number of phenomenological evidences (above all, the isotropy of the CMB temperature), on a quantum level, it appears very close to be just a ``toy model'' deprived of many basic features that more general cosmological models outline. We elucidate the reliability of this apparently strong claim in this subsection and in the following one. 

The main failure of the canonical quantum cosmology as depicted by the Wheeler--DeWitt (WDW) equation in the metric approach is that no removal of the initial singularity emerges in general when the nature of the Universe wave function is elucidated. Let~us now develop some simple technical considerations for the isotropic Universe (for a pioneering analysis, see \cite{Isham75}), limiting our attention to the spatially flat model and adopting as a configurational variable the cubed cosmic scale factor $v(t)=a^3(t)$ (for example, the Universe~volume).

In the Arnowitt--Deser--Misner (ADM) formulation \cite{ADM}, the RW line element reads as follows:
\begin{equation}
	ds^2=\text{N}^2dt^2-v^\frac{2}{3}\left(dx^2+dy^2+dz^2\right)\,, 
	\label{gioba1}
\end{equation}
where we set the speed of light equal to one, and $\text{N}=\text{N}(t)$ denotes the so-called lapse function. The action describing the Hamiltonian dynamics of the isotropic FLRW model takes the following form:
\begin{equation}
	S_{\text{FLRW}}=\int dt\left(P_v\,\dot{v}-\text{N}\mathcal{C}_{\text{FLRW}}\right)\,, 
	\label{gioba2}
\end{equation}
where we have set the space integration on a fiducial volume to unity, $P_v$ is the conjugate momentum to $v$, and the super-Hamiltonian $\mathcal{C}_{\text{FLRW}}$ reads as follows:
\begin{equation}
	\mathcal{C}_{\text{FLRW}}\equiv-\frac{9}{4}\,v\,P_v^2+\frac{1}{3}v\rho(v)\,.
	\label{gioba3}
\end{equation}

When the equation of state for the cosmological fluid takes the form $P = w\rho$ (with $P$ being the pressure and  $w$ a constant parameter), the matter energy density $\rho$ reads as  follows:
\begin{equation}
	\rho(v)=\frac{\rho_0^{(w)}}{v^{1+w}}\,, 
	\label{gioba4}
\end{equation}
with $\rho_0^{(w)} > 0$. Clearly, varying the action with respect to $\text{N}$, we obtain the constraint $\mathcal{C}_{\text{FLRW}} = 0$, which reduces to the Friedmann equation for the isotropic Universe, using the Hamilton equation $\dot{v} = -9vP_v/2$ to express the momentum $P_v$. The existence of a Hamiltonian constraint reflects the possibility to freely choose the time variable (i.e., the form of the lapse function) to describe the system dynamics, according to the general relativity principle. 

The Dirac prescription for the canonical quantization of  a  constrained  theory  consists  of implementing  the phase space variables to canonical operators \cite{Matschull1996Dirac}, leading to the following WDW equation for the isotropic Universe \eqref{gioba3}:
\begin{equation}
	\partial^2_v\psi(v)+\frac{4}{27}\rho(v)\psi(v)=0\,,
	\label{gioba5}
\end{equation}
in which we adopt the natural operator ordering, where momenta are always at the right of coordinate variables.
This equation clearly resembles a time-independent Schr\"{o}dinger equation in the space-like coordinate $v$, and no evolution emerges for the Universe wave function $\psi(v)$. 

If we instead adopt the symmetric operator ordering of \cite{Lulli_Cianfrani_Montani}, i.e., $vP_v^2\rightarrow\hat{P}_v\,v\,\hat{P}_v$, and~introduce the variable $\xi \equiv \ln v$, we arrive to an equation of the  following form:
\begin{equation}
	\partial^2_{\xi}\psi=-\frac{4}{27}\rho_0^{(w)}e^{-(w-1)\xi}\psi\,.
	\label{gioba6}
\end{equation}

For the relevant cosmological case of a ``stiff matter'', corresponding to $w=1$ and de facto mimicking a massless free scalar field (the kinetic component of an inflaton field), we obtain the simple solutions as follows:
\begin{equation}
	\psi(v)\,\propto\,\exp\left(\pm\,i\,\sqrt{\frac{4\rho_0^{(1)}}{27}\,}\,\ln v\right)\,.
	\label{gioba7}
\end{equation}

Indeed, the potential term of many inflationary models can be neglected at the high temperatures of the Planckian regime \cite{WeinbergCosmology,KolbTurner} (we recall that the transition phase responsible for inflation takes place in a classical Universe). The stiff matter is the most rapidly increasing contribution allowed by a causal fluid when the zero volume limit is approached, and it is therefore expected to dominate during the Planckian era. However, we stress that the singularity is also present for all natural values of the parameter $w$ \cite{MontaniPrimordial,KolbTurner}.

The classical Universe has a singular behavior for $v\rightarrow0$ (the Big Bang singularity), and it regularly expands indefinitely for $v\rightarrow \infty$; here, we see that the Universe wave function singles out a qualitatively similar behavior in these two different regimes. Thus, no indications emerge from the WDW equation about the singularity removal, and this turns out to be a general feature of the canonical metric approach. 

\subsection{Internal Clock}
It is clear that it is not possible to construct a Hilbert space for the isotropic Universe discussed above, and therefore, any precise notion of probability density is forbidden in the absence of a well-defined time variable. 

In general, any component of a gravitational system could be identified as a time variable for the classical dynamics, as soon as a specific time gauge is assigned. Such a concept can be retained also at the quantum level in a fully covariant form. Indeed, in the WDW equation, it is possible to identify a given internal degree of freedom as a ``relational time'', by promoting it to the role of a physical clock for the quantum evolution of the remaining gravitational or matter degrees of freedom. The most natural relational time for the isotropic Universe dynamics, and in general for quantum cosmology, is a free massless scalar field $\phi=\phi(v)$, which is expected to be present in the primordial phases of the Universe because of the inflationary paradigm; its energy density increases as $\sim$$v^{-2}$ toward the singularity, which is the fastest growth allowed before the fluid acquires a superluminal sound speed. 

If we replace the stiff matter in 
Equation \eqref{gioba6} with the energy density of $\phi$, i.e.,
\begin{equation}
	\rho_{\phi}=\frac{P_{\phi}^2}{2v^2}\,
	\label{gioba8}
\end{equation}
($P_{\phi}$ being the conjugate momentum to $\phi$), then we arrive to the following $1+1$ Klein--Gordon-like equation:
\begin{equation}
	\left(\partial^2_{\xi}-\frac{2}{27}\partial^2_{\phi}\right)\psi(\xi,\phi)=0\,.
	\label{gioba9}
\end{equation}

The general solution of this equation reads as  follows:
\begin{equation}
	\psi=\int dk_{\xi}A(k_{\xi})\exp\left\{ik_{\xi}\left(\xi\pm\sqrt{\frac{27}{2}}\,\phi\right)\right\}\,.
	\label{gioba10}
\end{equation}

Now, it is possible to adopt $\phi$ as a physical clock for the Universe dynamics, so we can construct localized wave packets of the form \eqref{gioba10}, for instance with a Gaussian weight function $A(k_{\xi})$. If we compare the peak of the Klein--Gordon probability density $i(\psi^*\partial_{\phi}\psi-\psi\partial_{\phi}\psi^*)$ with the classical trajectory $v=v(\phi)$, it is easy to check that there is a very good correspondence, leading to the fact that the singularity is not removed by the canonical quantization of the model. 

Thus, the introduction of the concept of a relational internal matter clock provides a good solution to the problem of time since the zero-eigenvalue Schr\"{o}dinger equation can be interpreted as a Klein--Gordon-like operator in the configurational space. The similarities of the relativistic case and the present WDW equation allow to define a conserved probability density, which retains its positive nature when it is possible to perform the frequency separation (violated when a non-zero potential for the scalar field $\phi$ is present).

We also observe that there exists a clear correspondence between the quantity $\rho_0^1$ in Equation \eqref{gioba6} and the quantum number $k_\xi$, namely, the following:
\begin{equation}
	\rho^1_0 \equiv \frac{27}{4}k_{\xi}^2\, .
	\label{gioba11}
\end{equation}

We conclude by observing that the considerations above regarding the absence of a singularity removal when comparing the classical and quantum evolution are particularly reliable in the present case, in view of the linearity of the dispersion relation for the $1+1$ Klein--Gordon-like equation. Such a property allows to construct localized non-spreading wave packets up to the initial singularity. It is immediate to realize (see below) that a linear dispersion relation is a feature that clearly does not survive when a higher dimensional problem is faced.

\subsection{The Bianchi Universes}
\label{giobaBianchi}
A better understanding of the minisuperspace formulation of canonical quantum gravity in the metric approach is provided by the investigation of the Bianchi Universes. These models generalize the isotropic Universe by preserving the homogeneity constraint and allowing for three different independent scale factors along the three spatial directions. 

The ADM line element of the Bianchi Universes reads as follows:
\begin{equation}
	ds^2=\text{N}^2dt^2-e^{2\alpha}\left(e^{2\beta}\right)_{ab}\,\sigma^a\sigma^b\,, 
	\label{gioba12}
\end{equation}
where the variable $\alpha$ is related to the Universe volume $v$ by the relation $\alpha = (1/3)\ln v$, the matrix $\beta$ parametrizes the anisotropies and has the diagonal form $\beta=\text{diag}(\beta_++\sqrt{3\,}\,\beta_-\,,\,\beta_+-\sqrt{3\,}\,\beta_-\,,\,-2\beta_-)$, while $\sigma_a$ and $\sigma_b$ are the 1-forms describing the specific isometry group under which that Bianchi model is invariant. The variables $(\alpha,\beta_\pm)$ are known as Misner variables, and their usefulness lies in the fact that they make the kinetic term in the Hamiltonian diagonal.

The homogeneity of the space allows to deal with the functions $\text{N}$, $\alpha$, $\beta_+$ and $\beta_-$ as depending on time only. The isotropic limit is recovered for $\beta_+ \equiv \beta_-\equiv 0$, and it is possible only for the three Bianchi models of type I, type V and type IX, corresponding to the flat, negatively and positively curved FLRW model respectively. 

The action of the Bianchi Universes in vacuum reads as follows:
\begin{equation}
	S_\text{B}=\int dt\big(P_{\alpha}\dot{\alpha}+P_+\dot{\beta}_++P_- \dot{\beta}_--\text{N}\mathcal{C}_\text{B}\big)\,, 
	\label{gioba13}
\end{equation}
with
\begin{equation}
	\mathcal{C}_\text{B}\equiv \frac{e^{-3\alpha}}{3(8\pi)^2}\big[-P_{\alpha}^2+P_+^2+P_-^2+3(4\pi)^4e^{4\alpha}\,U_\text{B}(\beta_+,\beta_-)\big]\, .
	\label{gioba14}
\end{equation}

Above, we set to unity the space integral on the fiducial volume and denoted the conjugate momenta to the corresponding variables $\alpha$, $\beta_+$ and $\beta_-$ with $P_{\alpha}$, $P_+$ and $P_-$ respectively. The~potential $U_\text{B}$ is provided by the spatial curvature of the specific model, and it is identically zero for the Bianchi I model only. 
It is immediate to recognize that the WDW equation for the Bianchi Universes takes the following form:
\begin{equation}
	\left[\partial_{\alpha}^2-\partial_{\beta_+}^2-\partial_{\beta_-}^2\right]\psi+3(4\pi)^4e^{4\alpha}\,U_\text{B}(\beta_+,\beta_-)\psi = 0\,, 
	\label{gioba15}
\end{equation}
with $\psi = \psi (\alpha ,\beta_+,\beta_-)$. In this case, there is no need to add a free massless scalar field to identify an internal time variable since the variable $\alpha$ (related to the three-metric determinant) has a different signature with respect to the two anisotropy degrees of freedom $\beta_+$ and $\beta_-$. This is a very general feature of the WDW equation, first investigated in \cite{deWitt1,deWitt2,deWitt3}. 

It is worth noting that the same signature of $\alpha$ and the same wave equation could be found by using the Universe volume $v=e^{3\alpha}$ and adopting the symmetric operator ordering, as done above in Equation \eqref{gioba6}. Hence, we can realize how misleading it was to use the volume of the Universe as a space-like coordinate in the isotropic Universe. This interpretation is clearly possible, but as far as we introduce $\beta_+$ and $\beta_-$ (the real physical degrees of freedom of the cosmological gravitational field), we are naturally led to consider $v$ or $\alpha$ as the most natural internal time variables to describe the quantum Universe evolution. 

The first classical Hamilton equation $\dot{\alpha}=-2\text{N}P_{\alpha}e^{-3\alpha}$ shows that we must require that $P_{\alpha}<0$ in order to deal with the expanding Universe ($\dot{\alpha}>0$). On the contrary, for the collapsing Universe, i.e., $\dot{\alpha}<0$, we need $P_{\alpha}>0$. However, we note that $P_{\alpha}$ is a constant of motion only when the potential term is negligible or when it is exactly zero, as in the Bianchi I model. 

If we set $U_\text{B}\equiv 0$ in the WDW Equation \eqref{gioba15}, we can easily perform the frequency separation, and we obtain the following general solutions:
\begin{equation}
	\psi^{\pm}(\alpha,\beta_\pm)=\int dk_+dk_-\,A(k_+,k_-)\,e^{i(k_+\beta_++k_-\beta_-\mp\sqrt{k_+^2+k_-^2\,}\,\alpha)}\,, 
	\label{gioba16}
\end{equation}
where the suffixes $(+)$ and $(-)$ refer to positive and negative frequency wave functions, respectively.

Since the mean value of the operator $\hat{P}_{\alpha}$ is negative for the positive frequency solution and positive for the negative one (see the sign of its eigenvalues), we are led to identify the expanding Universe with the positive frequency wave packet and vice-versa for the collapsing one. 

Actually, if we consider Gaussian weight $A(k_+,k_-)$, it is possible to construct localized wave packets, both representing the expanding ($\psi^+$) or collapsing ($\psi^-$) dynamics of the Universe. The localized states follow the classical trajectories $\beta_+(\alpha)$ and $\beta_-(\alpha)$, so we are naturally led to claim that the initial singularity is not removed by the canonical quantization of the system also for a Bianchi I model. However, now the dispersion relation contains a square root; therefore, it is no longer linear as it was for the isotropic Universe. As a result, the wave packet spreads toward the singularity (for $\alpha \rightarrow -\infty$), and the localized state cannot be extrapolated asymptotically. This fact prevents a definitive word on what the initial singularity resembles in such a non-localized picture of the Universe. However, we can surely claim that in the Planckian era, the Universe unavoidably becomes a fully quantum system.

\subsubsection*{The Bianchi IX Model}
\label{giobaBIX}
The peculiarity of the Bianchi IX model lies in the chaotic dynamics near the singularity; this has earned it the nickname of the \emph{Mixmaster} model.

The explicit form of the potential $U_{\text{BIX}}(\beta_\pm)$ is the following:
\begin{equation}
\label{V}
U_\text{BIX}(\beta_\pm)=2e^{4\beta_+}\big(\cosh(4\sqrt{3}\beta_-)-1\big)-4e^{-2\beta_+}\cosh(2\sqrt{3}\beta_-)+e^{-8\beta_+}\,.
\end{equation}

The potential walls are steeply exponential and define a closed domain with the symmetry of an equilateral triangle \cite{MisnerMixmaster}. These walls move outwards while approaching the cosmological singularity due to the term $e^{4\alpha}$ in front of the potential in \eqref{gioba14} that increases for $\alpha\rightarrow-\infty$.

The implementation of the ADM reduction allows to describe the Mixmaster dynamics by means of the motion of a pinpoint particle, named the point-Universe, moving in the triangular potential well. So, we solve the constraint \eqref{gioba14} with respect to the momentum conjugate to the chosen time coordinate, here $\alpha$, and then we obtain the reduced ADM~Hamiltonian:
\begin{equation}
\label{HADM}
\mathcal{C}^{\text{ADM}}_\text{BIX}:=-P_{\alpha}=\sqrt{P_+^2+P_-^2+e^{4\alpha}3(4\pi)^4U_\text{BIX}(\beta_\pm)}\,.
\end{equation}

Because of the steepness of the walls, we can consider the point-Universe as a free particle for most of its motion, except when a rebound against one of the three walls occurs. So, by using the free particle approximation $U_\text{BIX}(\beta_\pm)\sim 0$, we can derive the velocity of the point-Universe as follows:
\begin{equation}
\label{beta}
\beta'\equiv\sqrt{{\beta'_+}^2+{\beta'_-}^2}=1\,.
\end{equation}
where in this picture, the anisotropies have the role of the coordinates of the point-Universe.
On the other hand, it can be shown that the potential walls move outwards with velocity $|\beta'_{\text{wall}}|=\frac{1}{2}$, so a rebound is always possible. In particular, every rebound occurs according to the following reflection law:
\begin{equation}
\label{legge}
\frac{1}{2}\sin(\theta^i+\theta^f)=\sin\theta^i-\sin\theta^f\,,
\end{equation}
where $\theta^i$ and $\theta^f$ are the incidence angle and the reflection one to the potential wall normal, respectively. The maximum incidence angle results to be the following:
\begin{equation}
\theta_{\text{max}}\equiv\arccos\bigg(\frac{1}{2}\bigg)=\frac{\pi}{3}\,,
\end{equation}
so the point-Universe always experiences a rebound against one of the three potential walls, thanks to the triangular symmetry of the system.

In conclusion, the ADM reduction procedure in the Misner parametrization maps the dynamics of the Mixmaster Universe into the motion of a pinpoint particle inside a closed two-dimensional domain.  The particle undergoes an infinite series of rebounds against the potential walls while approaching the singularity, and the motion between two subsequent rebounds is a uniform rectilinear one. Once the particle is reflected off one of the walls, the values assumed by the constants of motion change, as well as, thus, the direction of the particle. This way, the trajectory of the point-Universe assumes all possible directions regardless of the initial conditions, giving rise to the chaotic behavior of the Bianchi IX dynamics  near the singularity.

\section{Loop Quantum Cosmology}
\label{secLQC}
The name \emph{loop quantum cosmology} refers to a specific quantum cosmological model, i.e., the quantization of the FLRW spacetime, according to the methods of LQG \mbox{\cite{BojowaldOriginalLQC,Ashtekar0_2006,Ashtekar_2006,SzulcFLRW_2007,VanderslootFLRW_2007,Ashtekar2_2006,Ashtekar_2008,Ashtekar_2011Review,PawowskiRadiationFLRW_2014}}. More in general, it is often also used to indicate all cosmological models that are quantized through LQG procedures \cite{Bojowald_2003HomogeneousLQC,Bojowald_2004HomogeneousLQCSpinConnection,Bojowald_2004ChaosSuppression,Bojowald_2004BianchiIXLQC,Wilson-Ewing_BianchiI,Wilson-Ewing_BianchiII,Wilson-Ewing_BianchiIX,Wilson-Ewing_BianchiIcomplete,InhomogeneousGowdyLQC_2010,Wilson_Ewing_2016,CorichiBianchiIXLQC_2016,Wilson_Ewing_2019}. Note, however, that this implies that LQC is \emph{not} the cosmological sector of LQG: the internal symmetries of the formalism used to derive the Loop quantization of general relativity do not allow the usual reduction of the Wheeler Superspace to the cosmological minisuperspaces. However, it is possible to implement the quantization procedure of LQG to a spacetime that is already reduced to a minisuperspace model; this is, indeed, the scope of LQC.

In this section, we briefly introduce the formalism of kinematical LQG and show in detail its implementation to the isotropic Universe in both the old (``standard'') and new (``improved'') prescriptions of LQC; we also present the work that was done on the Loop quantization of anisotropic models and then conclude with a short description of critiques and shortcomings.

\subsection{Loop Quantum Gravity}
LQG was developed in the 1990s \cite{Rovelli_Smolin_LQG,Rovelli_1995DiscretenessAreaVolume,Rovelli_1995SpinNetworks,Ashtekar_1995LQG,Thiemann_1998QSD1,Thiemann_1998QSD2,QuantumGeomI,QuantumGeomII} and remains today the best attempt at a background-independent quantization of general relativity (GR) (for recent, more~comprehensive reviews, see \cite{LQG30years,LQGReview_2021}). The requirement of background independence calls for a reformulation of GR through new formalisms that allow this quantization process: the~formalisms of geometrodynamics and of Gauge theories.

GR was reformulated as a $SU(2)$ Gauge theory by Ashtekar \cite{AshtekarVariables,AshtekarHamiltonianGR} by performing a $3+1$ splitting of spacetime and using as fundamental conjugate variables a connection $A^i_a$ and an electric field $E^a_i$, which take values in the Lie algebra $su(2)$ of $SU(2)$. The symmetry group is generated by the local $SU(2)$ gauge transformations that leave points of the manifold invariant, and the theory is covariant with respect to diffeomorphisms. The constraints represent the simplest covariant functions that contain $(A^i_a,E^a_i)$ at most, quadratically, and that do not reference any background quantity:
\begin{subequations}
\begin{equation}
    \mathcal{G}_i=\mathcal{D}_aE^a_i=0,\quad\mathcal{D}_aE^a_i=\partial_aE^a_i+\epsilon_{ij}^kA^j_aE^a_k,
    \label{cgauss}
\end{equation}
\begin{equation}
    \mathcal{C}_a^\text{LQC}=E^b_iF^i_{ab}=0,\quad F^i_{ab}=2\partial_{[a}A^i_{b]}+\epsilon^i_{jk}A^j_aA^k_b,
    \label{cdiffeo}
\end{equation}
\begin{equation}
    \mathcal{C}^\text{LQC}=\epsilon^{ij}_kE^a_iE^b_jF^k_{ab}=0,
    \label{cscalar}
\end{equation}
\label{LQGclassicalconstraints}
\end{subequations}
which are respectively the Gauss constraint (generator of the $SU(2)$ rotations), the diffeomorphism constraint (generator of spatial diffeomorphisms) and the scalar Hamiltonian constraint (generator of time evolution).

Before moving on to quantization, the canonical fields $(A^i_a,E^a_i)$ must be appropriately smeared, also because it is not possible to construct an operator corresponding to the connection \cite{Ashtekar_1995LQG}. This smearing is achieved, defining holonomies of the connections along an edge $\ell$ and fluxes of the electric field across a bidimensional surface $S$:
\begin{equation}
    h_\ell[A]=\mathbb{P}\,\exp\left(\int_\ell A^i_a\tau_id\ell^a\right),\quad\Phi_S[E]=\int_SE^a_i\tau^idS_a,
    \label{LQGholofluxaction}
\end{equation}
where $\tau^i$ are the $SU(2)$ generators. Note that the holonomies have a one-dimensional support; their trace for a closed edge results in the so-called Wilson loop that gives the theory its name.

Now, the quantum kinematics is obtained by promoting these objects to operators and defining their commutator; a very important consequence of the requirement of background independence, i.e., of diffeomorphism invariance, is that the holonomy-flux algebra results in having a unique representation and, therefore, a unique Hilbert space $\ham^{\text{kin}}$. This is called a \emph{spin network} space, defined as a graph $\Gamma$, made of a finite number $L$ of edges (each with a half-integer spin-quantum number $j_L$) and a finite number $n$ of nodes (each with an intertwiner $i_n$). The basis vectors of this Hilbert space are, therefore, \emph{spin network states} denoted as $\ket{S}=\ket{\Gamma,j_L,i_n}$; wave functions on the spin network are cylindrical functionals $\Psi_\Gamma[A]=\psi(h_{\ell_1}[A],h_{\ell_2}[A],\dots,h_{\ell_L}[A])$, which depend on the connections only through holonomies and are square-integrable with respect to the Haar measure.

A key result of the kinematical framework of LQG is the quantization of the geometrical operators of area and volume. For example, the area operator and its action on a functional can be defined through the flux operator \eqref{LQGholofluxaction}, and the eigenvalues result in being dependent on how many edges of $\Gamma$ intersect the considered surface. In particular, the~smallest non-zero eigenvalue of the area operator is a constant quantity depending on fundamental constants and on the Immirzi parameter only; it is called the \emph{area gap} $\Delta$, and is a key parameter of the theory. Note that this result is purely kinematical \cite{QuantumGeomI,Rovelli_1995DiscretenessAreaVolume,Rovelli_1995SpinNetworks}.

The dynamics is derived through the implementation of the operators corresponding to the constraints \eqref{LQGclassicalconstraints}; in order to do this, they must first be expressed in terms of the fundamental variables, i.e., holonomies and fluxes, and then quantized, usually through the Dirac procedure \cite{Matschull1996Dirac}. We will not implement the dynamics here, but will show the procedure directly in the cosmological sector of the following sections.

\subsection{Standard Loop Quantum Cosmology \label{standardLQC}}
We now introduce the ``old'' procedure to implement the quantization methods of LQG on the homogeneous and isotropic FLRW model \cite{BojowaldOriginalLQC,Ashtekar0_2006,Ashtekar_2006}. Note that the Gauss constraint \eqref{cgauss} and the diffeomorphism one \eqref{cdiffeo} are automatically satisfied by the symmetries of the model; therefore, we have to deal only with the scalar constraint \eqref{cscalar} which will be given the suffix ``grav'' to distinguish it from the matter Hamiltonian~$\mathcal{C}_\phi$.

\subsubsection{Classical Phase Space} 
The standard classical procedure in a flat, isotropic, open model is to introduce an elementary cell $\mathcal{V}$ and restrict all integrations to its volume $V_0$ calculated with respect to a fiducial metric $^0q_{ab}$. Given the symmetries of the model, the gravitational phase space variables $(A^i_a,E^a_i)$ can be expressed as follows:
\begin{equation}
A^i_a=c\, V_0^{-\frac{1}{3}}\,\,^0\omega^i_a,\quad E^a_i=p\, V_0^{-\frac{2}{3}}\sqrt{\text{det}(^0q_{ab})\,}\,\,^0e^a_i,
\label{definecp}
\end{equation}
where $(^0\omega^i_a, ^0e^a_i)$ are a set of orthonormal co-triads and triads adapted to $\mathcal V$ and compatible with $^0q_{ab}$. Therefore, the gravitational phase space becomes two-dimensional with fundamental variables $(c, p)$, defined to be insensitive to (positive) rescaling transformations of the fiducial metric and whose physical meaning is obtained through their relation with the cosmic scale factor $a(t)$: $c\propto\dot{a},\,\abs{p}\propto a^2$. The fundamental Poisson brackets are independent on the fiducial volume $V_0$ and are given by the following:
\begin{equation}
\pb{c}{p}=\frac{\gamma}{3}.
\label{poissoncp}
\end{equation}

This is the classical cosmological phase space that constitutes the starting point of LQC.

\subsubsection{Kinematics} The quantum theory is constructed, following Dirac, by firstly giving a kinematical description through the identification of elementary observables that have unambiguous operator analogs. LQC can be constructed following the procedure of the full theory: the~elementary variables of LQG are holonomies of the connections and fluxes of the fields, and their natural equivalent in this setting are holonomies $h^{\lambda}$ along straight edges $(\lambda\,\,^0e^a_k)$ and the momentum $p$ itself. Since the holonomy along the $i$th edge is given by
\begin{equation}
h^{\lambda}_i(c)=\cos\frac{\lambda c}{2}\,\mathbb I+2\sin\frac{\lambda c}{2}\,\tau_i,
\label{holonomyc}
\end{equation}
where $\mathbb I$ is the identity matrix, the elementary configurational variables can be taken to be the almost periodic functions $N_\lambda(c)=e^{i\frac{\lambda c}{2}}$ and the momentum $p$.

The Hilbert space $\ham_{\text{grav}}^{\text{kin}}$ is the space $L^2(\mathbb{R}_B, d\mu_H)$ of square integrable functions on the Bohr compactification of the real line endowed with the Haar measure.  It is convenient to work in the $p$-representation, in which eigenstates of $\hat p$ are kets $\ket{\mu}$ labeled by a real number and are orthonormal; the fundamental variables are promoted to operators acting~as follows:
\begin{subequations}
\begin{equation}
\hat{N}_\lambda(c)\ket{\mu}=\widehat{e^{i\frac{\lambda c}{2}}}\ket{\mu}=\ket{\mu+\lambda},
\label{actionofec}
\end{equation}
\begin{equation}
\hat{p}\ket{\mu}=\frac{\gamma}{6}\,\mu\,\ket{\mu}.
\label{actionofp}
\end{equation}
\label{actionofecp}
\end{subequations}

\subsubsection{Dynamics} The dynamics is defined by the introduction of an operator on $\ham_{\text{grav}}^{\text{kin}}$ corresponding to the Hamiltonian constraint $\mathcal{C}_{\text{grav}}^\text{LQC}$ shown in \eqref{cscalar}. Given the absence of the operator $\hat{c}$, this~must be done by returning to the integral expression of the constraint and expressing it as function of our fundamental variables before quantization. The gravitational Hamiltonian constraint of GR in the flat case becomes the following:
\begin{equation}
\mathcal{C}_{\text{grav}}^\text{LQC}=-\frac{1}{\gamma^2}\int_{\mathcal{V}}d^3x\,\text{N}\,\epsilon_i^{jk}\,F^i_{ab}\,e^{-1}\,E^a_j\,E^b_k,
\label{LQCclassicalconstraint}
\end{equation}
where $e=\sqrt{\abs{\text{det}\,E}}$, $\text{N}_i=0$ due to isotropy, and $\text{N}$ does not depend on spatial coordinates so it can be set to $1$ without loss of generality. Using the Thiemann strategy \cite{Thiemanntrick}, the term $\epsilon_i^{jk}\,e^{-1}\,E^a_j\,E^b_k$ can be written as follows:
\begin{equation}
\sum_k\frac{4\text{sgn}(p)}{\gamma\lambda V_0^{\frac{1}{3}}}\,\,^0\epsilon^{abc}\,\,^0\omega^k_c\,\Tr\left(h^\lambda_k\,\pb{(h^\lambda_k)^{-1}}{V}\,\tau_i\right),
\end{equation}
where $V=\abs{p}^{\frac{3}{2}}$ is the volume function on the phase space; for the field strength $F^i_{ab}$ we follow the standard strategy used in gauge theory of considering a square of side $\lambda\,V_0^{\frac{1}{3}}$ in the $ij$ plane spanned by two of the triad vectors and defining the curvature component as follows:
\begin{equation}
F^k_{ab}=-2\lim_{\lambda\to0}\,\Tr\left(\frac{h^\lambda_{ij}-1}{\lambda^2\,V_0^{\frac{2}{3}}}\right)\,\tau^k\,^0\omega^i_a\,^0\omega^j_b,
\end{equation}
where the holonomy around the square is simply the product of the holonomies along its sides: $h^\lambda_{ij}=h^\lambda_i h^\lambda_j (h^\lambda_i)^{-1} (h^\lambda_j)^{-1}$.

Given these expressions, the gravitational constraint can be written as the limit of a $\lambda$-dependent constraint that is now expressed entirely in terms of holonomies and $p$, and~can therefore be now promoted to the operator as follows:
\begin{gather}
\mathcal{C}_{\text{grav}}^\text{LQC}=\lim_{\lambda\to0}\mathcal{C}^{\lambda}_{\text{grav}},
\label{lambdalimit}
\\
\mathcal{C}^{\lambda}_{\text{grav}}=-\frac{4\,\text{sgn}(p)}{\gamma^3\lambda^3}\sum_{ijk}\epsilon^{ijk}\Tr\left(h^\lambda_{ij}\,\pb{(h^\lambda_k)^{-1}}{V}\right),
\\
\hat{\mathcal{C}}^{\lambda}_{\text{grav}}=\frac{24i\,\text{sgn(p)}}{\gamma^3\lambda^3}\,\sin[2](\lambda c)\,\hat{O}(\lambda),
\label{Clambda}
\\
\hat{O}(\lambda)=\sin\frac{\lambda c}{2}\hat{V}\cos\frac{\lambda c}{2}-\cos\frac{\lambda c}{2}\hat{V}\sin\frac{\lambda c}{2},
\end{gather}
where the action of the volume operator (acting simply as $\widehat{\abs{p}^{\frac{3}{2}}}$), of the holonomy operators and of sine and cosine functions can be easily derived from \eqref{actionofecp}. Note that in the promotion of the Hamiltonian constraint to a quantum operator, a specific discretization choice is made among many possibilities. This is a delicate point for the derivation of LQC, and as explained in later sections, it is addressed in \cite{Bojowald_2020BLAST,Haro}.

Now, in LQC, the limit $\lambda\to0$ does not exist by construction. This can be interpreted as a reminder of the underlying quantum geometry, where the area operator has a discrete spectrum with a smallest non-zero eigenvalue corresponding to the area gap $\Delta$. It is, therefore, incorrect to let $\lambda$ go to zero because in full LQG, the area of the $ij$ square cannot be zero; as a consequence, $\lambda$ must be set to a fixed positive value $\mu_0$ that can be appropriately related to the area gap by considering that the holonomies are eigenstates of the area operator $\hat{A}=\widehat{\abs{p}}$ and demanding that the eigenvalue be exactly equal to $\Delta$:
\begin{equation}
\hat{A}\,h^{\mu_0}_{ij}(c)=\frac{\gamma\mu_0}{6}\,h^{\mu_0}_{ij}(c)=\Delta\,h^{\mu_0}_{ij}(c).
\end{equation}

The operator corresponding to the Hamiltonian constraint can be now defined as the $\lambda$-dependent operator \eqref{Clambda} with $\lambda=\mu_0$:
\begin{equation}
\hat{\mathcal{C}}_\text{grav}^\text{LQC}=\hat{\mathcal{C}}^{\mu_0}_\text{grav}.
\end{equation}

The final step is to make this operator self-adjointed by either taking its self-adjoint part or by symmetrically redistributing the sine operator as follows:
\begin{subequations}
\begin{equation}
\hat{\mathcal{C}}_{\text{grav}}^{\text{LQC(1)}}=\frac{1}{2}\Big(\hat{\mathcal{C}}_{\text{grav}}^\text{LQC}+(\hat{\mathcal{C}}_{\text{grav}}^\text{LQC})^{\dagger}\Big),
\end{equation}
\begin{equation}
\hat{\mathcal{C}}_{\text{grav}}^{\text{LQC(2)}}=\frac{24i\,\text{sgn(p)}}{\gamma^3\mu_0^3}\,\sin(\mu_0 c)\,\hat{O}(\mu_0)\,\sin(\mu_0 c).
\end{equation}
\end{subequations}

The Ashtekar school uses the second one, but both are equivalent and yield similar results.

Now we can introduce matter in the form of a massless scalar field $\phi$ obeying an Hamiltonian constraint of the  following form:
\begin{equation}
\hat{\mathcal{C}}_\phi=\widehat{\,\abs{p}^{-\frac{3}{2}}\,}\,\hat{P}_{\phi}^2,
\label{Cmatt}
\end{equation}
where $P_{\phi}$ is the momentum conjugate to $\phi$. Physical states $\Psi(\mu,\phi)$ are the solutions of the total constraint as  follows:
\begin{equation}
(\hat{\mathcal{C}}_{\text{grav}}^\text{LQC}+\hat{\mathcal{C}}_\phi)\,\Psi(\mu,\phi)=0.
\label{totalconstraint}
\end{equation}

In the classical theory, the field does not appear in the matter part of the Hamiltonian; this~leads to its conjugate momentum $P_\phi$ being a constant of motion and to $\phi$ being able to play the role of emergent internal time. In quantum cosmology in general, this choice of relational time is the most natural one because near the classical singularity, a monotonic behavior of $\phi$ as a function of the isotropic scale factor $a(t)$ always appears. The constraint \eqref{totalconstraint} can then be considered an evolution equation with respect to this internal time $\phi$ and can be recast in a Klein--Gordon-like form, thus allowing for the usual separation into positive and negative frequency subspaces. Once that is done, the procedure to extract physics from the model is: to introduce an inner product on the space of solutions of the constraint to obtain the physical Hilbert space $\ham^{\text{phy}}$; to isolate classical Dirac observables to be promoted to a self-adjoint operator on $\ham^{\text{phy}}$; to use them to construct wave packets that are semiclassical at late times; and to evolve them backwards in time using the constraint~itself.

After the internal time procedure, the constraint \eqref{totalconstraint} takes the  following form:\vspace{6pt}
\begin{equation}
\pdv[2]{\Psi}{\phi}=\frac{1}{B}\Bigg(\mathcal{C}^+(\mu)\Psi(\mu+4\mu_0,\phi)+\mathcal{C}^0(\mu)\Psi(\mu,\phi)+\mathcal{C}^-(\mu)\Psi(\mu-4\mu_0,\phi)\Bigg)=-\Theta(\mu)\Psi(\mu,\phi),
\label{reducedconstraint}
\end{equation}

\begin{subequations}
\begin{equation}
\mathcal{C}^+(\mu)=\frac{1}{72\abs{\mu_0}^3}\,\abs\Big{\abs{\mu+3\mu_0}^\frac{3}{2}-\abs{\mu+\mu_0}^\frac{3}{2}},
\end{equation}
\begin{equation}
\mathcal{C}^-(\mu)=\mathcal{C}^+(\mu-4\mu_0),
\end{equation}
\begin{equation}
\mathcal{C}^0(\mu)=-\mathcal{C}^+(\mu)-\mathcal{C}^-(\mu),
\end{equation}
\end{subequations}
where $B=B(\mu)$ is the eigeinvalue of the inverse volume operator appearing in the matter constraint \eqref{Cmatt}:
\begin{subequations}
\begin{equation}
\widehat{\abs{p}^{-\frac{3}{2}}}\Psi(\mu,\phi)=\left(\frac{6}{\gamma}\right)^{\frac{3}{2}}B(\mu),
\end{equation}
\begin{equation}
B(\mu)=\left(\frac{2}{3\mu_0}\right)^6\Big(\abs{\mu+\mu_0}^{\frac{3}{4}}-\abs{\mu-\mu_0}^{\frac{3}{4}}\Big)^6.
\label{Bofmu}
\end{equation}
\end{subequations}

The operator $\Theta(\mu)$ on the right-hand side of \eqref{reducedconstraint} is a difference operator, as opposed to the differential character of the operator that appears in the equivalent equation of the WDW theory \cite{deWitt1,deWitt2,deWitt3}. This allows for the space of physical states to be naturally superselected into different sectors that can be analyzed separately.

In the choice of observables, classical considerations are helpful: it is possible to choose the conjugate momentum to the field since it is a constant of motion, and the value of $p$ at a fixed instant $\phi_0$. The set $(P_\phi, p|_{\phi_0})$ uniquely determines a classical trajectory; therefore, it constitutes a complete set of Dirac observables in the quantum theory. The operators act~as follows:
\begin{subequations}
\begin{equation}
\widehat{\,\,\abs{p}_{\phi_0}}\Psi(\mu,\phi)=e^{i\sqrt{\Theta(\mu)\,}\,(\phi-\phi_0)}\,\abs{\mu}\,\Psi(\mu,\phi_0),
\end{equation}
\begin{equation}
\hat{P}_{\phi}\Psi(\mu,\phi)=-i\,\pdv{\Psi(\mu,\phi)}{\phi},
\end{equation}
\end{subequations}
where $\Psi(\mu,\phi_0)$ is the initial configuration, i.e., the wave function calculated at a fixed initial time $\phi_0$ and the absolute value $\abs{\mu}$ is due to the fact that states are symmetric under the action of the parity operator $\hat{\Pi}$.

\begin{figure}[H]
    \centering
    \includegraphics[width=0.8\linewidth]{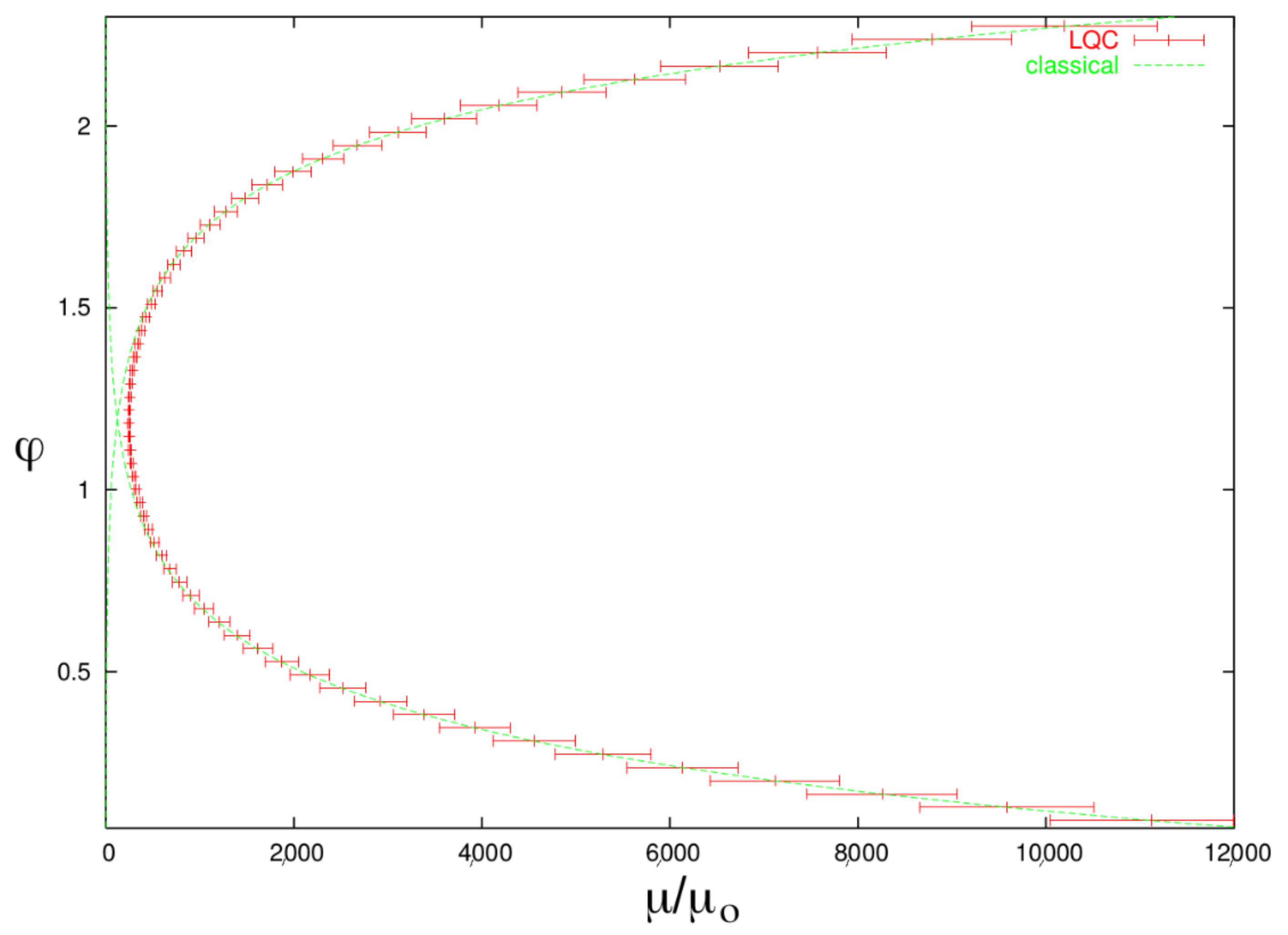}
    \caption{Expectation values and dispersion of $\abs{\mu}$ (red horizontal bars) in terms of $\mu_0$ as function of time $\varphi$ for a coherent state, compared with classical trajectories (green dashed lines). Image from \cite{Ashtekar_2006}.}
    \label{Ashtekarmu0}
\end{figure}

The evolution of wave packets is then carried out numerically. In the following, we briefly summarize the results that are relevant for the resolution of the singularity. For a more detailed analysis of all resulting properties, see \cite{Ashtekar0_2006,Ashtekar_2006}.
\begin{itemize}
\item Singularity resolution: an initially semiclassical state remains sharply peaked around the classical trajectories and the expectation values of the Dirac observables are in good agreement with their classical counterparts for most of the evolution when coherent states are considered. However, when the matter density approaches a critical value, the state bounces from the expanding branch to a contracting one with the same value of $\ev{\hat{P}_\phi}$, as shown in Figure \ref{Ashtekarmu0}. This occurs in every sector and for any choice of $P_\phi\gg 1$, universally solving the singularity by replacing the Big Bang with a Big~Bounce.
\item Critical density: the critical value of the matter density results in being inversely proportional to the expectation value $\ev{\hat{P}_\phi}$ and can, therefore, be made arbitrarily small by choosing a sufficiently large value for $P_\phi$. This fact, besides being physically unreasonable because it could imply departures from the classical trajectories well away from the Planck regime, becomes even more problematic in the case of a closed model: the point of maximum expansion depends on $P_\phi$ as well. In order to have a bounce density comparable with that of Planck, a very small value is needed, but in that case, the Universe would never become big enough to be considered classical; on the other hand, a closed Universe that grows to become classical needs a large value of $P_\phi$ but would have a bounce density comparable with, for example, that of water.
\end{itemize}

This framework, although it successfully solves the singularity, has, therefore, a very important drawback and needs to be substantially improved.

\subsection{Improved Loop Quantum Cosmology \label{improvedLQC}}
In this section, we present the new scheme introduced by the Ashtekar school in \cite{Ashtekar2_2006,Ashtekar_2008} that improves on the standard LQG procedure.

The idea is that the quantization of the area operator must refer to \emph{physical} geometries. Therefore, when performing the limit \eqref{lambdalimit} needed to construct the gravitational constraint, we should shrink the $ij$ square until its area reaches $\Delta$ as measured with respect to the physical metric instead of the fiducial one. The area of the faces of the elementary cell is simply $\abs{p}$, and each side of the square is $\lambda$ times the edge of the cell; with this consideration, the parameter $\lambda$ now becomes a function $\bar{\mu}(p)$ given by the following:
\begin{equation}
\bar{\mu}^2\,\abs{p}=\Delta.
\label{definemubar}
\end{equation}

This means that the curvature operator now depends both on the connection and the geometry, whereas with the previous $\mu_0$ scheme, it depended on the connection only. As~a consequence, more care is needed in the definition of the exponential operator because now, $e^{i\frac{\bar{\mu}c}{2}}$ depends also on $p$.

By using geometric considerations, we can make a comparison with the Schr\"odinger representation and set the following:
\begin{equation}
\widehat{e^{i\frac{\bar{\mu}c}{2}}}\,\Psi(\mu)=e^{\bar{\mu}\dv{\mu}}\,\Psi(\mu),
\end{equation}
i.e., the exponential operator translates the state by a unit affine parameter distance along the integral curve of the vector field $\bar{\mu}\dv{\mu}$. The affine parameter along this vector field is given by the following:
\begin{equation}
\nu=K\,\text{sgn}(\mu)\,\abs{\mu}^{\frac{3}{2}},
\label{definev}
\end{equation}
with $K=\frac{2\sqrt{2}}{3\sqrt{3\sqrt{3}\,}}$. Since $\nu(\mu)$ is an invertible and smooth function of $\mu$, the action of the exponential operator is well-defined; however, its expression in the $\mu$-representation is very complicated because the variable $\mu$ is not well-adapted to the vector field $\bar{\mu}\dv{\mu}$. It is therefore useful to change the basis from $\ket{\mu}$ to $\ket{\nu}$; in this representation, the action of the exponential operator takes the  following extremely simple form:
\begin{equation}
\widehat{e^{i\frac{\bar{\mu}c}{2}}}\,\Psi(\nu)=\Psi(\nu+1).
\label{vinv+1}
\end{equation}

The kets $\ket{\nu}$ still constitute an orthonormal basis on $\ham_\text{grav}^\text{kin}$ and are eigenvectors of the volume operator:
\begin{equation}
\hat{V}\ket{\nu}=\left(\frac{\gamma}{6}\right)^{\frac{3}{2}}\frac{\abs{\nu}}{K}\,\ket{\nu}.
\end{equation}

The gravitational constraint can now be constructed in the same way as before.

The matter constraint has the same form \eqref{Cmatt} of the standard case; therefore, it is sufficient to express the inverse volume eigenvalues \eqref{Bofmu} in terms of $\nu$:
\begin{equation}
B(\nu)=\left(\frac{3}{2}\right)^3K\,\abs{\nu}\,\,\abs\Big{\abs{\nu+1}^{\frac{1}{3}}-\abs{\nu-1}^{\frac{1}{3}}}^3.
\label{Bofv}
\end{equation}

Repeating the same steps of the standard case, the total constraint can again be expressed as a difference operator but this time in terms of $\nu$:
\begin{equation}
\pdv[2]{\Psi}{\phi}=\frac{1}{B}\Bigg(\mathcal{C}^+(\nu)\Psi(\nu+4,\phi)+\mathcal{C}^0(\nu)\Psi(\nu,\phi)+\mathcal{C}^-(\nu)\Psi(\nu-4,\phi)\Bigg)=-\Theta(\nu)\Psi(\nu,\phi),
\label{improvedconstraint}
\end{equation}
\begin{subequations}
\begin{equation}
\mathcal{C}^+(\nu)=\frac{3K}{64}\,\abs{\nu+2}\,\,\abs\Big{\abs{\nu+1}-\abs{\nu+3}},
\end{equation}
\begin{equation}
\mathcal{C}^-(\nu)=\mathcal{C}^+(\nu-4),
\end{equation}
\begin{equation}
\mathcal{C}^0(\nu)=-\mathcal{C}^+(\nu)-\mathcal{C}^-(\nu).
\end{equation}
\end{subequations}

The old operator $\Theta(\mu)$ in \eqref{reducedconstraint} involves steps that are constant in the eigenvalues of $\hat{p}$, while the new one $\Theta(\nu)$, called \emph{improved constraint}, involves steps that are constant in eigenvalues of the volume operator $\hat{V}$. In the $\ket{\mu}$ basis, these steps vary, becoming larger for smaller $\mu$ and diverging for $\nu=0$; however the constraint is well-defined since the operators acting on the state $\ket{\nu=0}$ are well-defined as well.

Regarding the Dirac observables, it is sufficient to substitute $p|_{\phi_0}$ with the volume $\nu|_{\phi_0}$, and the set $(P_\phi,\nu|_{\phi_0})$ is again complete. Therefore, the action of the correspondent operators is
\begin{subequations}
\begin{equation}
\widehat{\,\,\abs{\nu}_{\phi_0}}\Psi(\nu,\phi)=e^{i\sqrt{\Theta(\nu)\,}\,(\phi-\phi_0)}\,\abs{\nu}\,\Psi(\nu,\phi_0),
\end{equation}
\begin{equation}
\hat{P}_{\phi}\Psi(\nu,\phi)=-i\pdv{\Psi(\nu,\phi)}{\phi}.
\end{equation}
\end{subequations}

\begin{figure}[H]
    \centering
    \includegraphics[width=0.8\linewidth]{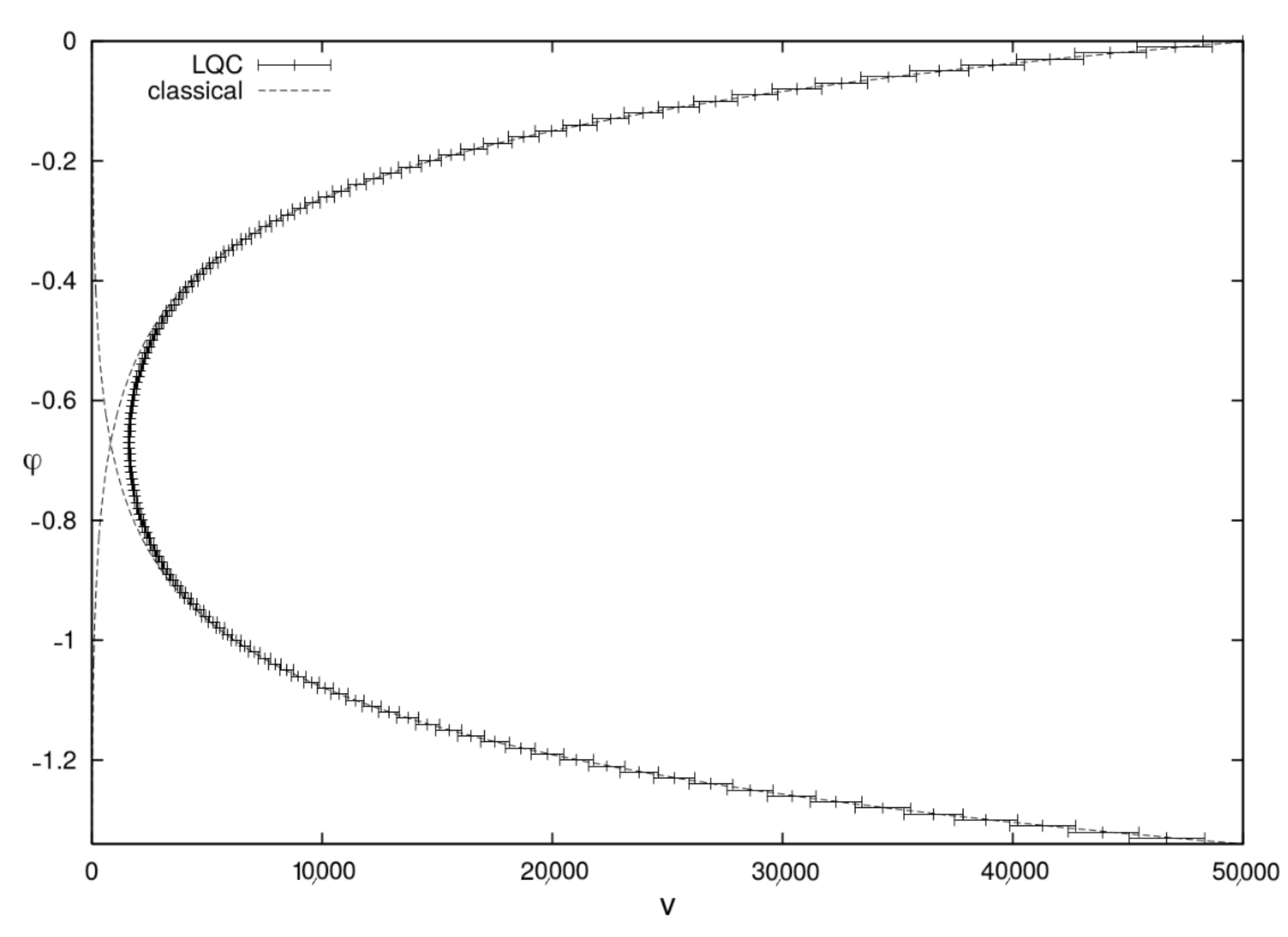}
    \caption{Expectation values and dispersion of $v$ as function of time $\varphi$ (horizontal bars) for a coherent state, compared with the classical trajectories (dashed lines). Image from \cite{Ashtekar2_2006}.}
    \label{Ashtekarmubar}
\end{figure}

After numerical calculations, the improved framework yields the following results.
\begin{itemize}
\item Singularity resolution: also in this case, the states remain sharply peaked throughout all the evolution, and the expectation values of the Dirac observables calculated on coherent states follow the classical trajectory up to a critical value of the energy density; when that value is approached, the states jump to a contracting branch and undergo a quantum bounce instead of following the classical trajectory into the singularity, as~shown in Figure \ref{Ashtekarmubar}.
\item Critical density: the real improvement of the new scheme is that the numerical value of the bounce density is independent of $\ev{\hat{P}_\phi}$ and is the same in all simulations, given by $\rho_\text{crit}\approx0.82\rho_P$. The behavior of the energy density was also studied independently from the evolution of wave packets by analyzing the evolution of the density operator defined as follows:
\begin{equation}
\hat{\rho}_\phi=\widehat{\left(\frac{P_\phi^2}{2V^2}\right)},
\label{densityoperator}
\end{equation}
and it was found that in all quantum solutions, the expectation value $\ev{\hat{\rho}_\phi}$ is bounded from above by the same value $\rho_\text{crit}$.
\end{itemize}

It is shown that the absolute value of the critical density is not modified even when a non-zero cosmological constant is included in the model \cite{Ashtekar2_2006}. The physical understanding of this phenomenon is given by an effective description obtained through a semiclassical~limit.

The improved scheme is able to overcome the main weakness of old standard LQC through a physically motivated modification in the construction procedure of the quantum gravitational constraint. This is the model currently referred to when talking about LQC and on which all subsequent literature is based. Indeed, this model has allowed for a series of phenomenological predictions; they are not part of the aim of this paper, but we present some examples below.

The $\bar{\mu}$ scheme made it possible to perform thermodynamical analyses where the Loop-quantized FLRW model is considered a thermodynamical system, and the energy density and pressure are given a precise thermodynamical meaning \cite{T1,T2}; the~computations for the duration of the inflationary de Sitter phase give results consistent with the minimum amount of e-folds necessary to solve the paradoxes (although slightly higher, depending on some parameters, such as the shear at the Bounce in anisotropic models), predict a phase of deflation in the contracting branch and also allow the extension of the standard inflationary paradigm from the Planck scale up to the onset of slow roll inflation, yielding novel effects, such as non-Gaussianities \cite{M2,I1,I2,I3,I4,I5}. Most importantly, the improved scheme allows for a computation of the primordial power spectrum through two main methods: in the first, the implementation of holonomy corrections as a deformed algebra yields a slightly blue-tilted scale invariant spectrum followed by oscillations and an exponential behavior in the ultraviolet \cite{D1,D2}; in the other, a test field approximation is used such that the evolution of tensor modes on any background quantum geometry is completely equivalent to that of the same modes propagating on a smooth but quantum-corrected metric called ``dressed metric”, and it is able to recover the red-titled ultraviolet power spectrum of classical cosmology \cite{M1,M2}. For a more detailed comparison between the two methods, see \cite{C1,C2,C3}.

\subsection{Effective Dynamics}
The semiclassical limit of LQC, i.e., the inclusion of quantum corrections in the classical dynamics, can be obtained through a geometric formulation of quantum mechanics where the Hilbert space is treated as an infinite-dimensional phase space \cite{Singh_2005EffectiveLQC}. In simpler cases with coherent states that are preserved by the full quantum dynamics, the resulting Hamiltonian coincides with the classical one; however, in more general systems it is possible to choose suitable semiclassical states that are preserved up to a desired accuracy (e.g., in~a $\hbar$ expansion), and the corresponding effective Hamiltonian preserving this evolution is generally different from the classical one \cite{Taveras_2008}.

\subsubsection{Effective $\mu_0$ Scheme} In our model with a massless scalar field, the leading order quantum corrections yield an effective Hamiltonian constraint for the $\mu_0$ scheme in the  following form:
\begin{equation}
\frac{\mathcal{C}_{\text{eff}}^{\mu_0}}{2}=-\frac{3}{\gamma^2\mu_0^2}\abs{p}^{\frac{1}{2}}\sin[2](\mu_0 c)+\frac{1}{2}B(\mu)P_{\phi}^2,
\end{equation}
where $B(\mu)$ is given by \eqref{Bofmu} and for $\mu\gg\mu_0$ can be approximated as follows:
\begin{equation}
B(\mu)=\left(\frac{6}{\gamma}\right)^{\frac{3}{2}}\abs{\mu}^{-\frac{3}{2}}\left(1+\frac{5}{96}\frac{\mu_0^2}{\mu^2}+O\Big(\frac{\mu_0^4}{\mu^4}\Big)\right).
\end{equation}

Since quantum corrections are significant only in the quantum region near $\mu=0$, we can ignore them and, through Hamilton equations, obtain a  modified Friedmann equation:
\begin{subequations}
\begin{equation}
 H^2=\left(\frac{\dot{p}}{2p}\right)^2=\frac{1}{3}\rho\left(1-\frac{\rho}{\rho_{\text{crit}}}\right),
 \end{equation}
 \begin{equation}
 \rho_{\text{crit}}^{\mu_0}=\left(\frac{3}{\gamma^2\mu_0^2}\right)^{\frac{3}{2}}\frac{\sqrt{2}}{P_\phi}.
\end{equation}
\label{Friedmodmu0}
\end{subequations}

As in the full quantum dynamics, the critical density at the bounce is inversely proportional to the value of the constant of motion $P_\phi$.

\subsubsection{Effective $\bar{\mu}$ Scheme} Applying the same procedure to the $\bar{\mu}$ scheme, the improved effective Hamiltonian reads as follows:
\begin{equation}
\frac{\mathcal{C}_\text{eff}^{\bar{\mu}}}{2}=-\frac{3}{\gamma^2\bar{\mu}^2}\abs{p}^{\frac{1}{2}}\sin[2](\bar{\mu}c)+\frac{1}{2}B(\nu)P_\phi^2,
\end{equation}
where $B(\nu)$ is the eigenvalue of the inverse volume operator expressed in terms of $\nu$ as given by \eqref{Bofv}. Again, for $\abs{\nu}\gg1$, $B(\nu)$ quickly approaches its classical value:
\begin{equation}
B(\nu)=\left(\frac{6}{\gamma}\right)^{3/2}\frac{K}{\abs{\nu}}\left(1+\frac{5}{9}\frac{1}{\abs{\nu}^2}+O\Big(\frac{1}{\abs{\nu}^4}\Big)\right).
\end{equation}
Neglecting the higher order quantum corrections as before and given that the Poisson bracket between $\nu$ and $c$ is easily derived from \eqref{poissoncp}, the modified Friedmann equation in this case is as follows:
\begin{subequations}
\begin{equation}
H^2=\left(\frac{\dot{\nu}}{3\nu}\right)^2=\frac{\rho}{3}\left(1-\frac{\rho}{\rho_{\text{crit}}}\right),
\end{equation}
\begin{equation}
\rho_{\text{crit}}^{\bar{\mu}}=\frac{4\sqrt{3}}{\gamma^3}.
\end{equation}
\label{Friedmodmubar}
\end{subequations}

The critical density does not depend on $P_\phi$ anymore, and that is the main reason for which the improved model is much more appealing than the standard one.

\subsection{Loop Quantization of the Anisotropic Sector\label{anisotropicLQC}}
Let us now show the work that was done on the implementation of the LQG quantization procedures to the anisotropic sector of cosmology, i.e., to the Bianchi models. The classical phase space in this case is six-dimensional since there are three spatial directions that evolve independently (i.e., three different scale factors $a_1(t)$, $a_2(t)$, $a_3(t)$); therefore, in the Hamiltonian formulation, the fundamental variables will be $(c_i,p_i)$ with $i=1,2,3$ where the momenta $c_i$ are dependent on the velocities $\dot{a}_i$ while the variables $p_i$ are proportional to the (comoving) areas perpendicular to the direction $i$: $\abs{p_i}\propto a_ja_k$ with $i\neq j\neq k$. We will briefly present the results obtained by Ashtekar and Wilson-Ewing on the improved method of Loop quantization of the Bianchi type I, II and IX models \cite{Wilson-Ewing_BianchiI,Wilson-Ewing_BianchiII,Wilson-Ewing_BianchiIX}; their work simplifies and improves the previous analyses on the Loop quantum homogeneous models by Bojowald \cite{Bojowald_2003HomogeneousLQC,Bojowald_2004HomogeneousLQCSpinConnection,Bojowald_2004BianchiIXLQC,Bojowald_2004ChaosSuppression}.

\subsubsection{Bianchi Type I} The Bianchi type I model corresponds to the simplest anisotropic model; its classical Hamiltonian constraint in the Ashtekar variables reads as follows:
\begin{equation}
    \mathcal{C}_{\text{BI}}=-\frac{1}{\gamma^2V}\sum_{i\neq j}c_ip_ic_jp_j,
    \label{BianchiIclassicalconstraint}
\end{equation}
where $V=a_1a_2a_3=\sqrt{\abs{p_1p_2p_3}}$ is the Universe volume. The Hamiltonian equations yield a system of six coupled differential equations that can be easily solved by recognizing that the quantities $\mathcal{K}_i=c_ip_i$ are constants of motion. The solution in the void case is the famous Kasner solution where $a_i(t)\propto t^{k_i}$, where $k_i$ are the constant Kasner indices that obey $\sum_ik_i=\sum_ik_i^2=1$ \cite{KasnerSolution}; usually, these indices are parametrized through a variable $u\in(1,+\infty)$. Repeating the procedure of the isotropic sector, we introduce matter in the form of a scalar field $\phi$ obeying a Hamiltonian similar to \eqref{Cmatt} and playing the role of relational time; then, we quantize the system according to the Dirac procedure, following \cite{Wilson-Ewing_BianchiI}.

Now, when implementing the $\bar{\mu}$ scheme, we are naturally induced to use three different parameters $\bar{\mu}_i$ relating to the three different directions. An analogous reasoning to the previous section yields the following:
\begin{equation}
    \bar{\mu}_i\bar{\mu}_j=\frac{\Delta}{\abs{p_k}}\quad\implies\quad\bar{\mu}_i=\sqrt{\Delta\frac{\abs{p_i}}{\abs{p_jp_k}}}
    \label{mujbar}
\end{equation}
with $i\neq j\neq k$. This implies that, when constructing the holonomies and extracting the operators corresponding to the quasi-periodic functions $\widehat{e^{i\frac{\bar{\mu}_ic_i}{2}}}$, their action would depend on all the $p_j$ and be unmanageable. A solution can be obtained using a generalization of \eqref{definemubar}, i.e., $\bar{\mu}_i=\sqrt{\frac{\Delta}{\abs{p_i}}}$ so that it could be possible to define volume-like variables $\nu_i\propto \abs{p_i}^{\frac{3}{2}}$ and implement the operators as in \eqref{vinv+1}: $\widehat{e^{i\frac{\bar{\mu}_ic_i}{2}}}\Psi(\nu_i,\nu_j,\nu_k)=\Psi(\nu_i+1,\nu_j,\nu_k)$. This process allows to find the quantum dynamics of the three spatial directions separately, and each results to be a copy of the isotropic model; however, the description of the evolution of the whole model is not viable in this framework.

In order to keep the correct expression \eqref{mujbar} of the parameters $\bar{\mu}_i$, a new representation was developed through the introduction of dimensionless variables $q_i\propto\text{sgn}(p_i)\sqrt{\abs{p_i}}$; this allows for the definition of a new basis in $\ham_\text{grav}^\text{kin}$ comprised of vectors that are still eigenvectors of the operators $p_i$, and the action of the exponential operators depend on these variables:
\begin{equation}
    \widehat{e^{i\frac{\bar{\mu}_ic_i}{2}}}\Psi(q_i,q_j,q_k)=\Psi(q_i+\frac{\text{sgn}(q_i)}{q_jq_k},q_j,q_k).
\end{equation}

Note how the shift along the $q_i$ direction depends on $q_j$ and $q_k$. In order to make this more manageable, there is a further possible substitution: to define a volume variable $\nu\propto q_1q_2q_3$ and use a basis $\ket{q_1,q_2,\nu}$ (note that it is possible substitute any of the $q_i$ with $\nu$ and obtain the same results). This way, after some calculations, it is possible to write the action of the Hamiltonian gravitational constraint as dependent only on the volume: it will give a combination of shifted wave functionals of the following form:
\begin{equation}
    \Psi(f_1(\nu)q_1,f_2(\nu)q_2,\nu\pm4)
\end{equation}
where $f_1(\nu)$, $f_2(\nu)$ are simple rescaling functions of $\nu$. This is more easily comparable with the dynamics of the isotropic model; indeed, it is possible to construct a projection that maps the anisotropic wave functional $\Psi(q_1,q_2,\nu)$ into the isotropic state $\Psi(\nu)$ of previous sections, as well as making the gravitational constraints become the same.

Finally, let us address the issue of singularity resolution. It is possible to decompose the Hilbert space in the two subspaces $\ham^{\text{grav}}_\text{singular}$ and $\ham^\text{grav}_\text{regular}$, where the first contains all states with support on points with $\nu=0$, and the second contains the states without; since all terms in the gravitational constraint contain a factor proportional to a power of $\nu$ (depending on the chosen factor ordering), the two subspaces are invariant under time evolution and remain decoupled. Therefore, a state that starts as regular will remain regular throughout all its evolution; in this sense, the singularity is avoided. This behavior is again captured by the effective dynamics that predicts the classical Kasner solution away from the singularity and a Bounce in the Planck regime that jumps to the contracting branch in a similar way to the isotropic model. However, in this case, we need to keep track also of the three Hubble functions $H_i$ that undergo different Bounces separately; still, for~conclusive evidence that this effective evolution correctly reproduces the exact quantum dynamics, one would need numerical simulations of the exact quantum model.

\subsubsection{Bianchi Type II} The Bianchi type II model augments the Bianchi I with a curvature term (coming from the full expression of the connection) and the introduction of a potential along only one direction (here the direction 1). Its classical Hamiltonian constraint is as  follows:
\begin{subequations}
\begin{equation}
    \mathcal{C}_\text{BII}=\mathcal{C}_\text{BI}+\mathcal{C}^\text{curv}_\text{BII}+U_\text{BII},
\end{equation}
\begin{equation}
    \mathcal{C}^\text{curv}_\text{BII}=-\frac{1}{\gamma^2V}\,\epsilon\,p_2p_3c_1,
\end{equation}
\begin{equation}
    U_\text{BII}=\frac{1+\gamma^2V}{\gamma^2}\,\frac{p_2^2p_3^2}{4p_1^2},
\end{equation}
\end{subequations}
where $\epsilon=\pm1$ is an internal pseudoscalar parametrizing the orientation of the triad. The~classical dynamics is comprised of two different Kasner epochs bridged by a transition that changes the value of the Kasner indices \cite{KasnerSolution}; in terms of the variable $u$ introduced before, the second Kasner epoch is parametrized by $u_2=u_1-1$, where $u_1$ is the value of the first~epoch.

When implementing the quantization procedure, it is possible to follow the same steps of Bianchi I, but more care is needed for the new terms that appear in the constraint \cite{Wilson-Ewing_BianchiII}. In particular, after the implementation of the $\bar{\mu}$ scheme through \eqref{mujbar} and the definition of the variables $q_i$, the term $\mathcal{C}^\text{curv}_\text{BII}$ contains a power $\abs{p_1}^{-\frac{1}{2}}$, that usually becomes $\abs{p_1}^{-\frac{1}{4}}$ after a symmetric factor ordering; this is handled through a variation of the Thiemann inverse triad identities \cite{Thiemanntrick}, that in the $q_i$ representation yield the following:
\begin{subequations}
\begin{equation}
    \widehat{\abs{p_1}^{-\frac{1}{4}}}\ket{q_1,q_2,q_3}\propto\text{sgn}(q_1)\sqrt{\abs{q_2q_3}}\,g(\nu,q_2,q_3)\,\ket{q_1,q_2,q_3},
\end{equation}
\begin{equation}
    g(\nu,q_2,q_3)=\sqrt{\nu+\text{sgn}(q_2q_3)}\,-\sqrt{\nu-\text{sgn}(q_2q_3)}.
\end{equation}
\label{p1^-1/4}
\end{subequations}

On the other hand, the term $U_\text{BII}$ contains only a power $\widehat{\abs{p_1}^{-2}}$ whose action can be simply defined as the eighth power of the operator \eqref{p1^-1/4}. This form suggests that, in this case, the simplest representation is to substitute $\nu$ to $q_1$ and use the basis $\ket{\nu,q_2,q_3}$.

The quantum dynamics of the Bianchi II model is analogous to that of Bianchi I (and therefore of FLRW) because the regular and singular Hilbert spaces decouple in this model as well, and a state that starts away from $\nu=0$ will never reach it. Far from the singularity, the classical dynamics of the two bridged Kasner-like solutions is recovered, while near the Planck regime, there is a Bounce that joins with the contracting branch.

\subsubsection{Bianchi Type IX} The Bianchi type IX is the most complex homogeneous model (together with type VIII). The Hamiltonian is as follows:
\begin{subequations}
\begin{equation}
    \mathcal{C}_\text{BIX}=\mathcal{C}_\text{BI}+\mathcal{C}^\text{curv}_\text{BIX}+U_\text{BIX},
\end{equation}
\begin{equation}
    \mathcal{C}^\text{curv}_\text{BIX}=-\frac{\epsilon}{\gamma^2V}\left(c_1p_2p_3+c_2p_3p_1+c_3p_1p_2\right),
\end{equation}
\begin{equation}
    U_\text{BIX}=\frac{1}{\gamma^2V}\left(\frac{p_2^2p_3^2}{4p_1^2}+\frac{p_1^2p_3^2}{4p_2^2}+\frac{p_1^2p_2^2}{4p_3^2}-\frac{p_1^2+p_2^2+p_3^2}{2}\right);
\end{equation}
\end{subequations}
the classical dynamics is a chaotic evolution from one Kasner solution to the next, each~changing the Kasner indices and each closer in time to the next, until the singularity is reached. In~terms of $u$, each transition from one Kasner epoch (with a specific value of $u$) to another one lowers the previous value by $1$, until a value $u_{1_\text{final}}<1$ is reached that puts an end to the first Kasner era; at that point, the following epoch starts a second era with $u_{2_\text{init}}=u_{1_\text{final}}^{-1}$ and again each transition lowers the previous value of $u$ by 1. The cycle continues indefinitely toward the singularity, giving rise to the same chaotic evolution that appears in the point-Universe description outlined in Section \ref{giobaBianchi}. However, chaos is tamed when introducing matter in the form of a massless scalar field $\phi$ (which is exactly what is done in order to implement the quantization procedure): the Friedmann equation becomes asymptotically velocity-term dominated (AVTD), so that a standard Kasner-like dynamics is recovered, and the singularity is approached through a single, stable Kasner~epoch.

The quantization procedure is the same as the other models. The scalar field plays the role of relational time. The $\bar{\mu}_i$ are defined as in \eqref{mujbar}, and we introduce the variables $(q_1,q_2,q_3)$ and then change to $(q_1,q_2,\nu)$ so that the action of the gravitational constraint involves constant shifts in $\nu$ and rescalings of $q_i$ dependent only on $\nu$ (note that in this model the symmetries are in place again, so any $q_i$ could be substituted with $\nu$); the singular and regular Hilbert spaces decouple and the singularity is still solved.

We must now address the issue of chaoticity. It can be shown that already for the classical dynamics, the presence of the scalar field tames the chaos \cite{BelinskiScalarFieldChaos}; this is true also in the quantum model, as long as the AVTD regime is reached before the quantum gravitational effects become relevant, but if the value of the momentum conjugate to the scalar field is too small, this will not happen. However one could argue that the quantum gravity effects giving rise to the Bounce make it so that the model evolves away from the high curvature region toward the classical contracting branch, and there is not a sufficient number of Kasner epochs for chaos to appear before the AVTD regime is unavoidably reached, even~if this happens after the Bounce. Further arguments for the removal of chaos in the Loop quantum Bianchi type IX model are given in \cite{Bojowald_2004ChaosSuppression,Bojowald_2004HomogeneousLQCSpinConnection,Bojowald_2004BianchiIXLQC}. Still, numerical simulations of the complete quantum system are needed in order to give a definitive word on the chaoticity of these models.

\subsection{Criticisms and Shortcomings of LQC}
Over the years, many criticisms have been made on the LQC framework, mainly about the following points: whether the Bounce can be regarded as a semiclassical phenomenon or must be considered a purely quantum effect; the fact that the quantum dynamics is not derived by a symmetry reduction of the full LQG theory, but by quantizing cosmological models that are reduced before quantization; the use of the area gap as a parameter to construct the dynamics of the reduced theory and its effective description. In this section, we briefly summarize these issues.

As already mentioned, LQC is not the cosmological sector of LQG, but rather the implementation of the latter's quantization procedures on a cosmological spacetime; a good symmetry reduction of full LQG would require that some degrees of freedom be frozen out, but this procedure conflicts with the quantum character of the $SU(2)$ variables. The spatial geometry of a cosmological spacetime is fixed, and during the quantization on invariant variables, any kind of spatial structure, such as the possibility to perform local $SU(2)$ transformations, is lost. Furthermore, as it is well known, in LQG the implementation of the scalar constraint is not yet a viable task \cite{ThiemannBook}, and it is worth noting how this problem is somehow bypassed in LQC, where the dynamics for the cosmological models is constructed; however, this procedure is far from being completely clear. Another way to see the problem of the $SU(2)$ symmetry is that the resulting algebra on the reduced model is different from the holonomy-flux algebra of the full theory; therefore, LQC is not equivalent to LQG~\cite{Cianfrani_Montani_Gaugefixing}. Taking this a step further, in \cite{Bojowald_2020BLAST}, it is claimed that the resulting algebra of LQC has several different representations, among which the Ashtekar school implicitly chooses the one that favors bouncing solutions, while in \cite{Haro}, it is argued that the mechanism for the resolution of the singularity lies in the regularization of the constraint rather than in the quantization procedure itself (and indeed the singularity in LQC is avoided already at a semiclassical level). Alternatively, in \cite{Cianfrani_Montani_Gaugefixing2} an $SU(2)$-invariant gauge fixing is considered, which yields a modified holonomy-flux algebra that reproduces the original one of LQG only when holonomies are evaluated along the triad vectors; the quantization procedure is then performed according to the full theory, and the resulting model is a quantum cosmology that manages to better preserve the $SU(2)$ structure. A different approach to derive a consistent description of the Loop quantum cosmological sector is provided in \cite{Bojowald_2009Consistent}, where through the introduction of local patches, it is possible to define local cosmological variables that properly take into account the presence and the properties in full LQG of both holonomy corrections and inverse-volume operators. Another interesting and more developed approach that tries to solve this problem is that of quantum reduced loop gravity (QRLG). In this approach, inspired by the criticisms in \cite{Cianfrani_Montani_Gaugefixing,Cianfrani_Montani_Gaugefixing2}, some gauge-fixing conditions are implemented on the kinematical Hilbert space of LQG that restrict the full gravitational model to a diagonal metric tensor and to diagonal triads; then, the cosmological reduction is performed by considering only that part of the scalar constraint that generates the evolution of the homogeneous part of the metric. Finally, the dynamics is obtained by performing a cubation (instead of a triangulation) on the reduced spin-foam graphs. This way, QRLG gives a quantum description of the Universe in terms of a cuboidal graph and it provides a framework for deriving the cosmological setting from full LQG. For~a more detailed presentation, see \cite{Alesci1,Alesci2,Alesci3,Alesci4,Alesci5,Alesci6}. In this context, also the formalism of group field theory (GFT) for quantum gravity contributes to clarify the link between the effective cosmological equations and LQC when applied to the cosmological sector. In this approach, the effective cosmological dynamics emerges as an hydrodynamic-like approximation of the multi-condensate quantum states, i.e., the fundamental quantum gravitational degrees of freedom. In particular, a second order quantization and the basis to the idea of lattice refinement are provided, showing the dependence of the effective cosmological connection on the number of spin network vertices (a quantity of a purely quantum origin) and~thus, on the scale factor. For more details regarding the GFT cosmology and the emergent bouncing dynamics, see \cite{GFT1,GFT2,GFT3,GFT4,GFT5}.

Another problem that is often raised, linked to the previous one, is that an external parameter fixing the discretization scale must be introduced from the full theory by hand because LQC is derived independently from LQG; this leads to some issues. For example, in \cite{Bojowald_2020BLAST} it is stated that an effective description will have a scale of validity (given by the area gap itself), while the Ashtekar school uses the same effective equations across very different regimes (namely, it follows the evolution of a wave packet from the classical regime up to the Planck region near the singularity; this is connected also to the issue about the nature of the Bounce).

LQG and LQC attempt to provide a promising framework for a quantum mechanical description of general relativity and of cosmological models, but as outlined in this section, both---the latter in particular---need to be substantially improved.

A good achievement toward this goal is the formulation of polymer quantum mechanics (PQM), a new quantum mechanical framework that is able to reproduce LQC effects but can be derived independently from LQG and is much more versatile and easily applicable to any Hamiltonian system. Its implementation on the cosmological minisuperspaces is the focus of the next~sections.


\section{Polymer Cosmology\label{secPOL}}
The power of the polymer formulation stands in its capability of introducing regularization effects typical of the LQG quantum gravity approach by means of a simpler mathematical framework with respect to the LQC theory. Therefore, its employment in the cosmological sector has great relevance in trying to overcome the singularity issue of GR and making also a comparison with the LQC main results regarding the presence of an initial Big Bounce and its properties. In this sense, the principal feature of the polymer approach is making clear that the properties of the cosmological dynamics are strongly dependent on the set of variables on which the polymer quantization is implemented~\cite{Federico,Silvia,CrinoPintaudi,Lecian_2013,Giovannetti_2019,stefano}.

In this section, we focus on the main applications of the polymer formulation to the FLRW \cite{Federico}, Bianchi I \cite{Silvia} and Bianchi IX models \cite{Giovannetti_2019,stefano} that represent the main cosmological scenarios on which the polymer-modified dynamics of the primordial Universe are tested. We will start with a discussion on the main results obtained by treating the polymer quantization of these models in the Ashtekar variables, that constitute the setting more connected to the original LQC formulation from which the polymer formulation is derived. Indeed, as originally affirmed by Ashtekar in \cite{ASHTEKARpol}, at the Planck scale, the implementation of LQG shows that quantum geometry has a close similarity with polymers and that the continuum picture arises only upon a coarse-graining procedure by means of suitable semiclassical states. Then, we will proceed by applying the polymer quantization to the volume-like variables, obtained after doing a canonical transformation from the Ashtekar connections to new generalized coordinates. In particular, we will implement both a semiclassical and a quantum treatment for the FLRW and the Bianchi I models, i.e., the homogeneous and isotropic model and its simplest anisotropic generalization. Finally, we will apply the semiclassical polymer framework in the Misner-like variables (the isotropic variable $\alpha$ or the Universe volume $V$ plus the anisotropies) to the Bianchi IX model that represent the most general candidate for the primordial Universe from which even the properties of the general cosmological solution can be extrapolated.

\subsection{Polymer Quantum Mechanics\label{POL}}
In this section, we introduce PQM as described in the main paper written by Corichi in 2007 \cite{Corichi_2007}, where a complete mathematical framework is developed. PQM is an alternative representation of the canonical commutation relations non-unitarily connected to the ordinary Schr\"{o}dinger one. In fact, it can be introduced as a limit of the Fock representation, where the continuity hypothesis of the Stone--Von Neumann theorem is violated. However, PQM can also be derived without recurring to its connection with the Schr\"{o}dinger representation as follows.

\subsubsection{Polymer Kinematics}
In order to introduce the polymer representation without any reference to the Schr\"{o}dinger one, let us consider the abstract kets $\ket{b}$ labeled by the real parameter $b\in\mathbb{R}$ and taken from the Hilbert space $\mathcal{H}_{\text{poly}}$. 

A generic cylindrical state can be defined through the finite linear combination as follows:
\begin{equation}
	\label{cyl}
	\ket{\psi}=\sum_{i=1}^Nn_i\ket{b_i},
\end{equation}
where $b_i\in\mathbb{R},\,i=1,\dots,N\in\mathbb{N}$. We choose the inner product so that the fundamental kets are orthonormal as follows:
\begin{equation}
\label{ip}
	\braket{b_i}{b_j}=\delta_{ij}\,.
\end{equation}

From this choice, it follows that the inner product between two cylindrical states $\ket{\psi}=\sum_{i}n_i\ket{b_i}$ and $\ket{\phi}=\sum_{j}m_j\ket{b_j}$ is as follows:
\begin{equation}
	\braket{\phi}{\psi}=\sum_{i,j}m^*_in_j\delta_{ij}=\sum_{i}m^*_in_i.
\end{equation}

It can be demonstrated that the Hilbert space $\mathcal{H}_{\text{poly}}$ is the Cauchy completion of the finite linear combination of the form \eqref{cyl} with respect to the inner product \eqref{ip} and that it results to be non-separable. 

Two fundamental operators can be defined on this Hilbert space: the symmetric label operator $\hat{\epsilon}$ and the shift operator $\hat{s}(\zeta)$ with $\zeta\in\mathbb{R}$. They act on the kets $\ket{b}$ as follows:
\begin{equation}
	\hat{\epsilon}\ket{b}:=b\,\ket{b}\,,\quad\hat{s}(\zeta)\ket{b}:=\ket{b+\zeta}\,.
\end{equation}

The shift operator defines a one-parameter family of unitary operators on $\mathcal{H}_{\text{poly}}$. However, since the kets $\ket{b}$ and $\ket{b+\zeta}$ are orthogonal for any $\zeta\neq0$, the shift operator $\hat{s}(\zeta)$ is discontinuous in $\zeta$, and there is no Hermitian operator that can generate it by exponentiation.

Now, the abstract structure of the Hilbert space is described, so we can proceed by defining the physical states and operators. In the following, we will consider a one-dimensional system identified by the phase-space coordinates $(Q,P)$, and we will separate the discussion into two cases referred to the two possible polarizations for the wave function. We suppose also that the configurational coordinate $Q$ has a discrete character, due to the relation that it often possess with geometrical quantities. This is a way to investigate the physical effects of discreteness at a certain scale, for example, when introducing quantum gravity effects on the cosmological dynamics.

\paragraph{$P$-Polarization}
In the momentum polarization, the wave function is written as follows:
\begin{equation}
	\psi(P):=\braket{P}{\psi}
\end{equation}
where
\begin{equation}
	\psi_{b}(P):=\braket{P}{b}=e^{ib P}.
\end{equation} 

The shift operator $\hat{s}(\zeta)$ is identified with the multiplicative exponential operator $\hat{T}(\zeta)$:
\begin{equation}
	\hat{T}(\zeta)\psi_{b}(P)= e^{i\zeta P}e^{ibP}=e^{i(b+\zeta) P}=\psi_{b+\zeta}(P);
\end{equation}
$\hat{T}(\zeta)$ is discontinuous by definition and as a result, the momentum $P$ cannot be promoted to a well-defined operator. On the other hand, $\hat{Q}$ corresponds to the label operator $\hat{\epsilon}$ and in this polarization acts differentially:
\begin{equation}
	\hat{Q}\psi_{b}(P)= -i\frac{\partial}{\partial P} \psi_{b}(P)= b \psi_{b}(P).
\end{equation}

Additionally, it has to be considered as a discrete operator since $\psi_b(P)$ are orthonormal for all $b$, even though $b$ belongs to a continuous set.

By means of the $C^*$-algebra it can be seen that $\mathcal{H}_{\text{poly}}$ is isomorphic to the following:
\begin{equation}
	\mathcal{H}_{\text{poly},P}:=L^2(\mathbb{R}_B,d\mu_H)
\end{equation}
where $\mathbb{R}_B$ is the Bohr compactification of the real line, i.e., the dual group of the real line equipped by the discrete topology, and $d\mu_H$ the Haar measure. Moreover, the wave functions are  quasi-periodic with the inner product as follows:
\begin{equation}
	\braket{\psi_{b_i}}{\psi_{b_j}}:=\int_{\mathbb{R}_B}d\mu_H\,\psi_{b_i}^\dag(P)\psi_{b_j}(P)=\lim_{L\to\infty}\,\frac{1}{2L}\int_{-L}^LdP\,\psi_{b_i}^\dag(P)\psi_{b_j}(P)=\delta_{ij}.
	\label{eq:internopoly}
\end{equation}

\paragraph{$Q$-Polarization}
In the position polarization, the wave functions depend on the configurational variable $Q$ and a generic state, written as follows:
\begin{equation}
	\psi(Q):=\braket{Q}{\psi}
\end{equation}
where the basis functions can be derived using a Fourier-like transform as follows: \vspace{6pt}
\begin{equation}
	\tilde{\psi}_b(Q):=\braket{Q}{b}=\bra{Q}\int_{\mathbb{R}_B}d\mu_H\,\ket{P}\braket{P}{b}=\bra{Q}\int_{\mathbb{R}_B}d\mu_H\,\ket{P}\psi_b(P)=\int_{\mathbb{R}_B}d\mu_H e^{-iQP}e^{ib P}=\delta_{Qb}
\end{equation}
\noindent through which we can easily see that the $\hat{P}$ operator does not exist since the derivative of the Kronecker delta is not well defined. However, for the operator $\hat{T}(\zeta)$ we have the following:
\begin{equation}
	\hat{T}(\zeta)\psi(Q)=\psi(Q+\zeta).
\end{equation}

As in the previous case, the $\hat{Q}$ operator corresponds to $\hat{\epsilon}$, but in this polarization, it acts in a multiplicative way, i.e., the following:
\begin{equation}
	\hat{Q}\tilde{\psi_b}(Q):=b \tilde{\psi_b}(Q).
\end{equation}

The Hilbert space has analogous features as before:
\begin{equation}
	\mathcal{H}_{\text{poly},Q}:=L^2(\mathbb{R}_d, d\mu_c)
\end{equation}
where $\mathbb{R}_d$ is the real line equipped with the discrete topology and $d\mu_c$ is the counting measure. The inner product is as follows:
\begin{equation}
	\big<\tilde{\psi_{b_i}}(Q),\tilde{\psi_{b_j}}(Q)\big>=\delta_{ij}
\end{equation}
so, it is clear how the $\hat{Q}$ operator is discrete also in this polarization.

\subsubsection{Polymer Dynamics}{\label{Polymer}}
In the previous section, the polymer kinematic Hilbert space was introduced. In particular, the discussion above has highlighted that it is not possible to well define the $\hat{Q}$ and $\hat{P}$ operators simultaneously in the polymer framework. So, it is necessary to understand how to implement the dynamics in order to apply the polymer framework to a physical system.

Let us consider a one-dimensional system described by the Hamiltonian as follows:
\begin{equation}
	\mathcal{C}=\frac{P^2}{2m}+U(Q)
\end{equation}
in the $P$-polarization. If we assume that $\hat{Q}$ is a discrete operator, we have to find an approximate form for $\hat{P}$. For this reason, the required regularization procedure consists of introducing a lattice with a constant spacing $b_0$:
\begin{equation}
	\gamma_{b_0}=\{Q \in \mathbb{R} : Q=Nb_0, \; \forall\;  N \in \mathbb{Z}\}\,.
\end{equation}

In order to remain in the lattice, the only states permitted are as follows:
\begin{equation}
\ket{\psi}=\sum_{N}n_N\ket{b_N}\in\mathcal{H}_{\gamma_{b_0}}
\end{equation}
where $b_N=Nb_0$ and $\mathcal{H}_{\gamma_{b_0}}$ is a subspace of  $\mathcal{H}_{\text{poly}}$ that contains all the functions $\psi$ so that $\sum_{N}\abs{b_N}^2<\infty$. 

Now, we have to find an approximate form for $\hat{P}$ in order to have a well-defined Hamiltonian operator through which implement the dynamics in both the polarizations. We notice that the operator $\widehat{e^{i\zeta P}}$ is well defined and acts as the shift operator on the kets $\ket{b}$. In particular, its action is restricted to the lattice only if $\zeta$ is a multiple of $b_0$ and the simplest choice corresponds to $\zeta=b_0$, so that its actions reads as follows:
\begin{equation}
\hat{T}(b_0)\ket{b_N}:=\ket{b_N+b_0}=\ket{b_{N+1}}.
\end{equation}

Therefore, it is possible to use the shift operator to introduce the following approximation:
\begin{eqnarray}
	\label{psub}P\sim\frac{1}{b_0}\sin(b_0P)=\frac{1}{2ib_0}(e^{ib_0P}-e^{-ib_0P}),
\end{eqnarray}
valid in the limit $b_0P\ll1$, so that the regularized $\hat{P}$ operator acts as follows:
\begin{equation}
	\hat{P}_{b_0}\ket{b_N}=\frac{1}{2ib_0}[\hat{T}(b_0)-\hat{T}(-b_0)]\ket{b_N}=\frac{1}{2ib_0}(\ket{b_{N+1}}-\ket{b_{N-1}}).
\end{equation}

It is possible to introduce also an approximate version of $\hat{P}^2$ as follows:
\begin{equation}
	\label{sin2}
	\hat{P}^2_{b_0}\ket{b_N}\equiv\hat{P}_{b_0}\cdot\hat{P}_{b_0}\ket{b_N}=\frac{1}{4b_0^2}[-\ket{b_{N-2}}+2\ket{b_{N}}-\ket{b_{N+2}}]=\frac{1}{b_0^2}\sin^2(b_0p)\ket{b_{N}}\,.
\end{equation}

We remind that $\hat{Q}$ is a well-defined operator, so the regularized version of the Hamiltonian is written as follows:
\begin{equation}
	\hat{\mathcal{C}}^\text{poly}:=\frac{1}{2m}\hat{P}_{b_0}^2+\hat{U}(Q)
\end{equation}
that represents a symmetric and well-defined operator on $\mathcal{H}_{\gamma_{b_0}}$.

We notice that this Hamiltonian provides an effective description at the given scale $b_0$. More specifically, the question about the consistency between the effective theories at different scales and the existence of the continuum limit is deeply investigated in \cite{Corichi2_2007}. In particular, it is demonstrated that the continuum Hamiltonian can be represented in a Hilbert space unitarily equivalent to the ordinary $L^2$
space of the Schr{\"o}dinger theory by means of a renormalization procedure that involves coarse graining as well as rescaling, following Wilson’s renormalization group ideas.

When implementing PQM on the cosmological minisuperspaces, the geometrical variables (namely, the areas $p,p_i$ in the Ashtekar variables, the scale factor $\alpha$ and the anisotropies $\beta_i$ in the Misner variables, and both the volumes $v,v_i$ and the lengths $q_i$ in the volume-like variables) will be discretized and therefore will play the same role as the position $Q$; therefore, all the conjugate variables, i.e., $c,\tilde{c},c_i,\eta_i,P_\alpha,P_v,P_\pm$, will play the same role as the momentum $P$ and will be subjected to the polymer substitution \eqref{psub}.

\subsection{Polymer Cosmology in the Ashtekar Variables\label{A}}
In this section, we present the main results about the polymer semiclassical quantization of the FLRW and Bianchi I models in the Ashtekar variables, following \cite{Federico,Silvia}. In particular, the emergence of a Big Bounce regularizing the initial singularity is a solid prediction in the polymer framework, but the physical properties of the dynamics result in being dependent on the initial conditions on the motion. 

\subsubsection{The FLRW Universe in the Ashtekar Variables\label{semash}}

In this subsection, we treat the semiclassical and quantum polymer dynamics of the FLRW model \cite{Federico}. Additionally, some analogies with the original LQC scheme are~highlighted.

The classical Hamiltonian constraint for this configuration is as follows:
\begin{equation}
	\mathcal{C}_\text{FLRW}=-\frac{3}{\gamma^2}\sqrt{p}\hspace{4pt}c^2+\frac{P_\phi^2}{2\abs{p}^{\frac{3}{2}}}=0\,,
\end{equation}
where a massless scalar field is included so that $\phi$ can be chosen as the internal clock for the dynamics.

On a semiclassical level, the polymer paradigm is implemented by considering the variable $p$ as discrete in view of its geometrical character (i.e., it has the dimension of an area) and so a regularized version for the momentum $c$ in the form $c\to\frac{\sin(b_0c)}{b_0}$ is introduced, obtaining the following:
\begin{equation}
	\label{Cpoly}
	\mathcal{C}_\text{FLRW}^{\text{poly}}=-\frac{3}{\gamma^2b_0^2}\sqrt{p}\hspace{4pt}\mbox{sin}^2(b_0 c)+\frac{P_\phi^2}{2\abs{p}^{\frac{3}{2}}}=0\,,
\end{equation}
in which for the square of the momentum $c$, we have used the semiclassical version of \eqref{sin2}.

Given the Poisson brackets $\pb{c}{p}=\frac{\gamma}{3}$, we can obtain the equations of motion for $p$ and $c$ as follows:
\begin{subequations}
	\begin{equation}
		\label{pi}
		\dot p=-\frac{2\mbox{N}}{\gamma b_0}\sqrt{\abs{p}}\hspace{4pt}\mbox{sin}(b_0 c)\mbox{cos}(b_0 c),
	\end{equation}
	\begin{equation}
		\label{c}
		\dot c=\frac{\text{N}}{3}\Big(\frac{3}{\gamma b_0^2}\frac{1}{2\sqrt{p}}\mbox{sin}^2(b_0 c)+\frac{3\gamma}{4}\frac{P_\phi^2}{\abs{p}^{5/2}}\Big).
	\end{equation}
\end{subequations}

The analytical expression of the Friedmann equation can be derived as follows:
\begin{equation}
	H^2=\Big(\frac{\Dot{a}}{a}\Big)^2=\Big(\frac{\Dot{p}}{2p}\Big)^2=\frac{1}{\gamma^2 b_0^2}\frac{1}{\abs{p}}\mbox{sin}^2(b_0 c)\Big(1-\mbox{sin}^2(b_0 c)\Big);
\end{equation}
then, by using \eqref{Cpoly}, we obtain the following:
\begin{equation}
	H^2=\Big(\frac{\Dot{p}}{2p}\Big)^2=\frac{\rho}{3}\Big(1-\frac{\rho}{\rho_{\text{crit}}}\Big),
\end{equation}
where
\begin{equation}
	\rho_{\text{crit}}=\frac{3}{\gamma^2 b_0^2\abs{p}}.
	\label{rhocrit1}
\end{equation}

Let us now consider the scalar field $\phi$ as the internal time for the dynamics, by~requiring the lapse function to be as follows:
\begin{equation}
	\label{N}
	1=\Dot{\phi}=\mbox{N}\frac{\partial\mathcal{C}_\text{FLRW}^{\text{poly}}}{\partial P_\phi}=\mbox{N}\frac{P_\phi}{p^{\frac{3}{2}}},\quad\mbox{N}=\frac{\abs{p}^{\frac{3}{2}}}{P_\phi}=\frac{1}{\sqrt{2\rho}};
\end{equation}
therefore, the effective Friedmann equation in the ($p,\phi$) plane reads as follows:
\begin{equation}
	\Big(\frac{1}{\abs{p}}\frac{dp}{d\phi}\Big)^2=\frac{2}{3}\Big(1-\frac{\gamma^2 b_0^2}{6}\frac{P_\phi^2}{\abs{p}^2}\Big),
	\label{friedmannfirstcase}
\end{equation}
and it is analytically solvable after rewriting it in a dimensionless form. The expression of $p(\phi)$ can be written as follows:
\begin{equation}
	p(\phi)=\sqrt{\frac{\gamma^2 b_0^2}{6}}\,P_\phi\,\cosh\left(\sqrt{\frac{2}{3}}\,\phi\right).
	\label{pfi}
\end{equation}

\begin{figure}[H]
	\centering 
	\includegraphics[width=0.8\linewidth]{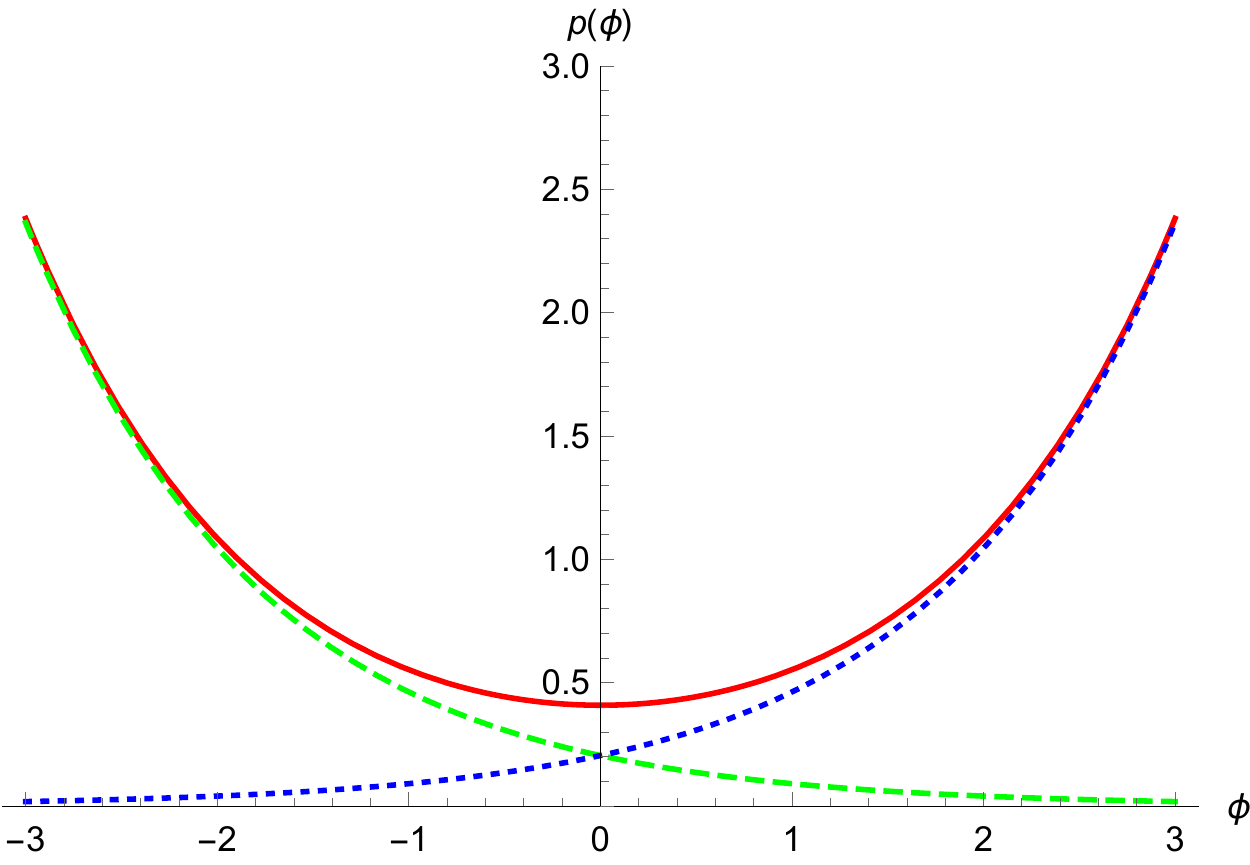} 
	\caption{The polymer trajectory (red continuous line) is compared to the classical ones for the flat FLRW model. The Big Bang solution is the blue dotted line and the Big Crunch solution is the green dashed line.}
	\label{confrontoclassicopolymer}
\end{figure}

As shown in Figure \ref{confrontoclassicopolymer} the polymer trajectory follows the classical one until it reaches a purely quantum era where the effects of quantum geometry become dominant and the resulting dynamics is that of a bouncing Universe replacing the classical Big Bang.

However, the critical energy density depends on the initial conditions on the momentum conjugate to the scalar field, as we can see in the following expression:
\begin{equation}
	\rho_{\text{crit}}=\Big(\frac{3}{\gamma^2 b_0^2}\Big)^{\frac{3}{2}}\frac{\sqrt{2}}{P_\phi},
	\label{rhocritsemiclpc}
\end{equation}
obtained by putting together the Equations \eqref{pfi} and \eqref{rhocrit1}. The non-universal character of the bouncing dynamics in this set of variables has the following consequence:
\begin{equation}
\rho_{\text{crit}}\xrightarrow{}\infty\,\quad \text{for}\quad P_\phi\xrightarrow{}0\,,
\end{equation}

\begin{figure}[H]
	\centering 
	\includegraphics[width=0.8\linewidth]{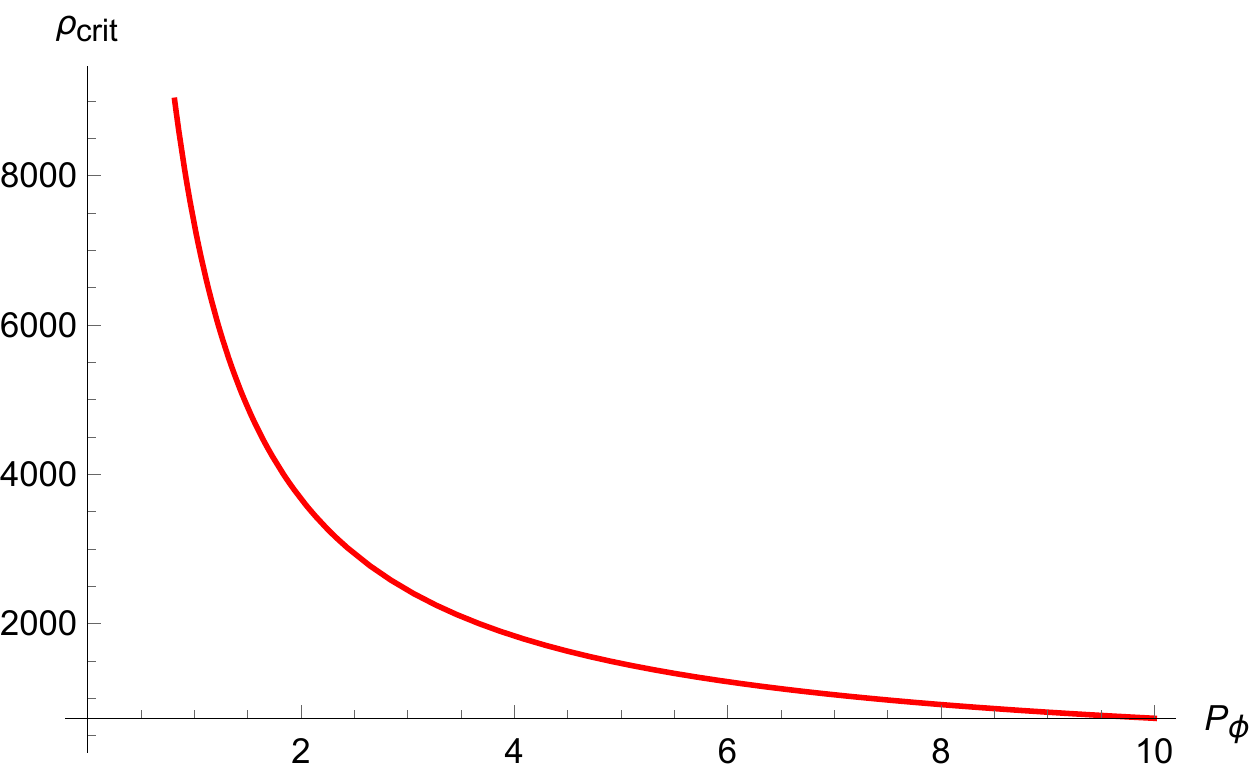}
	\caption{Dependence of the critical energy density on the momentum of the scalar field. For $P_\phi\rightarrow0$ the Bounce approaches the singularity.}
	\label{rhocritica}
\end{figure}

Thus, the initial singularity can be asymptotically approached, and the quantum corrections become irrelevant (see Figure \ref{rhocritica}).
Thus, in the Ashtekar variables, the non-diverging behavior of the energy density at the Bounce ensures the regularization of the singularity, due to its scalar nature, but it can assume arbitrarily large values and is not a fixed feature of the dynamics. This result is very similar to the $\mu_0$ scheme of LQC presented in \mbox{Section \ref{standardLQC}}.

Now we want to extend the semiclassical results obtained above to a full quantum level. In order to implement the Dirac quantization method \cite{Matschull1996Dirac}, we have to promote variables to quantum operators, i.e., the following:
\begin{equation}
\hat{p}=-\frac{i\gamma}{3}\frac{d}{dc},\quad\hat{c}=\frac{1}{b_0}\sin(b_0c),\quad\hat{P}_\phi=-i\frac{d}{d\phi};
\label{operatorspc}
\end{equation}

Thus, the Hamiltonian constraint operator in the momentum representation is as follows:
\begin{equation}
	\hat{\mathcal{C}}_\text{FLRW}^\text{poly}=\Big[-\frac{2}{3b_0^2}\frac{d^2}{dc^2}\mbox{sin}^2(b_0 c)+\frac{d^2}{d\phi^2}\Big]=0.
\end{equation}

In particular, the Hamiltonian constraint selects the physical states by annihilation, giving rise to the following WDW equation:
\begin{equation}
	\Big[-\frac{2}{3b_0^2}\Big(\mbox{sin}(b_0 c)\frac{d}{dc}\Big)^2+\frac{d^2}{d\phi^2}\Big]\Psi(c,\phi)=0,
	\label{wheelerpoly}
\end{equation}
where we have used a mixed factor ordering that will lead us to a solvable differential equation through the following substitution:
\begin{equation}
	x=\sqrt{\frac{3}{2}}\ln{\Big[\tan\Big(\frac{b_0 c}{2}\Big)\Big]}+\bar{x}.
	\label{cambiovariabili}
\end{equation}

Thus, \eqref{wheelerpoly} assumes the form of a massless Klein--Gordon-like equation:
\begin{equation}
	\label{KG}
	\frac{d^2}{dx^2}\Psi(x,\phi)=\frac{d^2}{d\phi^2}\Psi(x,\phi),
\end{equation}
where $\Psi$ is the wave function of the Universe. The solution can be written as follows:
\begin{equation}
	\Psi(x,\phi)=\displaystyle\int_0^{\infty}dk_\phi\hspace{4pt}\frac{e^{-\frac{\abs{k_\phi-\overline{k}_\phi}^2}{2\sigma^2}}}{\sqrt{4\pi\sigma^2}}\,k_\phi\,e^{ik_\phi x}e^{-ik_\phi\phi},
\end{equation}
where we have considered only positive energy-like eigenvalues $k_\phi$ and have used a Gaussian-like weighing function peaked on the initial value $\overline{k}_\phi$.

Now, in order to investigate the non-singular behaviour of the model, we can evaluate the expectation value of the energy density operator as  follows:
\begin{equation}
	\hat{\rho}=\frac{\hat{P}_{\phi}^2}{2\abs{\hat{p}}^3}
\end{equation}
using the basic operators \eqref{operatorspc} and the substitution \eqref{cambiovariabili} to compute the Klein--Gordon scalar product:
\begin{equation}
	\ev{\hat{O}}{\Psi}=\int_{-\infty}^{\infty}dx\,\,i\left(\Psi^*\,\partial_\phi(\hat{O}\Psi)-(\hat{O}\Psi)\,\partial_\phi\Psi^*\right),
	\label{KGev}
\end{equation}
where we consider $\Psi$ to be normalized.

\begin{figure}[H]
	\centering
	\includegraphics[width=0.8\linewidth]{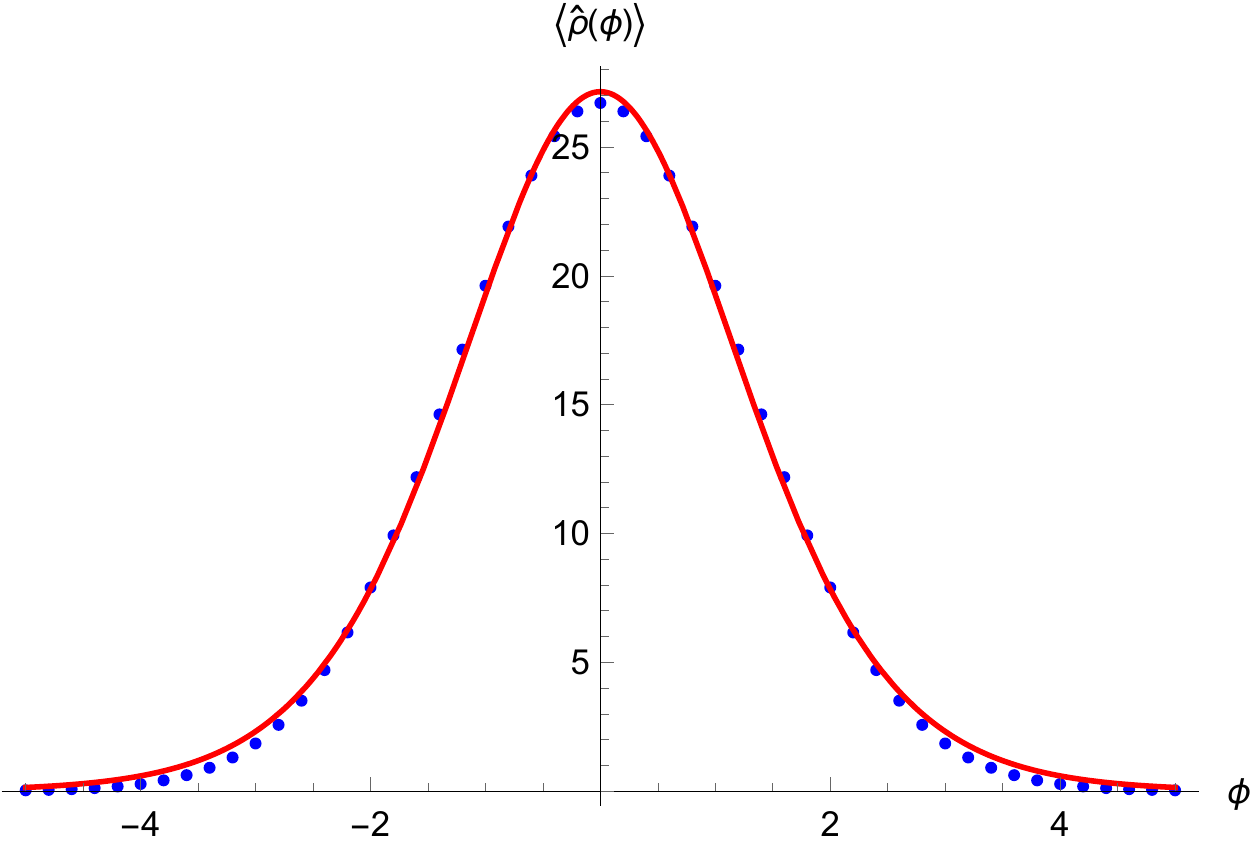}
	\caption{The expectation value of the energy density as function of time (blue dots), fitted with a function in accordance with the semiclassical evolution (continuous red line).}
	\label{rho(phi)pc}
\end{figure}

\begin{figure}[H]
	\centering
	\includegraphics[width=0.8\linewidth]{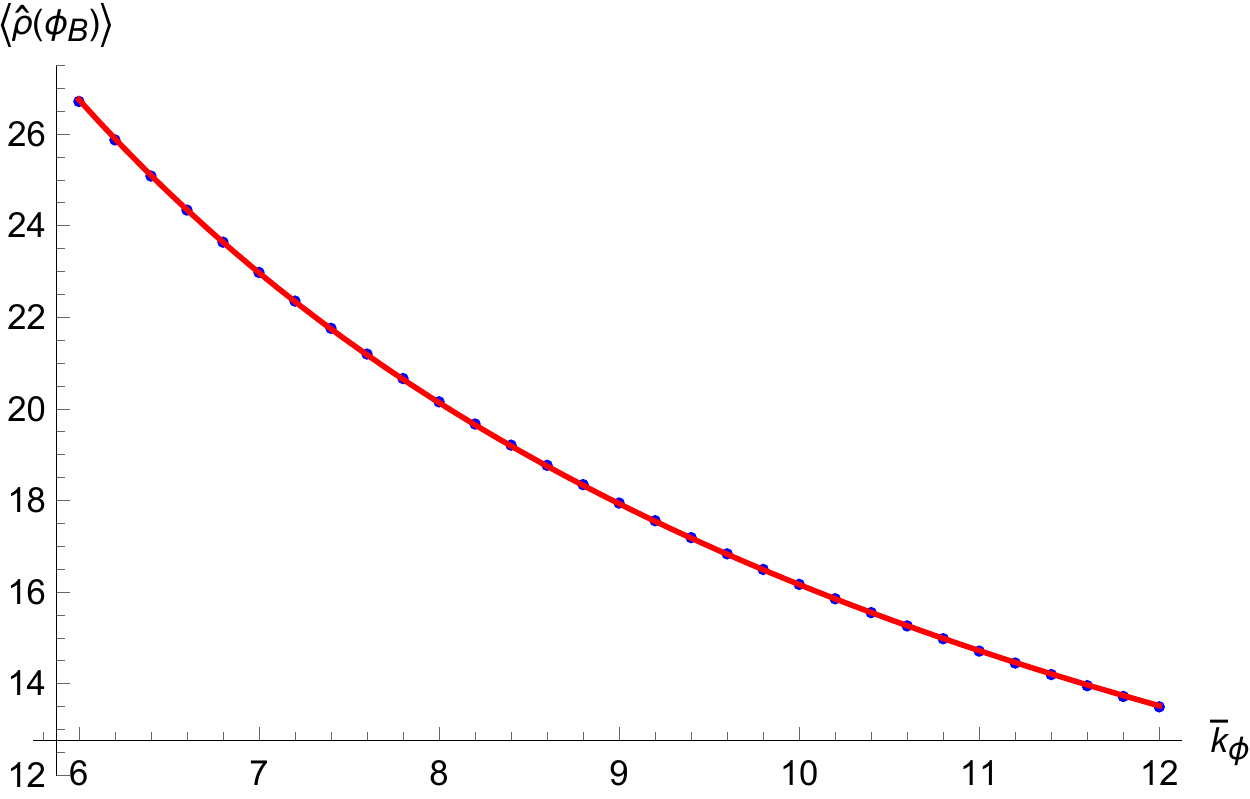}
	\caption{The expectation value of the energy density at the time $\phi_B$ of the Bounce as function of $\overline{k}_\phi$ (blue dots), fitted with a function in accordance with the semiclassical evolution (continuous red~line).}
	\label{rhomaxpc}
\end{figure}

In Figure \ref{rho(phi)pc}, we show the time dependence of $\ev{\hat{\rho}(\phi)}$ for a fixed value of $\overline{k}_\phi$, while the maximum $\ev{\hat{\rho}(\phi_B)}$, i.e., the expectation value of the density at the Bounce is presented in Figure \ref{rhomaxpc} from which we can appreciate the inversely proportional relation with $\overline{k}_\phi$ in~accordance with the semiclassical critical density given in \eqref{rhocritsemiclpc}. The points representing the quantum expectation values are obtained through numerical integration and are fitted with the continuous lines; they are in good accordance with the semiclassical trajectories when taking into account numerical effects and quantum fluctuations.

The quantum analysis performed here has highlighted the non-diverging nature (although strongly dependent on the initial conditions) of the energy density expectation value at the Bounce. This result ensures the existence of a minimum non-zero volume in view of the scalar and physical nature of the energy density, thus confirming the replacement of the singularity with a Big Bounce also at a quantum level. Clearly, a more precise assessment of the nature of the Bounce would require a non-trivial calculation of the expectation value of the Universe volume operator and also a careful analysis of the variance of the energy density operator on the Universe wave function (see for instance~\cite{BojowaldMoments}).


\subsubsection{The Bianchi I Universe in the Ashtekar Variables\label{sem1}}
In this section, we extend all the results obtained for the FLRW model by considering its simplest anisotropic generalization, i.e., the Bianchi I model, as done in \cite{Silvia}. Firstly, we~develop the polymer semiclassical analysis of the model in the Ashtekar variables, and then we study the dynamics of the Universe wave packet at a quantum level.

We proceed by discretizing the areas $p_i$ and imposing in \eqref{BianchiIclassicalconstraint} the polymer substitution for the connections $c_i$:
\begin{equation}
	c_i\to{1\over b_i}\sin({b_ic_i}),
\end{equation}
where $b_i$ ($i=1,2,3$) refer to the three independent polymer lattices and $\{c_{i},p_j\}=\gamma\delta_{ij}$. Then, the polymer Hamiltonian takes the following form:
\begin{align}
	\label{eqn:Hpoly}
	\mathcal{\mathcal{C}}^{\text{poly}}_\text{BI}=&-{1\over{\gamma^2V}}\sum_{i\neq j}{{\sin(b_ic_i)p_i\sin(b_jc_j)p_j}\over b_ib_j}+{P^2_{\phi}\over 2V}=0,
\end{align}
where $i,j=1,2,3$ and $V=\sqrt{p_1p_2p_3}$.

We derive the dynamics after choosing $\phi$ as the internal time, i.e., imposing the gauge $N={{\sqrt{p_1p_2p_3}}\over P_\phi}$, so the equations of motion take the following form:
\begin{subequations}
\begin{equation}
	{{dp_i}\over{d\phi}}=-{{p_i\cos(b_ic_i)}\over\gamma P_{\phi}}\Big[{p_j\over{b_j}}\sin(b_jc_j)+{p_k\over{b_k}}\sin(b_kc_k)\Big],
\end{equation}
\begin{equation}
    {{dc_i}\over{d\phi}}={\sin(b_ic_i)\over{\gamma b_i P_{\phi}}}\Big[{p_j\over{b_j}}\sin(b_jc_j)+{p_k\over{b_k}}\sin(b_kc_k)\Big],
\end{equation}
\label{Bsystem}
\end{subequations}
for $i,j,k=1,2,3$, $i\neq j\neq k$. It is possible to solve this system by establishing the initial conditions on the variables $(c_{i},p_{i})$, which also satisfy the Hamiltonian constraint \eqref{eqn:Hpoly}. In~this respect, we make the following choice:
\begin{equation}
	\label{condin}
	\begin{aligned}
	c_1(0)=&c_2(0)=c_3(0)={\pi\over 2 b_0},\\ p_1(0)=&\bar{p}_1,\quad p_2(0)=\bar{p}_2,\\
	p_3(0)=&\bar{p}_3=\frac{P_\phi^2\gamma^2 b_0-2\bar{p_1}\bar{p}_2}{2(\bar{p}_1+\bar{p}_2)},
	\end{aligned}
\end{equation}
where $b_1=b_2=b_3=b_0$ without loss of generality. Moreover, it can be easily seen that the momentum conjugate to the scalar field is a first integral since the variable $\phi$ is cyclic in \eqref{eqn:Hpoly}. Other constants of motion can be obtained by combining the Hamilton Equations \eqref{Bsystem}, so we obtain the following:
\begin{equation}
	{p_{i}\sin(b_0 c_i)\over{b_0}}=\mathcal{K}_{i},\hspace{0.5cm}P_{\phi}=\mathcal{K}_{\phi}\,,
\end{equation}
where the considered values of $\mathcal{K}_{\phi}$ and $\mathcal{K}_{i}$ depend on the initial conditions. Identifying these first integrals allows to transform the six-equations system shown in \eqref{Bsystem} in three closed systems along the three spatial directions as follows:
\begin{subequations}
\label{Bsystemdecoupled}
\begin{equation}
	{{dp_i}\over{d\phi}}=-{{p_i\cos(b_0 c_i)}\over\gamma P_{\phi}}\Big[\mathcal{K}_j+\mathcal{K}_k\Big]\,,
\end{equation}
\begin{equation}
	{{dc_i}\over{d\phi}}={\sin(b_0 c_i)\over{\gamma b_0 P_{\phi}}}\Big[\mathcal{K}_j+\mathcal{K}_k\Big]\,,
\end{equation}
\end{subequations}
where $i\neq j \neq k$.

Thanks to this procedure, the equations of motion can be solved analytically, leading to the following solutions:

\begin{equation}
	\begin{aligned}
		&p_1(\phi)=\bar{p}_1\cosh\Big[\frac{(\gamma^2P_\phi^2b_0^2+2\bar{p}_2^2)\phi}{2\gamma P_\phi b_0(\bar{p}_1+\bar{p}_2)}\Big],\\
		&p_2(\phi)=\bar{p}_2\cosh\Big[\frac{(\gamma^2P_\phi^2 b_0^2+2\bar{p}_1^2)\phi}{2\gamma P_\phi b_0(\bar{p}_1+\bar{p}_2)}\Big],\\
		&p_3(\phi)=\bar{p}_3\cosh\Big[\frac{(\bar{p}_1+\bar{p}_2)\phi}{\gamma P_\phi b_0}\Big].\\
	\end{aligned}
\end{equation}

\begin{figure}[H]
	\centering
	\includegraphics[width=0.8\linewidth]{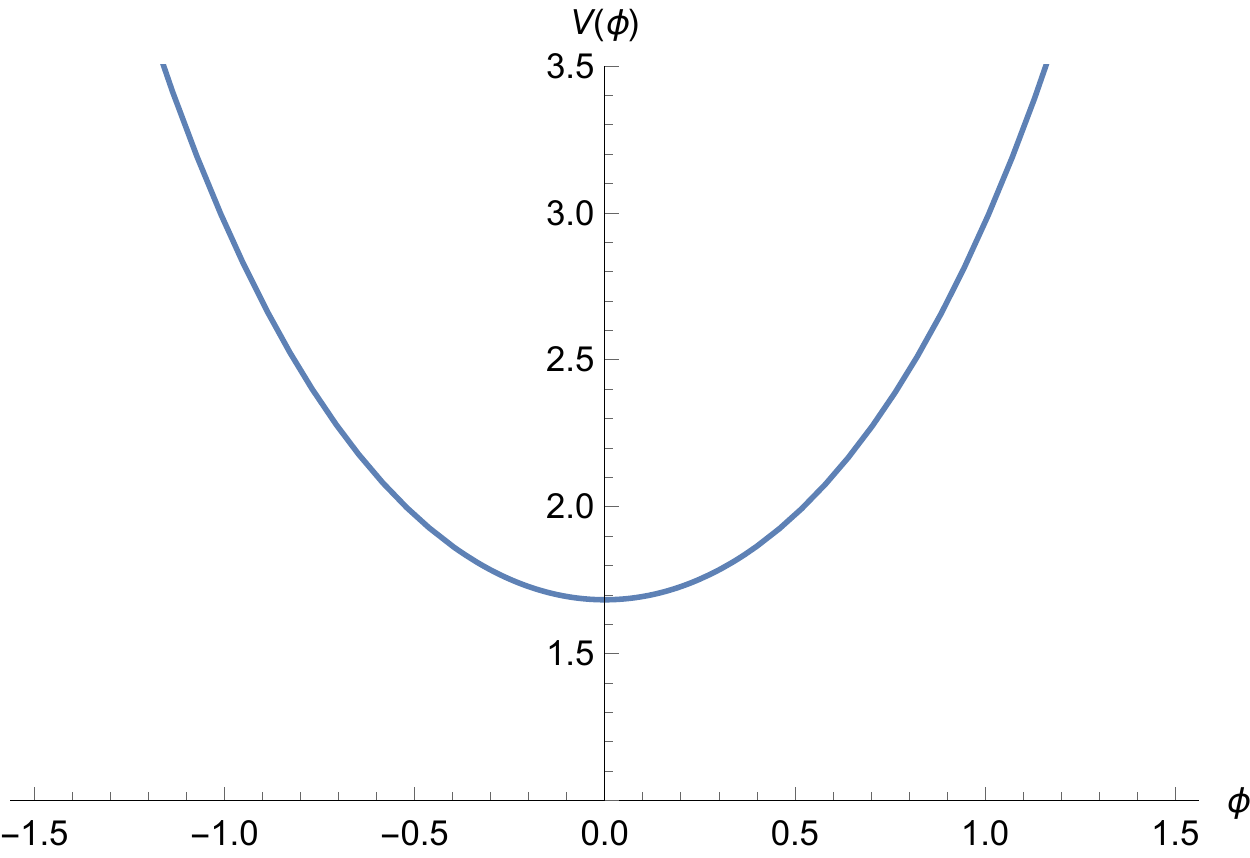}
	\caption{Polymer trajectory of the Universe volume $V=\sqrt{p_{1}p_{2}p_{3}}$ as function of $\phi$: the Big Bounce replaces the classical singularity of the Bianchi I model.}
	\label{V(phi)}
\end{figure}

We have reported only the explicit expression of the variables $p_i$ as functions of $\phi$ since we are interested in the Universe volume behavior $V(\phi)=\sqrt{p_1(\phi)p_2(\phi)p_3(\phi)}$ that is shown in Figure \ref{V(phi)}. The resulting trajectory highlights that a quantum Big Bounce replaces the classical Big Bang thanks to the polymer effects, which are expected to become dominant near the Planckian region. 

Now, we focus our attention on the analysis of the critical energy density of matter at the Bounce:
\begin{equation}
	\rho_\text{crit}^\phi=\frac{P_\phi^2}{2V(\phi)^2}\Bigg|_{\phi_B}=\frac{P_\phi^2}{2\bar{p}_1\bar{p}_2\bar{p}_3}=\frac{P_\phi^2(\bar{p}_1+\bar{p}_2)}{\bar{p}_1\bar{p}_2(\gamma^2 P_\phi^2b_0^2-2\bar{p}_1\bar{p}_2)},
	\label{rhocritash} 
\end{equation}
i.e., the energy density of the matter scalar field in correspondence of the minimum Universe volume at the time of the Bounce $\phi_B=0$. As we can see from \eqref{rhocritash}, $\rho_\text{crit}^\phi$~clearly depends on the initial conditions on the motion. Moreover, in the simplest case $\mathcal{K}_{1}=\mathcal{K}_{2}=\mathcal{K}_{3}$ it reduces to the  following:
\begin{equation}
	\rho_\text{crit}^\phi=\Big(\frac{3}{\gamma^2 b_0^2}\Big)^{\frac{3}{2}}\frac{\sqrt{2}}{P_\phi},
\end{equation}
reproducing a consistent behavior with that one obtained in the FLRW model in the same variables; see \eqref{rhocritsemiclpc}. Indeed, it is possible to see that for $P_{\phi}\ll1$ the critical matter energy density increases until it diverges, while on the other hand, it approaches zero when $P_{\phi}\gg1$, highlighting that in the Ashtekar formulation of the Bianchi I model, the Big Bounce has no universal features.
		
Let us now study this model on a quantum level by implementing the Dirac quantization. In the momentum representation of the polymer formulation, the fundamental operators act as follows:
\begin{align}
	\hat{p}_{i}:=-i\gamma {d\over dc_{i}}, \hspace{0.5cm} \hat{c}_i:={\sin(b_{i}c_{i})\over b_{i}},\hspace{0.5cm}
	\hat{P}_{\phi}:=-i{d\over d\phi}.
\end{align}

Before quantization, we rewrite the WDW equation in the form of a Schr\"odinger one in the attempt of defining a positive and conserved probability density. Therefore, we~recall the semiclassical scalar constraint and perform an ADM reduction of the variational~principle:
\begin{equation}
	\label{scalar}
	P_{\phi}^{2}-\Theta=0,
\end{equation}
where\vspace{6pt}
\begin{equation}
	\Theta={2\over\gamma^{2}}\Big[ {\sin(b_1c_1)p_1\sin(b_2c_2)p_2\over b_1b_2}+{\sin(b_1c_1)p_1\sin(b_3c_3)p_3\over b_1b_3}+{\sin(b_2c_2)p_2\sin(b_3c_3)p_3\over b_2b_3}\Big].
\end{equation}

After choosing the scalar field $\phi$ as the temporal parameter, we derive the ADM Hamiltonian by solving the scalar constraint \eqref{scalar} with respect to the momentum associated to the scalar field:
\begin{equation}
	P_{\phi}\equiv\mathcal{C}_\text{BI}^{\text{ADM-poly}}=\sqrt{\Theta}.
	\label{ADM}
\end{equation}
where we choose the positive root in order to guarantee the positive character of the lapse function. Thanks to this procedure, the WDW equation can be rewritten in the form of a Schr\"odinger one by promoting the ADM Hamiltonian to a quantum operator:
\begin{equation}
	\label{eqn:schro}
	-i\partial_{\phi}\Psi=\sqrt{\hat{\Theta}\,}\,\Psi,
\end{equation}
where the operator $\sqrt{\hat{\Theta}}$, that we assume well-defined, can be written as follows:
\begin{align}
	\sqrt{\hat{\Theta}}= \Big[{2\over\gamma^{2}}\Big(\partial_{x_{1}}\partial_{x_{2}}+\partial_{x_{1}}\partial_{x_{3}}+\partial_{x_{2}}\partial_{x_{3}}\Big)\Big]^{1/2},
\end{align}
where the new variables $x_i$ are defined from the connections $c_i$ as follows:
\begin{equation}
\label{xiS}
x_i=\ln\bigg[{\tan\bigg({\frac{\mu c_i}{2}}}\bigg)\bigg]+\bar{x}_i.
\end{equation}

Now, we are allowed to introduce the probability density as  follows:
\begin{equation}
	\label{J0density}
	\mathcal{P}(\vec x,\phi)=\Psi^{*}(\vec x,\phi)\Psi(\vec x,\phi),
\end{equation}
where 
\begin{equation}
	\Psi(\vec x,\phi)=\int_{-\infty}^{\infty}dk_1\,dk_2\,dk_3\,A(k_1,k_2,k_3)\,e^{i(k_1x_1+k_2x_2+k_3x_3+\sqrt{2\abs{k_1k_2+k_1k_3+k_2k_3}\,}\,\phi)},
	\label{packbis} 
\end{equation}
\begin{equation}
A(k_1,k_2,k_3)=\text{exp}\left(\sum_{i=1}^{3}\,-\frac{(k_i-\overline{k}_i)^2}{2\sigma_{k_i}^2}\right).
\end{equation}

This quantity is positive everywhere by definition, and its spatial integral remains constant through time.

Hence, we have discarded the covariant formulation of the WDW equation in favor of a Schr\"odinger-like one in order to define a probability density through which we analyze the quantum dynamics of the Bianchi I wave packet presented in \eqref{packbis}. This way, we~have avoided all the issues regarding the sign of the probability density that would be encountered in the interpretation of the WDW formulation as a Klein--Gordon-like theory.

\begin{figure}[H]
	\begin{minipage}{3.9cm}
		\includegraphics[width=1.5\linewidth]{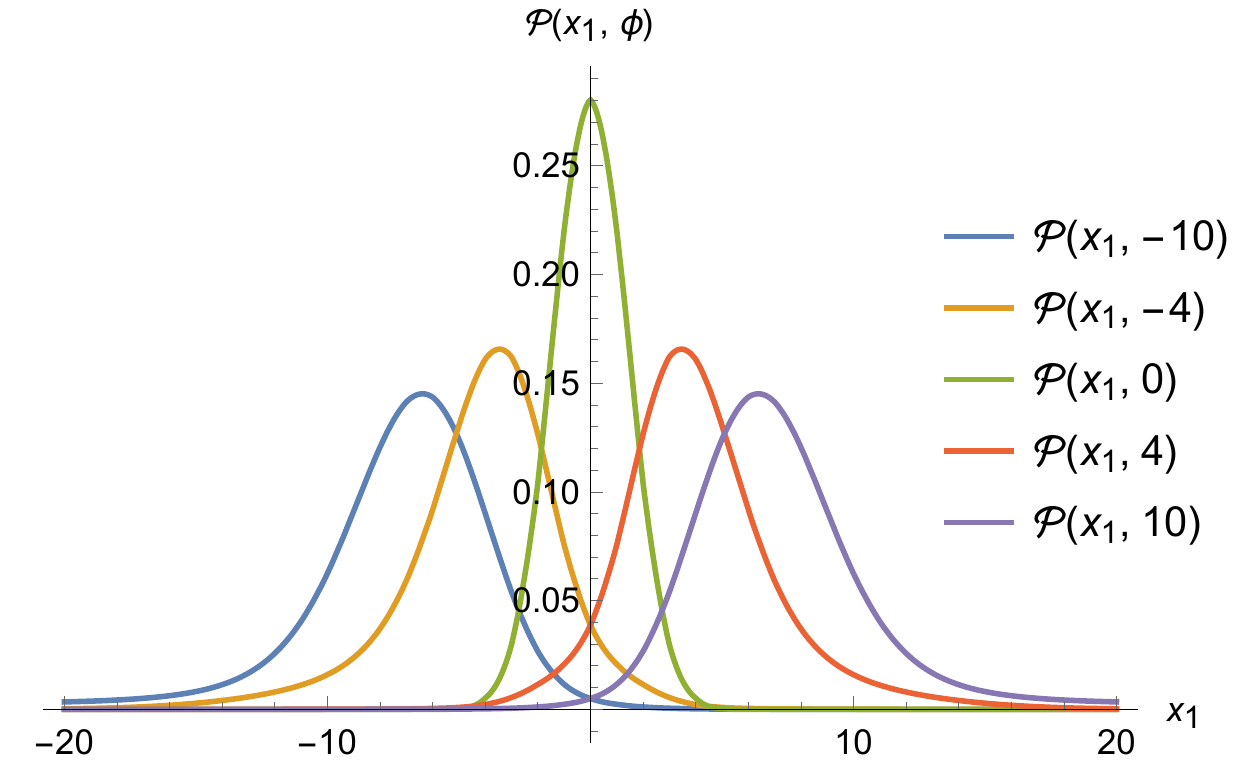}
	\end{minipage}
	\qquad\qquad\qquad
	\begin{minipage}{3.9cm}
		\includegraphics[width=1.5\linewidth]{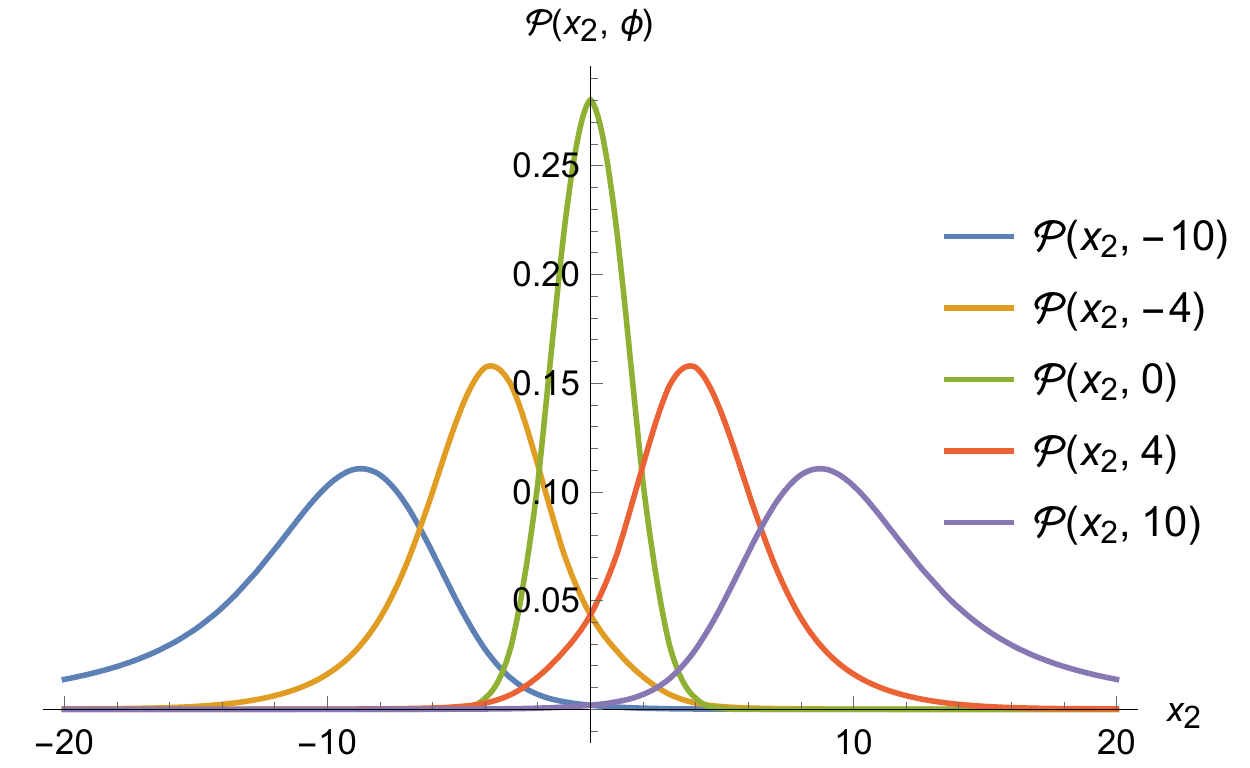}
	\end{minipage}
	\qquad\qquad\qquad
	\begin{minipage}{3.9cm}
		\includegraphics[width=1.5\linewidth]{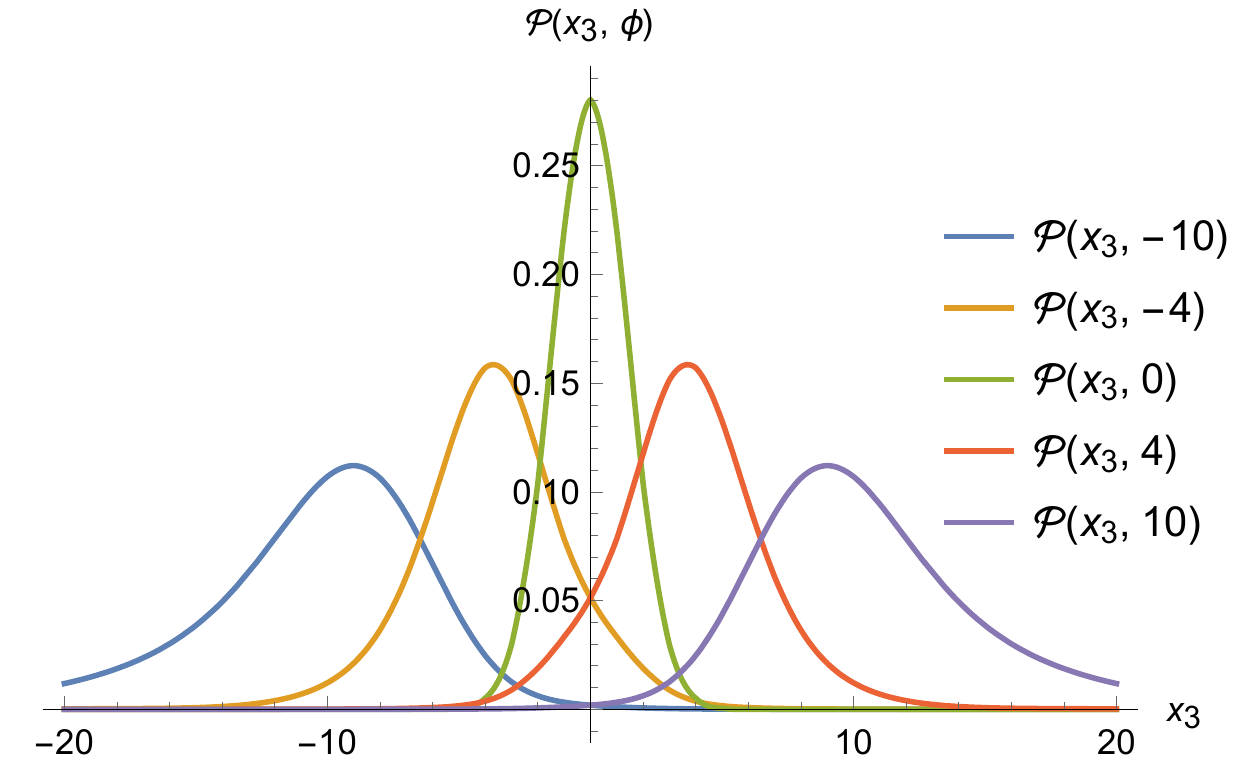}
	\end{minipage}
	\qquad\qquad\qquad\quad
	\caption{The normalized sections $\mathcal{P}(x_i,\phi)$ are shown in sequence for $i=1,2,3$ respectively at different times. Their spreading behavior over time is evident together with the Gaussian-like shape.}
	\label{qprob}
\end{figure}

\begin{figure}[H]
	\begin{minipage}{3.9cm}
		\includegraphics[width=1.5\linewidth]{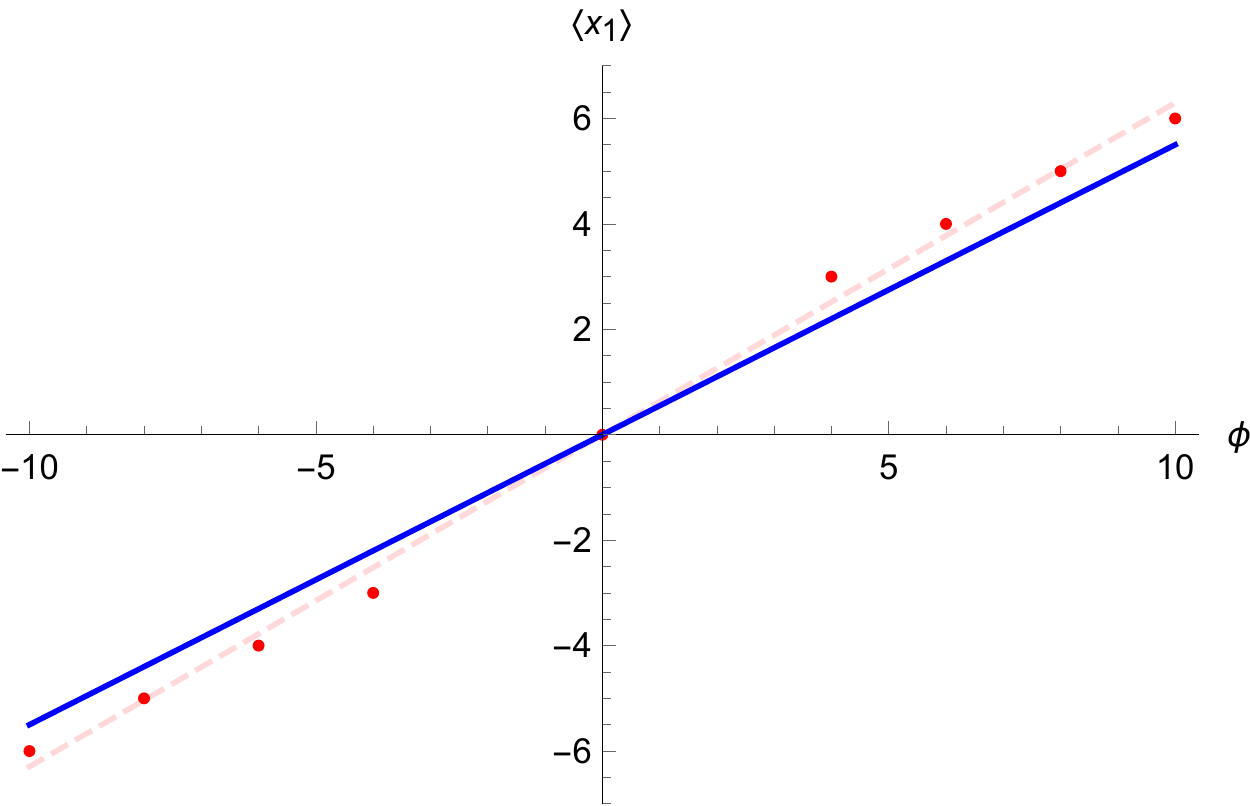}
	\end{minipage}
	\qquad\qquad\qquad
	\begin{minipage}{3.9cm}
		\includegraphics[width=1.5\linewidth]{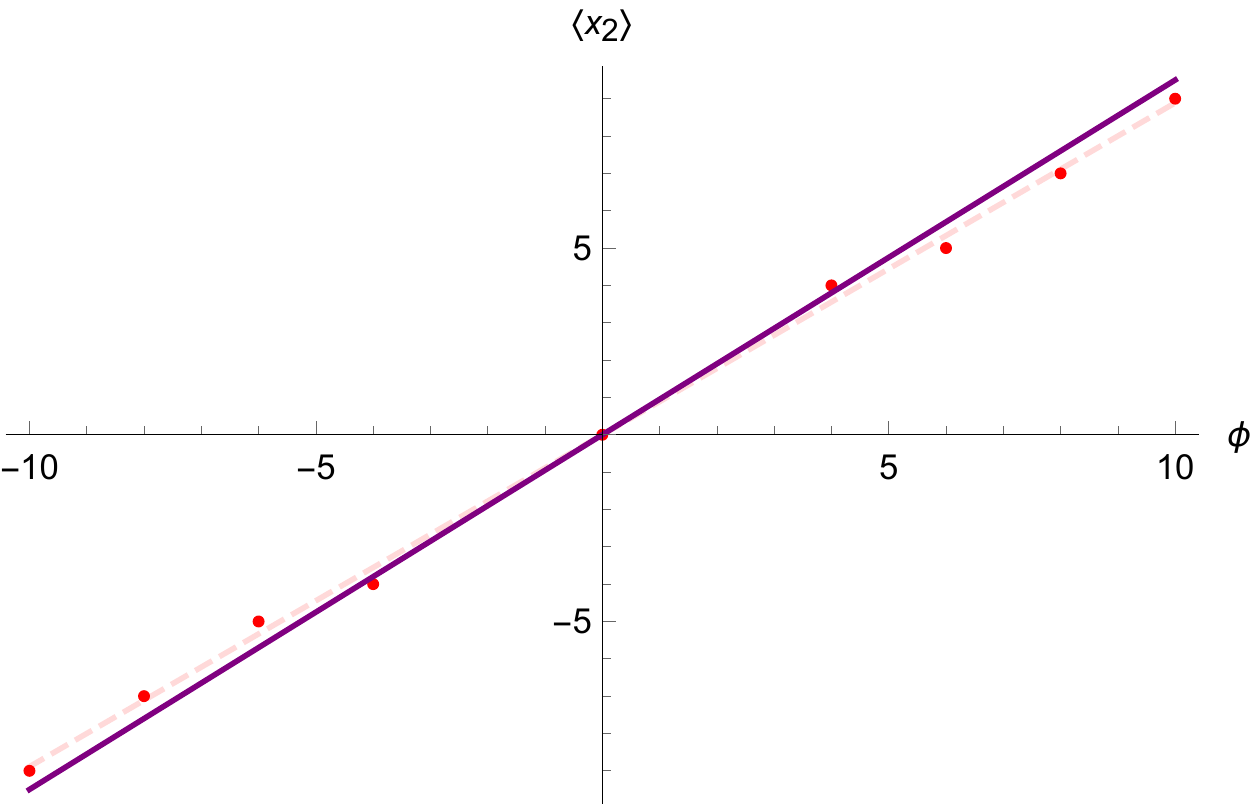}
	\end{minipage}
	\qquad\qquad\qquad
	\begin{minipage}{3.9cm}
		\includegraphics[width=1.5\linewidth]{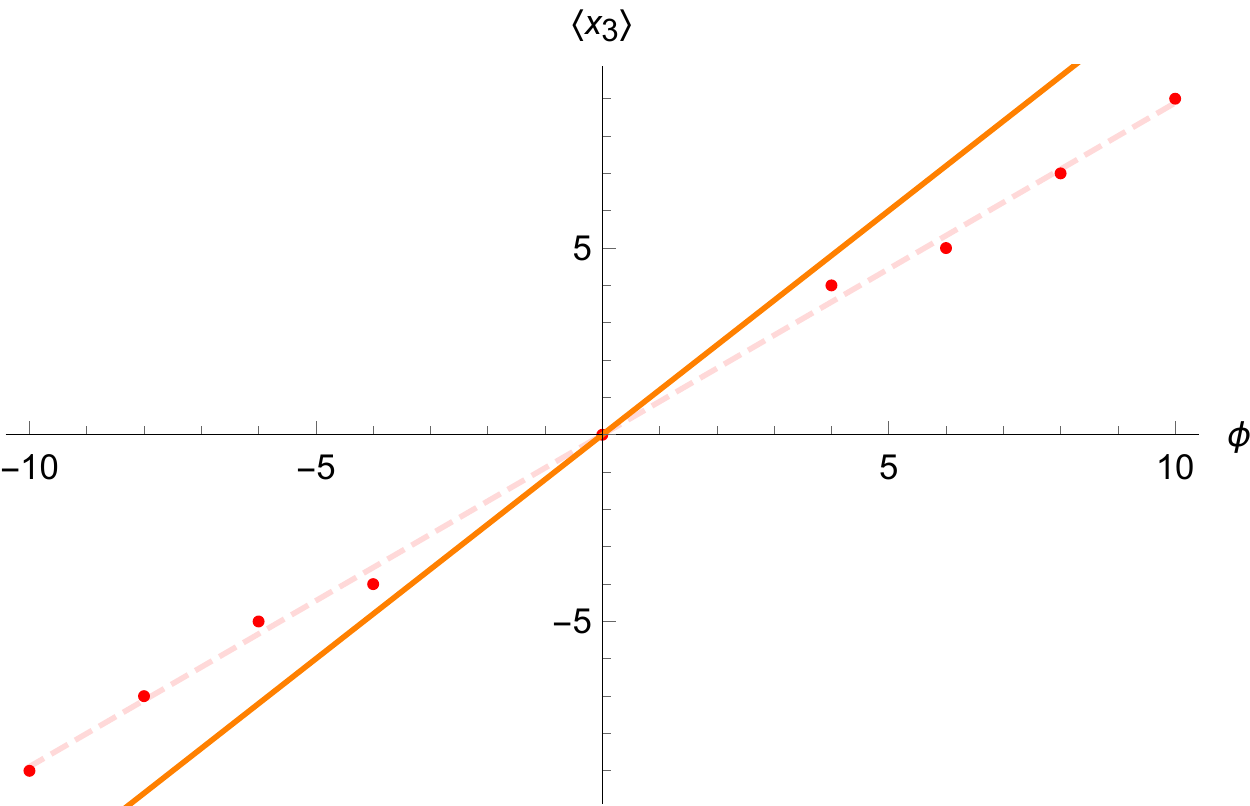}
	\end{minipage}
	\qquad\qquad\qquad\quad
	\caption{The three pictures show the position of the peaks of $\mathcal{P}(x_i,\phi)$ for $i=1,2,3$ respectively in function of time $\phi$ (red dots). The resulting fitting functions (red dashed straight lines) overlap the semiclassical trajectories (continuous lines) with a confidence level of the order of $3$ standard deviations.}
	\label{qpeak}
\end{figure}

In Figure \ref{qprob}, some different sections of the probability density $\mathcal{P}$ are shown at different times in order to verify how its shape and its maximum evolve. The present sections were obtained by fixing two of the three coordinates through the values that they assume in the semiclassical trajectories\footnote{In \cite{Silvia} it is shown that by combining the semiclassical solutions for $c_i(\phi)$ with $\eqref{xiS}$ we obtain that $x_i(\phi)\propto\phi$ with slopes depending on the initial conditions on the motion.}. As we can see from Figure \ref{qprob}, the normalized quantum distributions of $x_1$, $x_2$, $x_3$ are shown in sequence, and their spreading behavior over time is evident. Additionally, in Figure \ref{qpeak}, the position of the peaks of $\mathcal{P}$ is represented by the red dots that are fitted by means of a linear interpolation and compared with the semiclassical trajectories. We can affirm that there is a good correspondence between the quantum behavior of the wave packet and the solutions of the semiclassical dynamics since all the three slopes of the functions resulting from the fit of the red dots are consistent with the semiclassical ones with a confidence level of $3$ standard deviations (we have supposed a relative error of $10\%$, accounting for quantum fluctuations and numerical integration errors). Moreover, we want to highlight that this quantum analysis based on the sections of $\mathcal{P}$ is justified by the semiclassical decoupling of the equations of motion.  In conclusion, we~remark that this feature of our analysis in the Ashtekar variables suggests the presence of a bouncing dynamics with non-universal properties also at a quantum level.

\clearpage
\nointerlineskip

\vspace{-9pt}
\subsection{Polymer Cosmology in the Volume-Like Variables\label{VL}}
In this section, we describe the FLRW dynamics in the polymer framework when the polymer lattice is implemented on the Universe volume $v$. Then, we generalize the analysis by taking under consideration the Bianchi I model, which admits two different sets of volume-like variables thanks to its anisotropic structure. In particular, when the Universe volume itself is considered to be one of the configurational variables, it is possible to derive a polymer modified Friedmann equation for the Bianchi I model. We will see that the total critical energy density has universal features, as demonstrated in \cite{Mantero} for the isotropic case.

\subsubsection{The FLRW Universe in the Volume Variables\label{FLRW}}
Following \cite{Federico}, we outline the semiclassical dynamics of the model and then we perform a full quantum analysis by analyzing the properties  of the Universe wave packet. 
In~order to apply the polymer semiclassical framework to the flat FLRW model in the volume variables, we introduce the Hamiltonian constraint as follows:
\begin{equation}
	\mathcal{C}_\text{FLRW}=-\frac{27}{4\gamma^2}v\hspace{4pt}\Tilde{c}^2+\frac{P_\phi^2}{2v},
\end{equation} 
where the phase space variables $\Tilde{c},v$ are linked to the Ashtekar variables $c,p$ through the following canonical transformation:
\begin{equation}
	v=\abs{p}^{\frac{3}{2}}=a^3,\quad\Tilde{c}=\frac{2}{3}\frac{c}{\sqrt{\abs{p}}}\propto\frac{\dot a}{a}
	\label{variablevc}
\end{equation}
that preserves the Poisson brackets 	$\pb{\Tilde{c}}{v}=\frac{\gamma}{3}$.
Additionally, a massless scalar field has been added to the dynamics with the role of a relational time. 
Thus, the modified polymer Hamiltonian~is as follows:
\begin{equation}
	\label{Cpolytilde}
	\mathcal{C}_\text{FLRW}^{\text{poly}}=-\frac{27}{4\gamma^2b_0^2}\,v\,\mbox{sin}^2(b_0\Tilde{c})+\frac{P_\phi^2}{2v}=0,
\end{equation}
where the variable $v$ is defined as discrete and so the semiclassical polymer substitution \eqref{psub} $\tilde{c}\to\frac{\sin(b_0\tilde{c})}{b_0}$ is used for the generalized coordinate $\Tilde{c}$.

The equations of motion for these variables are as follows:
\begin{subequations}
	\label{Hamtilde}
	\begin{equation}
		\dot v=-\frac{9\mbox{N}}{2\gamma b_0}\,v\,\mbox{sin}(b_0\Tilde{c})\mbox{cos}(b_0\Tilde{c}),
		\label{Hamtilde1}
	\end{equation}
	\begin{equation}
		\Dot{\Tilde{c}}=\frac{\mbox{N}}{3}\Big(\frac{27}{4\gamma b_0^2}\sin[2](b_0\Tilde{c})+\gamma \frac{P_\phi^2}{2v^2}\Big).
	\end{equation}
\end{subequations}

As expected in the polymer framework, a modified Friedmann equation appears; in particular, we can obtain its analytical expression by using \eqref{Hamtilde1} and the vanishing Hamiltonian constraint \eqref{Cpolytilde}:
\begin{equation}
	\label{rho}
	H^2=\Big({\dot v\over 3v} \Big)^{2}=\frac{\rho}{3}\Big(1-\frac{\rho}{\rho_{\text{crit}}}\Big),\quad\rho_{\text{crit}}=\frac{27}{4\gamma^2b_0^2}.
\end{equation}

The presence of a bouncing point for the Hubble parameter represents the mechanism through which the initial singularity is regularized. In addition, the critical energy density at which the Big Bounce occurs is fixed by the spacing $b_0$ associated to the polymer lattice, giving the dynamics a universal character. This result is in strong contact with the analogous considerations made in the context of the $\bar{\mu}$ approach in the LQC formulation in \mbox{Section \ref{improvedLQC}}.

Considering now the scalar field $\phi$ as the internal time of the dynamics, we fix the time gauge, i.e.,
\begin{equation}
	1=\Dot{\phi}=\mbox{N}\frac{\partial\mathcal{C}_\text{FLRW}^{\text{poly}}}{\partial P_\phi}=\mbox{N}\frac{P_\phi}{v},\quad\mbox{N}=\frac{v}{P_\phi}=\frac{1}{\sqrt{2\rho}};
\end{equation}
thus the effective Friedmann equation in the ($v,\phi$) plane reads as follows:
\begin{equation}
	\label{nu}
	\Big(\frac{1}{v}\frac{dv}{d\phi}\Big)^2=\frac{3}{2}\Big(1-\frac{4\gamma^2b_0^2}{54}\frac{P_\phi^2}{v^2}\Big),
\end{equation}
whose analytical solution is as follows:
\begin{equation}
	v(\phi)=\sqrt{\frac{4\gamma^2b_0^2}{54}\,}\,P_\phi\,\cosh\left(\sqrt{\frac{3}{2}}\,\phi\right).
	\label{nufi}
\end{equation}

\begin{figure}[H]
	\centering 
	\includegraphics[width=0.8\linewidth]{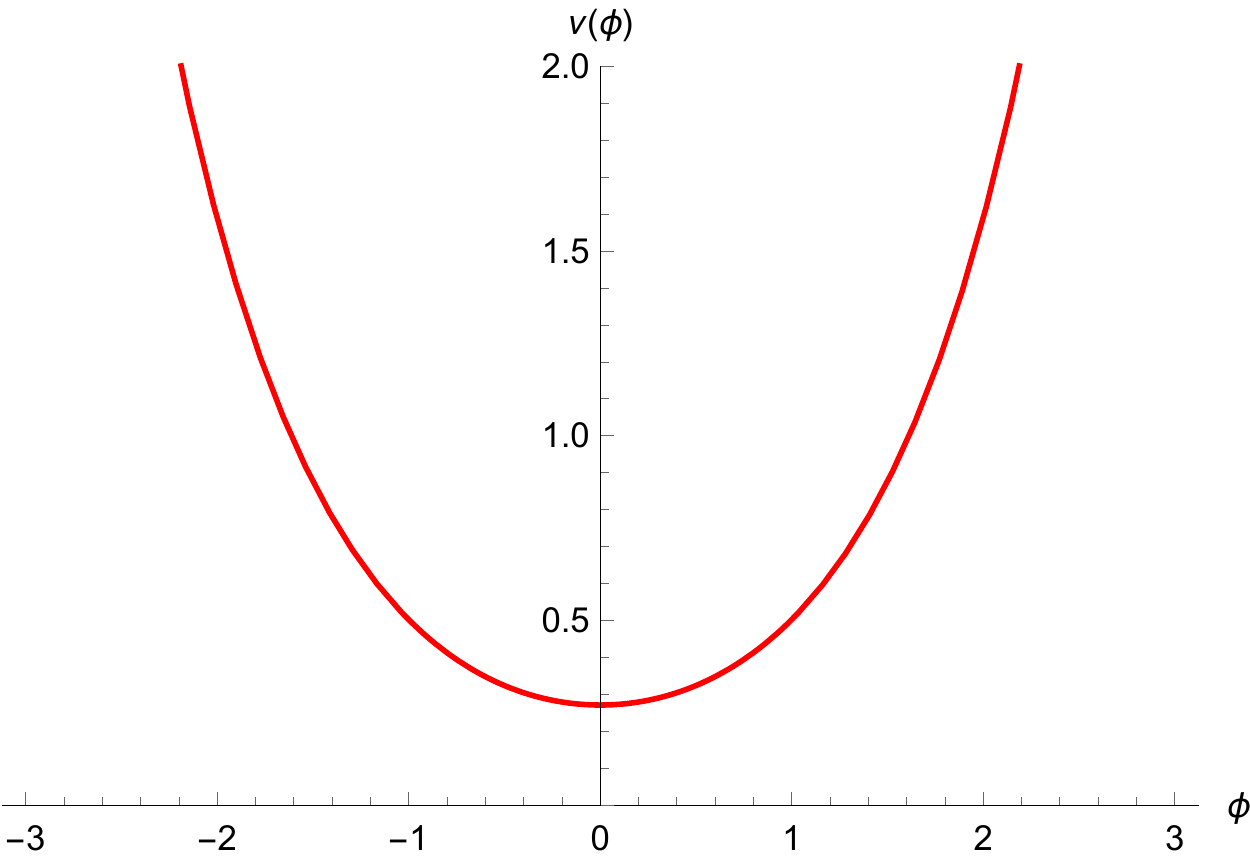} \caption{The polymer trajectory of the volume in $(v,\tilde{c})$ for the flat FLRW Universe. The volume presents a minimum, showing that a Big Bounce takes place.}
	\label{graficovpolymer}
\end{figure}

As shown in Figure \ref{graficovpolymer} the Bounce is clearly visible since the Universe possesses a minimum~volume.

Now, we promote the system to a full quantum level by applying the Dirac procedure~\cite{Matschull1996Dirac}. In particular, the Hamiltonian operator annihilates and selects the physical wave functions when applied to the generic quantum states, yielding the WDW equation:
\begin{equation}
	\hat{\mathcal{C}}_\text{FLRW}^{\text{poly}}\,\ket{\Psi}=0.
\end{equation}

If we promote the classical variables to a quantum level in the momentum representation, we obtain the following:
\begin{equation}
    \label{operatorsVc}
	\hat{v}=-\frac{i\gamma}{3}\frac{d}{d\tilde{c}},\quad\hat{\tilde{c}}=\frac{1}{b_0}\mbox{sin}(b_0\tilde{c}),\quad\hat{P}_\phi=-i\frac{d}{d\phi},
\end{equation}
recalling that $v$ is discrete and therefore $\Tilde{c}$ must be regularized. So, the quantum version of the Hamiltonian constraint \eqref{Cpolytilde} is as follows:
\begin{equation}
	\Big[-\frac{3}{2b_0^2}\Big(\mbox{sin}\Big(b_0 \Tilde{c}\Big)\frac{d}{d\Tilde{c}}\Big)^2+\frac{d^2}{d\phi^2}\Big]\Psi(\Tilde{c},\phi)=0;
	\label{wheelerpoly2}
\end{equation}

Using again a mixed factor ordering and the substitution 
\begin{equation}
	\tilde{x}=\sqrt{\frac{2}{3}}\ln{\Big[\tan\Big(\frac{b_0 \Tilde{c}}{2}\Big)\Big]}+\tilde{\bar{x}},
	\label{cambiovariabili2}
\end{equation}
we can recast recast the expression \eqref{wheelerpoly2} in the form of a massless Klein--Gordon-like equation:
\begin{equation}
	\frac{d^2}{d\tilde{x}^2}\Psi(\tilde{x},\phi)=\frac{d^2}{d\phi^2}\Psi(\tilde{x},\phi).
\end{equation}

This way, the wave packet representing the wave function of the Universe can be written as follows:
\begin{equation}
	\label{psi}
	\Psi(\Tilde{x},\phi)=\displaystyle\int_0^{\infty}dk_\phi\hspace{4pt}\frac{e^{-\frac{\abs{k_\phi-\overline{k}_\phi}^2}{2\sigma^2}}}{\sqrt{4\pi\sigma^2}}\,k_\phi\,e^{ik_\phi\tilde{x}}e^{-ik_\phi\phi}
\end{equation}
in the $\Tilde{x}$-representation. We have constructed a superposition of plane waves in $\phi$ by means of Gaussian-like coefficients and we have restricted the analysis only to the positive energy-like eigenvalues $k_\phi$.

In this case, to study the bouncing behavior of the model, we can evaluate the expectation values of both the volume and the energy density operators:
\begin{equation}
	\hat{V}=\hat{v},\quad\hat{\rho}=\frac{\hat{P}_\phi^2}{2\hat{v}^2}.
	\label{operatorsFRLWvc}
\end{equation}

\begin{figure}[H]
	\centering
	\includegraphics[width=0.8\linewidth]{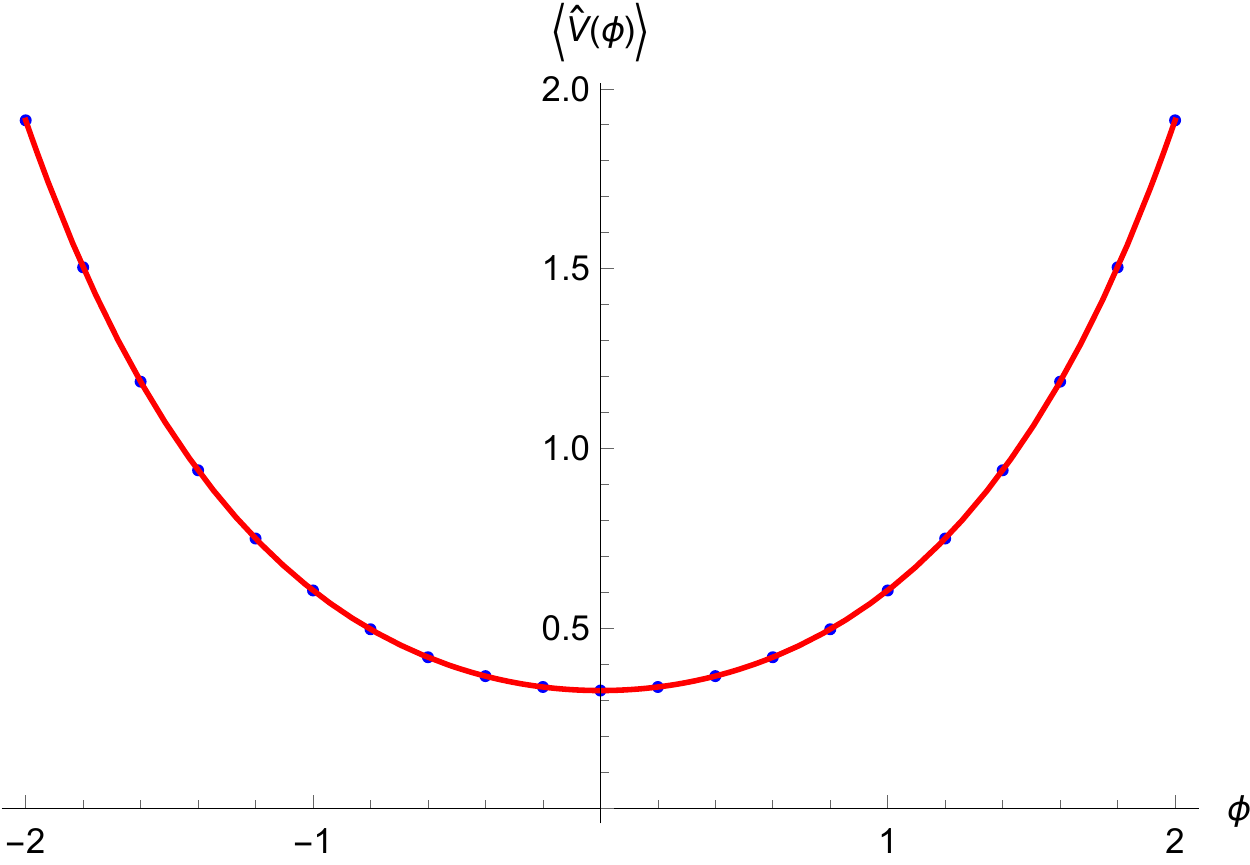}
	\caption{The expectation value of the Universe volume as a function of time (blue dots), fitted with a function in accordance with the semiclassical evolution (continuous red line).}
	\label{V(phi)Vc}
\end{figure}

\begin{figure}[H]
	\centering
	\includegraphics[width=0.8\linewidth]{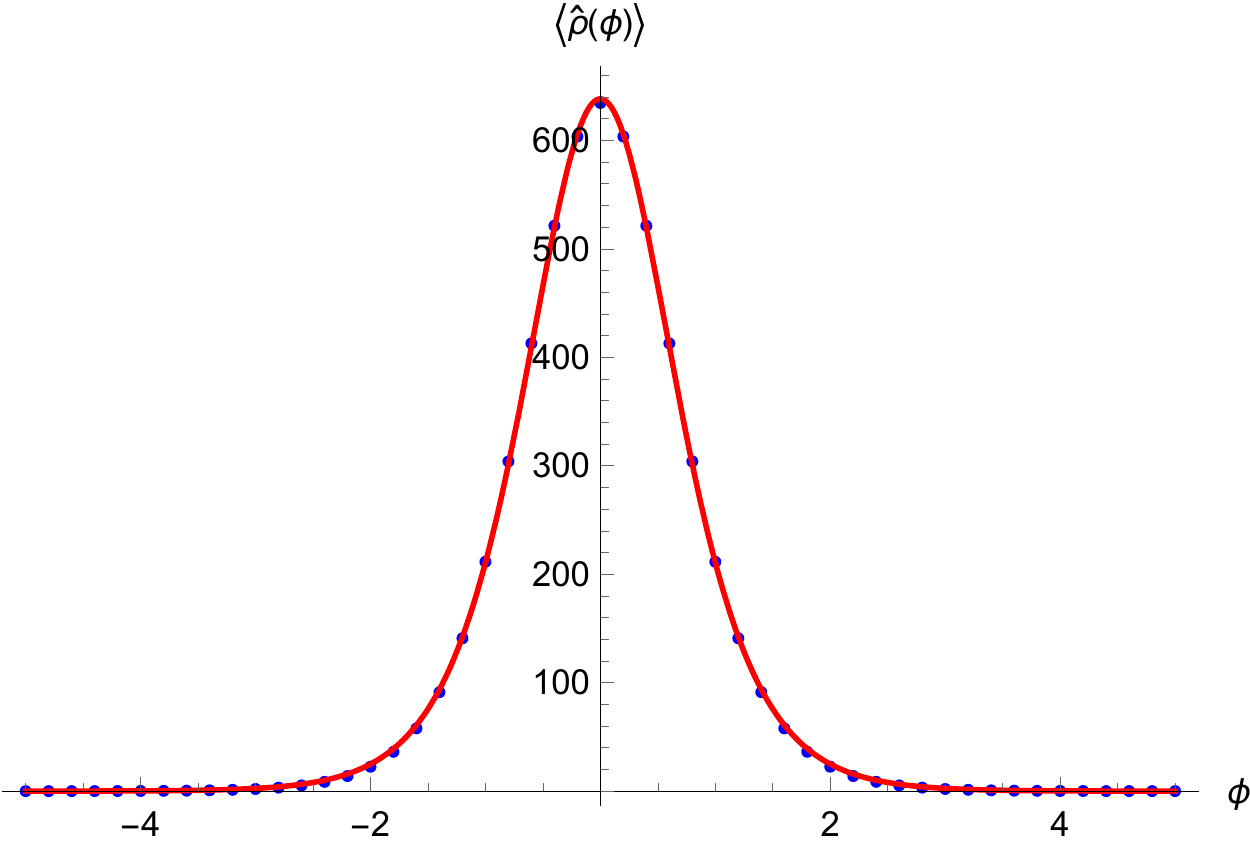}
	\caption{The expectation value of the energy density as a function of time (blue dots), fitted with a function in accordance with the semiclassical evolution (continuous red line).}
	\label{RhoVc}
\end{figure}

\begin{figure}[H]
	\centering
	\includegraphics[width=0.8\linewidth]{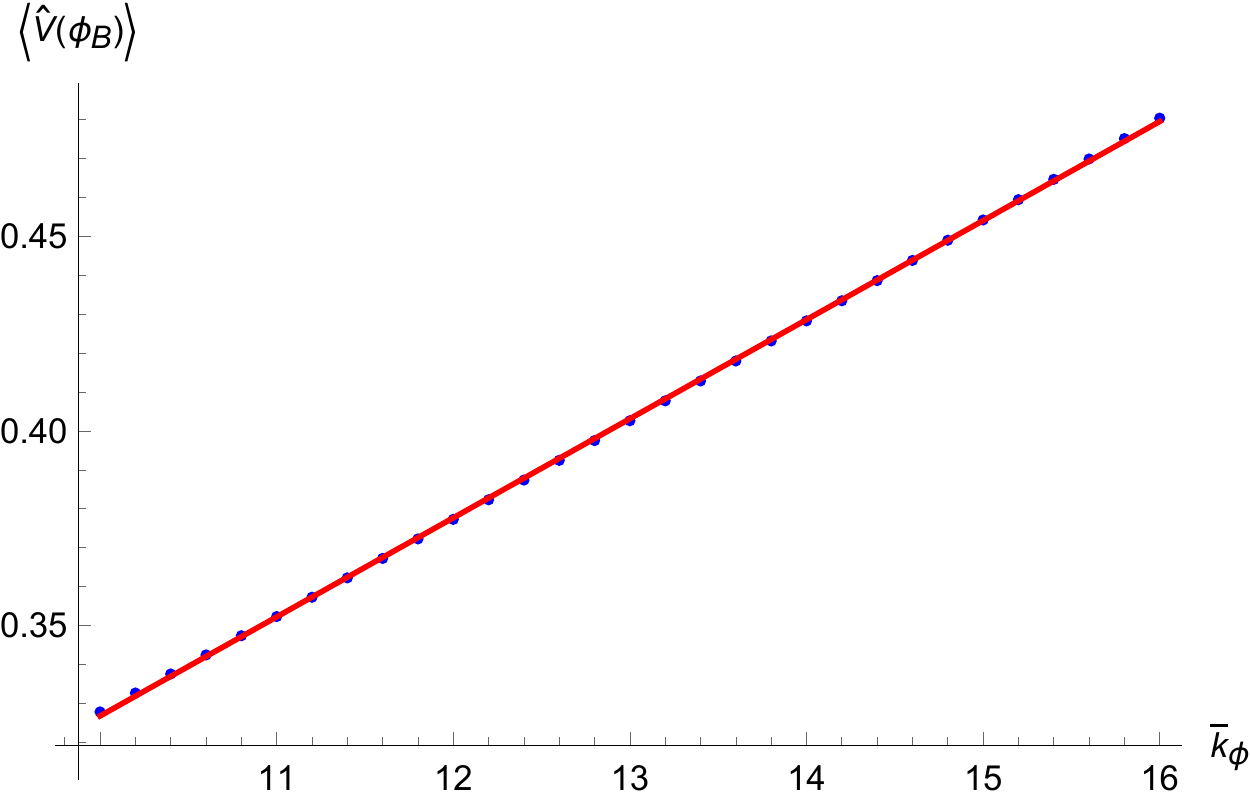}
	\caption{The expectation value of the Universe volume at the time $\phi_B$ of the Bounce as a function of $\overline{k}_\phi$ (blue dots), fitted with a function in accordance with the semiclassical evolution (continuous red~line).}
	\label{VminVc}
\end{figure}

\begin{figure}[H]
	\includegraphics[width=0.8\linewidth]{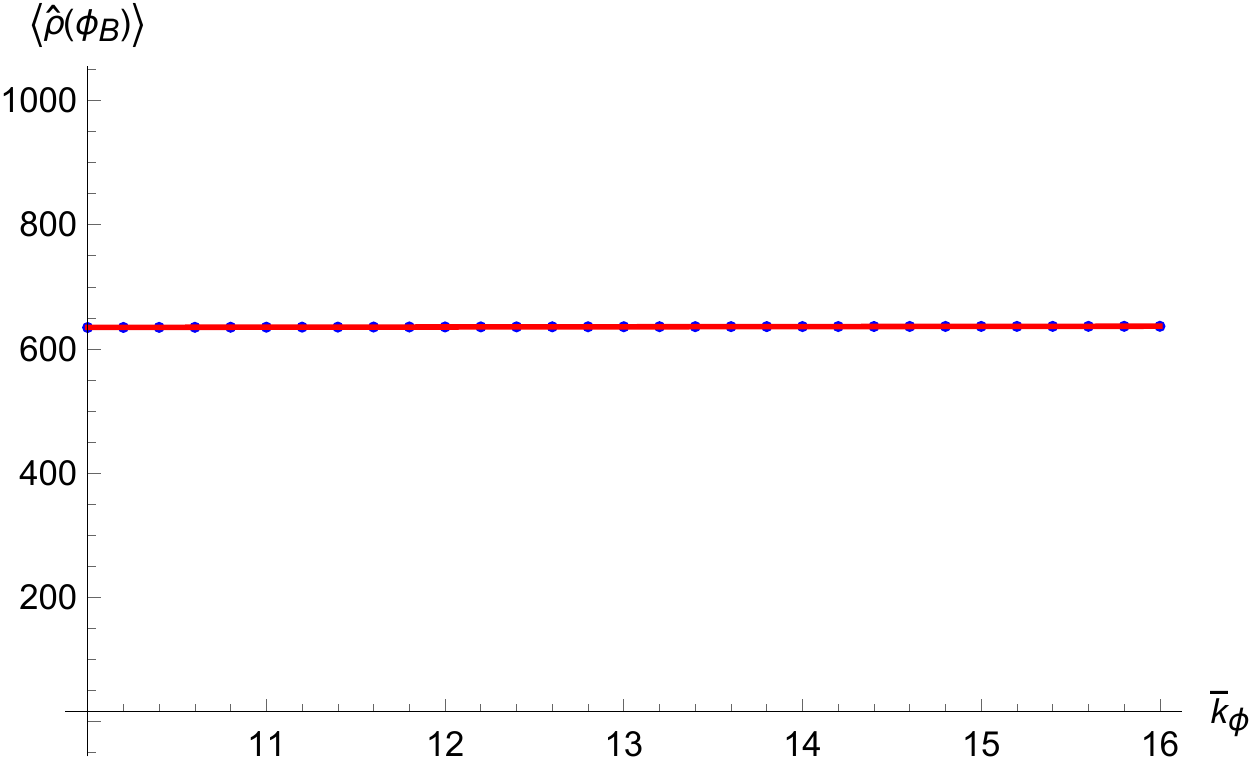}
	\caption{The expectation value of the energy density at the time $\phi_B$ of the Bounce as a function of $\overline{k}_\phi$ (blue dots), fitted with a function in accordance with the semiclassical evolution (continuous red~line).}
	\label{RhomaxVc}
\end{figure}

In what follows, we express all the quantities as functions of the new variable $\tilde{x}$ using \eqref{cambiovariabili2}, so that we can calculate the expectation values through the Klein--Gordon scalar~product as follows:
\begin{equation}
	\ev{\hat{O}}{\Psi}=\int_{-\infty}^{\infty}d\tilde{x}\,\,i\left(\Psi^*\,\partial_\phi(\hat{O}\Psi)-(\hat{O}\Psi)\,\partial_\phi\Psi^*\right),
	\label{KGv}
\end{equation}
where we assume normalized wave functions. The action of the operators \eqref{operatorsFRLWvc} is derived from the basic operators shown in \eqref{operatorsVc}.
In Figures \ref{V(phi)Vc} and \ref{RhoVc}, we show, respectively, the expectation values $\langle\hat{V}(\phi)\rangle$ and $\ev{\hat{\rho}(\phi)}$ as functions of time. In Figure \ref{VminVc}, the value $\ev{\hat{V}(\phi_B)}$ of the volume at the Bounce is shown as a function of the initial value $\overline{k}_\phi$; it is clear how the minimum volume scales linearly with the energy-like eigenvalue, in accordance with \eqref{nufi}. Then, in Figure \ref{RhomaxVc}, we show the Bounce density $\ev{\hat{\rho}(\phi_B)}$ for different values of $\overline{k}_\phi$. In~accordance with the semiclassical critical density \eqref{rho}, choosing the volume itself as the configurational variable for the quantization of the system makes the density at the Bounce independent from the initial conditions on the quantum solution \eqref{psi}. This quantum analysis allows a direct comparison with the $\bar{\mu}$ scheme of LQC, and it shows how polymer quantum cosmology is unable to reproduce the inverse triad corrections. We notice that in Figures \ref{V(phi)Vc}--\ref{RhomaxVc}, the numerically integrated points are fitted with the continuous lines to check for consistency with the semiclassical solutions.



\subsubsection{The Bianchi I Model in the Anisotropic Volume Variables: $(v_{1},v_{2},v_{3})$}
In this subsection, we study the polymer semiclassical dynamics of the Bianchi I model in complete analogy with the analysis performed for the FLRW model in the volume variables, as presented in \cite{Silvia}. More specifically, the anisotropic character of the Bianchi I model leads to the possibility of taking into account two different sets of volume-like variables that coincide in the case of the isotropic model. The first that we take under consideration consists of three equivalent generalized coordinates (see \cite{Szulc_2008}):
\begin{equation}
	v_{i}=\text{sgn}(p_{i})\abs{p_{i}}^{3\over2},\quad\eta_{i}={c_{i}\over\sqrt{\abs{p_{i}}}}
\end{equation}
for $i=1,2,3$, where $\eta_i$ are the conjugate momenta and the new symplectic structure for the system is characterized by the following Poisson brackets:
\begin{equation}
	\{\eta_{i},v_{j}\}={3\over 2}\gamma\delta_{ij}.
\end{equation}

In this case, we are not promoting one of the configurational variables to represent the Universe volume. On the contrary, we are imposing that the three independent coordinates are isomorphic to the isotropic volume for each direction so that $V=(v_{1}v_{2}v_{3})^{1\over3}$. In~particular, each coordinate $v_i$ reduces to the Universe volume in the isotropic limit. 

The Hamiltonian constraint for this framework in the polymer representation reads as follows:
\begin{equation}
\label{HVv}
	\mathcal{C}_\text{BI}^{\text{poly}}=-{1\over{4\gamma^{2}V}}\sum_{i\neq j}{v_{i}\sin(b_{i}\eta_{i})v_{j}\sin(b_{j}\eta_{j})\over b_{i}b_{j}}+{P_{\phi}^{2}\over 2 V}=0,
\end{equation}
where $i,j=1,2,3$ and the substitution $\eta_i\to\frac{\sin(b_i\eta_i)}{b_i}$ is performed.

If we fix the gauge through the choice $N=\frac{V}{P_\phi}$, the Hamilton equations will describe the dynamics of the model with respect to $\phi$ playing the role of the time variable:
\begin{subequations}
\begin{equation}
    \frac{dv_i}{d\phi}=-{{v_i\cos(b_i\eta_i)}\over 4\gamma P_{\phi}}\bigg[{v_j\over{b_j}}\sin(b_j\eta_j)+{v_k\over{b_k}}\sin(b_k\eta_k)\bigg]
\end{equation}
\begin{equation}
    \frac{d\eta_i}{d\phi}={\sin(b_i\eta_i)\over{4\gamma b_iP_{\phi}}}\bigg[{v_j\over{b_j}}\sin(b_j\eta_j)+{v_k\over{b_k}}\sin(b_k\eta_k)\bigg]
\end{equation}
\end{subequations}
for $i,j,k=1,2,3$ and $i\neq j\neq k$. In analogy with the previous treatment, we can identity the following constants of motion:
\begin{equation}
	{v_{i}\sin(b_0\eta_i)\over{b_0}}=\mathcal{K}_{i},\hspace{0.5cm}P_{\phi}=\mathcal{K}_{\phi}
\end{equation}
where $b_1=b_2=b_3=b_0$. Taking general initial conditions that satisfy the constraint \eqref{HVv}, the analytical solutions for the anisotropic volume coordinates read as follows:
\begin{equation}
	\label{aniV}
	\begin{aligned}
		&v_1(\phi)=\bar{v}_1\cosh\Big[\frac{(2\gamma^2P_\phi^2b_0^2+\bar{v}_2^2)\phi}{4\gamma P_\phi b_0(\bar{v}_1+\bar{v}_2)}\Big],\\
		&v_2(\phi)=\bar{v}_2\cosh\Big[\frac{(2\gamma^2P_\phi^2b_0^2+\bar{v}_1^2)\phi}{4\gamma P_\phi b_0(\bar{v}_1+\bar{v}_2)}\Big],\\
		&v_3(\phi)=\bar{v}_3\cosh\Big[\frac{(\bar{v}_1+\bar{v}_2)\phi}{4\gamma P_\phi b_0}\Big],
	\end{aligned}
\end{equation}
where
\begin{equation}
    \begin{aligned}
    v_1(0)&=\bar{v}_1,\quad v_2(0)=\bar{v}_2,\\
    v_3(0)=&\bar{v}_3=\frac{(2\gamma^2b_0^2P_\phi^2-\bar{v}_1\bar{v}_2)}{(\bar{v}_1+\bar{v}_2)}.
    \end{aligned}
\end{equation}

\begin{figure}[H]
	\centering
	\includegraphics[width=0.8\linewidth]{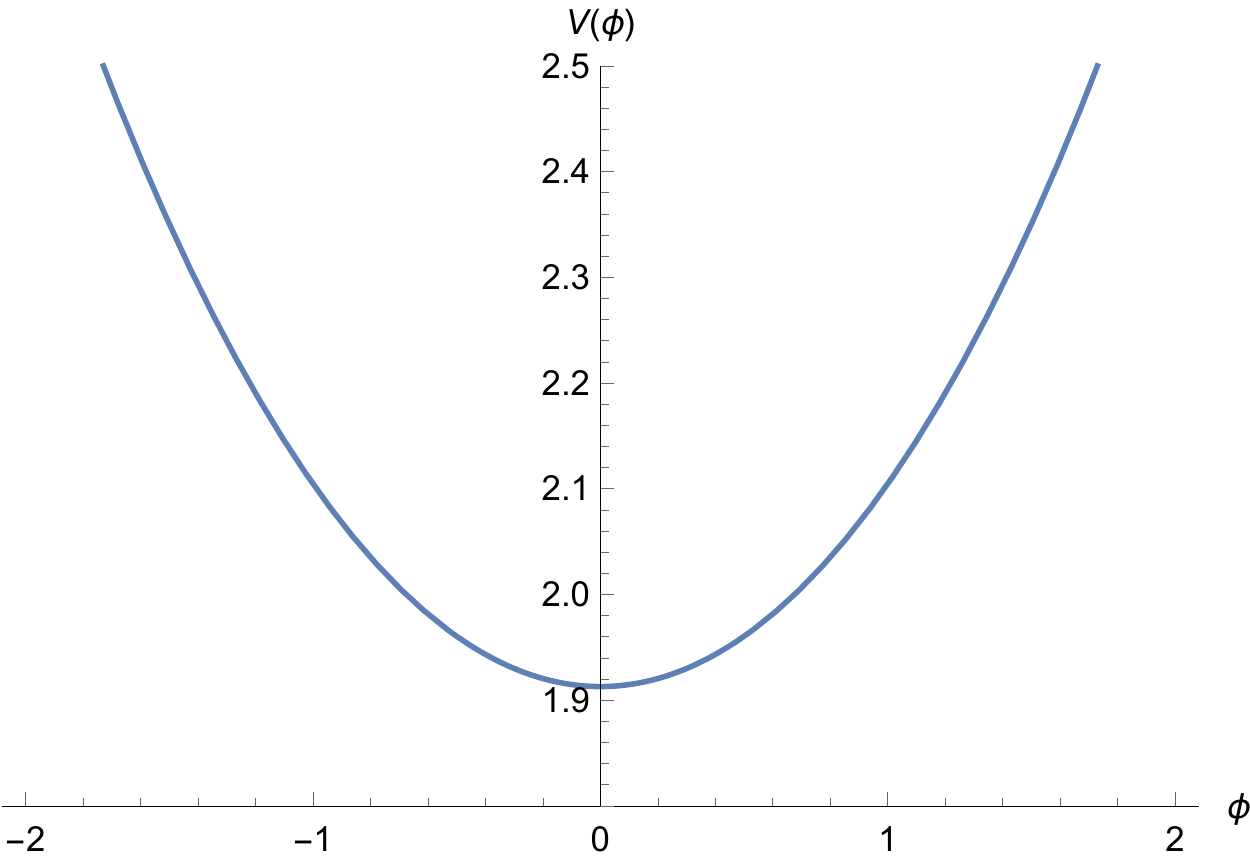}
	\caption{Bouncing trajectory of the Universe volume $V=(v_1v_2v_3)^{1/3}$ as function of $\phi$ in the semiclassical polymer framework.}
	\label{Vv(phi)}
\end{figure}

By combining the solutions \eqref{aniV}, we can find the Universe volume behavior in function of $\phi$ as $V(\phi)=(v_1(\phi)v_2(\phi)v_3(\phi))^{1/3}$. 
As shown in Figure \ref{Vv(phi)}, the Big Bounce appears as a polymer regularization effect in place of the classical Big Bang. Analogously, it is possible to investigate the properties of the critical energy density of matter by computing the energy density of the scalar field in correspondence of the minimum volume at the time $\phi_B=0$ of the Bounce, i.e., the following:
\begin{equation}
	\rho_\text{crit}^\phi=\frac{P_\phi^2}{2V(\phi)^2}\Bigg|_{\phi_B}=\frac{P_\phi^2}{2(\bar{v}_1\bar{v}_2\bar{v}_3)^{2/3}}=\frac{P_\phi^2(\bar{v}_1+\bar{v}_2)^{2/3}}{2[\bar{v}_1\bar{v}_2(2\gamma^2P_\phi^2b_0^2-\bar{v}_1\bar{v}_2)]^{2/3}};
	\label{rhocritVv}
\end{equation}
this results to be dependent on the initial conditions also in this set of anisotropic volume variables. In the next section, we implement directly the Universe volume as a generalized coordinate, trying to overcome this issue.
\subsubsection{The Bianchi I Model in the Volume Variables: $(v,q_{1},q_{2})$ \label{Vl}}
Now, we consider the set of volume variables introduced in \cite{Silvia}:
\begin{equation}
	q_{i}=\text{sgn}(p_{i})\sqrt{\abs{p_{i}}\,}, \hspace{0.5cm} v=q_{1}q_{2}q_{3}
\end{equation}
where $\eta_{i}$ (with $i=1,2,3$) are the  conjugate momenta and $v$ represents the Universe volume. The Poisson brackets are as follows:
\begin{equation}
	\{\eta_{i},	q_{j}\}=\gamma\delta_{ij}.
\end{equation}
and the Hamiltonian takes the following form:
\begin{equation}
	\mathcal{C}_\text{BI}=-{1\over{4\gamma^{2}V}}(q_{1}\eta_{1}q_{2}\eta_{2}+q_{1}\eta_{1}v\eta_{3}+q_{2}\eta_{2}v\eta_{3})+{P_{\phi}^{2}\over 2 V}=0.
\end{equation}

When we rewrite this expression using the polymer substitution $\eta_i\to\frac{\sin(b_i\eta_i)}{b_i}$, we obtain the following:
\vspace{6pt}
\begin{equation}
\label{HV}
\mathcal{C}_\text{BI}^{\text{poly}}=-{1\over{4\gamma^{2}V}}\Big(\sum_{i=1,2}{q_{i}\sin(b_{i}\eta_{i})v\sin(b_{3}\eta_{3})\over b_{i}b_{3}}+{q_{1}\sin(b_{1}\eta_{1})q_{2}\sin(b_{2}\eta_{2})\over b_{1}b_{2}}\Big)+{P_{\phi}^{2}\over 2 V}=0.
\end{equation}

In order to derive the dynamics of the model in function of the relational time $\phi$, we~impose the  following:
\begin{equation}
	\label{shiftf}
	\dot \phi:=\text{N}{{\partial \mathcal{C}_\text{BI}^{\text{poly}}}\over{\partial P_\phi}}=1,\quad \text{N}=\frac{V}{P_\phi},
\end{equation} so the Hamilton equations for the couple ($v,\eta_{3}$) take the following expressions:
\begin{subequations}
\begin{equation}
\label{volumev}
	\frac{dv}{d\phi}=-{{v\cos(b_3\eta_3)}\over 4\gamma P_{\phi}}\bigg[{q_1\over{b_1}}\sin(b_1\eta_1)+{q_{2}\over{b_2}}\sin(b_2\eta_2)\bigg],
\end{equation}
\begin{equation}
	\frac{d\eta_3}{d\phi}={\sin(b_3\eta_3)\over{4\gamma b_3P_{\phi}}}\bigg[{q_1\over{b_1}}\sin(b_1\eta_1)+{q_{2}\over{b_2}}\sin(b_2\eta_2)\bigg],
\end{equation}
\end{subequations}
while for the conjugate variables $(q_1,\eta_1)$, $(q_2,\eta_2)$ we have the following:
\begin{subequations}
\begin{equation}
	\frac{dq_i}{d\phi}=-{{q_i\cos(b_i\eta_i)}\over 4\gamma P_{\phi}}\bigg[{v\over{b_3}}\sin(b_3\eta_3)+{q_{j}\over{b_j}}\sin(b_j\eta_j)\bigg],
\end{equation}
\begin{equation}
    \frac{d\eta_i}{d\phi}={\sin(b_i\eta_i)\over{4\gamma b_iP_{\phi}}}\bigg[{v\over{b_3}}\sin(b_3\eta_3)+{q_{j}\over{b_j}}\sin(b_j\eta_j)\bigg],
\end{equation}
\end{subequations}
for $i,j=1,2$ and $i\neq j$.

Once fixed the initial conditions on the variables $(q_1,\eta_1)$, $(q_2,\eta_2)$, $(v,\eta_3)$ according to \eqref{HV}, we can solve this system analytically since the 3D motion is decoupled in three one-dimensional trajectories, thanks to the use of the constants of motion as follows:
\begin{equation}
	{q_{i}\sin(b_0\eta_i)\over{b_0}}=\mathcal{K}_{i},\quad P_{\phi}=\mathcal{K}_{\phi}
\end{equation}
where $i=1,2,3$ and $b_1=b_2=b_3=b_0$. Differently from the previous analyses, we fix the constants of motion as follows:
\begin{equation}
	\begin{aligned}
		\label{cond}
			&\mathcal{K}_1=\sqrt{\frac{2\gamma^2P_\phi^2+\mathcal{K}^2}{3}}+\mathcal{K},\\
			&\mathcal{K}_2=\sqrt{\frac{2\gamma^2P_\phi^2+\mathcal{K}^2}{3}}-\mathcal{K},\\
			&\mathcal{K}_3=\sqrt{\frac{2\gamma^2P_\phi^2+\mathcal{K}^2}{3}},
	\end{aligned}
\end{equation}
where $\mathcal{K}$ is an arbitrary constant. As we will see below, this choice allows to write a convenient form for the Friedmann equation in order to develop a more rigorous analysis of the bouncing behavior of the model. Indeed, the existence of a non-trivial solution to the equation $H^2=0$ identifies the expression of the critical energy density for which the scale factor velocity becomes null. More precisely, this information allows to identify also the anisotropy contribution to the total critical energy density added to the standard one associated to the matter fields. This way, it is possible to rigorously analyze the physical properties of the critical point.

In this set of volume variables, the Hubble parameter can be written as follows:
\begin{equation}
\begin{aligned}
	&H^2=\Big(\frac{1}{3v}\frac{dv}{dt}\Big)^2=\frac{(\mathcal{K}_1+\mathcal{K}_2)^2}{144\gamma^2v^2}\cos^2(b_0\eta_3)=\\
	=&\frac{(\mathcal{K}_1+\mathcal{K}_2)^2}{144\gamma^2v^2}[1-\sin^2(b_0\eta_3)]=\frac{(\mathcal{K}_1+\mathcal{K}_2)^2}{144\gamma^2v^2}(1-\frac{b_0^2}{v^2}\mathcal{K}_3^2)
	\label{BF}
\end{aligned}
\end{equation}
where we restored the synchronous time-gauge $\text{N}=1$ in the equation for the volume written in \eqref{volumev}. Now, if we substitute the conditions expressed above in \eqref{cond}, we obtain the following:
\begin{equation}
	\label{FRIEDMANN}
	H^2=\frac{P_\phi^2+\bar{\mathcal{K}}^2}{54v^2}\Big[1-\frac{4\gamma^2b_0^2}{3}\Big(\frac{P_\phi^2+\bar{\mathcal{K}}^2}{2v^2}\Big)\Big],\quad\bar{\mathcal{K}}={\mathcal{K}\over{\sqrt{2\gamma^2\,}}}.
\end{equation}

This expression represents the polymer modified Friedmann equation for the Bianchi I model in the volume variables that, written in this convenient form, allows to derive the total critical energy density of the model. In particular, the additional term $\bar{\mathcal{K}}^2/2v^2$ reasonably mimics the anisotropic contribution $\rho^{\text{aniso}}$, so that we can compute $\rho^\text{tot}_{\text{crit}}$ as  follows:
\begin{equation}
\label{tced}
\rho^\text{tot}_\text{crit}=\rho^\phi_\text{crit}+\rho^\text{aniso}_\text{crit},
\end{equation}
where the term regarding the matter scalar field takes the usual expression $P_\phi^2/(2v^2)$.
Moreover, from \eqref{FRIEDMANN} the total critical energy density results to be as follows:
\begin{equation}
\rho^\text{tot}_{\text{crit}}=\frac{3}{4\gamma^2b_0^2}\,.
\label{tcedv}
\end{equation}

\begin{figure}[H]
	\centering
	\includegraphics[width=0.8\linewidth]{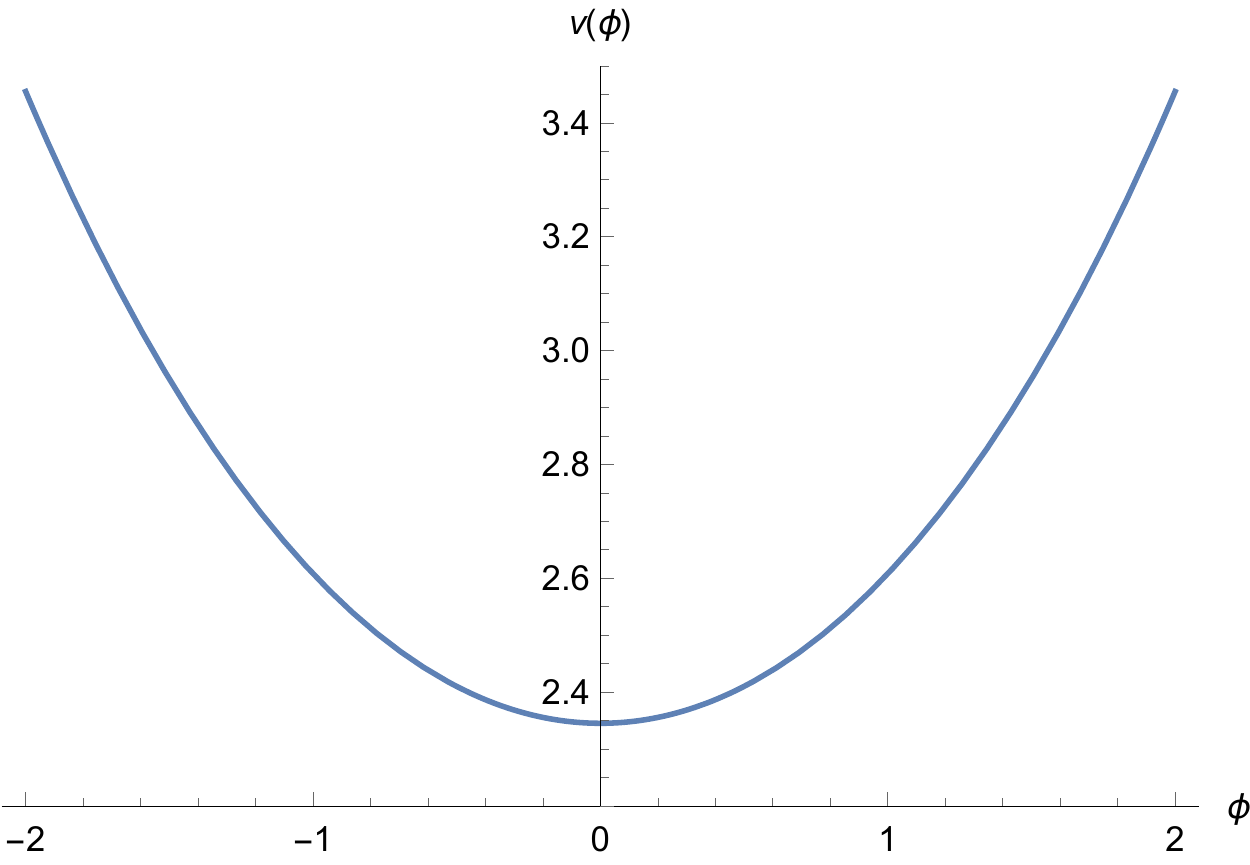}
	\caption{Semiclassical polymer trajectory of the Universe volume $v(\phi)$.}
	\label{vol}
\end{figure}

The solution for the Universe volume $v(\phi)$ when the initial conditions on the motion satisfy \eqref{cond} is as follows:
\begin{equation}
v(\phi)=\sqrt{\frac{2\gamma^2P_\phi^2+\mathcal{K}^2}{3}\,\,}\,b_0\,\cosh(\frac{\sqrt{2\gamma^2P_\phi^2+\mathcal{K}^2\,}\,\phi}{2\sqrt{3\,}\,\gamma P_\phi})\,,
\end{equation}
clearly resembling a bouncing behavior as shown in Figure \ref{vol}.

Now, we can verify whether the total critical energy density computed as \eqref{tced} acquires the expression \eqref{tcedv}. In particular, we have the following:
\begin{equation}
\rho^\text{tot}_\text{crit}=\frac{P_\phi^2}{2v(\phi)^2}\Bigg|_{\phi_B}+\frac{\bar{\mathcal{K}}^2}{2v(\phi)^2}\Bigg|_{\phi_B}=\frac{3P_\phi^2}{2b_0^2(2\gamma^2P_\phi^2+\mathcal{K}^2)}+\frac{3\mathcal{K}^2}{4\gamma^2b_0^2(2\gamma^2P_\phi^2+\mathcal{K}^2)}\,,
\end{equation}
which clearly reduces to \eqref{tcedv}; therefore, the total critical density computed from the laws of motion for the Universe volume is consistent with the expression derived from the Friedmann Equation \eqref{FRIEDMANN}. Imposing a cut-off on the volume variables makes the critical energy density independent from the initial conditions on the scalar field, and therefore, produces a Big Bounce with universal properties. This result shows that taking the Universe volume itself as a configurational variable makes the Big Bounce acquire universal physical properties, in agreement with the behavior obtained for the set $(v,\tilde{c})$ in the FLRW model. 

We notice that, even if the polymer-modified Friedmann equation in this convenient form is derived by considering the particular choice \eqref{cond} for the constants of motion, the physical properties of the bouncing behavior of our model as derived from \eqref{FRIEDMANN} have a general meaning since they are independent from the value assigned to the constant $\mathcal{K}$.

\subsection{Polymer Cosmology in the Misner-Like Variables\label{M}}

In this section, we treat the semiclassical polymer dynamics of the most general anisotropic Universe, i.e., the Bianchi IX model. First, we implement the polymer paradigm in the Misner variables \cite{Giovannetti_2019}, and then we generalize the analysis by considering a inhomogeneous extension of the model in Misner-like variables, i.e., the anisotropies together with the Universe volume $v\propto e^{3\alpha}$ \cite{stefano}.

\subsubsection{The Bianchi IX Model in the Misner Variables}
\label{Semiclassical}
The aim of this section is to discuss the main features of the Mixmaster semiclassical dynamics in the polymer representation \cite{Giovannetti_2019}. We choose to define the Misner variables $(\alpha,\beta_\pm)$ as discrete, so we perform the following formal substitutions:
\begin{subequations}
\begin{equation}
\label{pol1}
P_\pm^2\rightarrow\frac{1}{b^2}\sin^2(b P_\pm)\,,
\end{equation}
\begin{equation}
\label{pol2}
P_\alpha^2\rightarrow\frac{1}{b_\alpha^2}\sin^2(b_\alpha P_\alpha)\,,
\end{equation}
\end{subequations}
where $b$ is the polymer parameter for the anisotropies, while $b_\alpha$ is that one related to the isotropic variable $\alpha$.

The super Hamiltonian constraint \eqref{gioba14} becomes the following:\vspace{6pt}
\begin{equation}
\label{super}
\mathcal{C}_\text{BIX}^{\text{poly}}=\frac{\text{N}}{3(8\pi)^2}e^{-3\alpha}\bigg[-\frac{1}{b_\alpha^2}\sin^2(b_\alpha P_\alpha)+\frac{1}{b^2}\sin^2(b P_+)+\frac{1}{b^2}\sin^2(b P_-)+\frac{\mathcal{U}_\text{BIX}(\beta_\pm)}{b^2_\alpha}\bigg]=0\,,
\end{equation}
while the ADM Hamiltonian has the following expression:
\begin{equation}
\label{Ham}
\mathcal{C}_\text{BIX}^{\text{ADM-poly}}=\frac{\arcsin\sqrt{\frac{b_\alpha^2}{b^2}[\sin^2(b P_+)+\sin^2(b P_-)]+\mathcal{U}_{\text{BIX}}}}{b_\alpha}\,,
\end{equation}
with the condition $0\leq \frac{b_\alpha^2}{b^2}[\sin^2(b P_+)+\sin^2(b P_-)]+\mathcal{U}_\text{BIX}(\beta_\pm)\leq1$, due to the presence of the inverse sine function. In both \eqref{super} and \eqref{Ham}, we have performed the substitution $\mathcal{U}_\text{BIX}(\beta_\pm)=b_\alpha^23(4\pi)^4e^{4\alpha}U_\text{BIX}(\beta_\pm)$.

The dynamics of the model is described by the following Hamilton equations:
\begin{subequations}
\label{EqHam}
	\begin{equation}
	\label{b'}
	\beta_\pm'=\frac{\partial\mathcal{C}_\text{BIX}^{\text{ADM-poly}}}{\partial P_\pm}=\frac{b_\alpha}{b}\frac{\sin(2b P_\pm)}{\sin(2b_\alpha\mathcal{C}_\text{BIX}^{\text{ADM-poly}})},
\end{equation}
\begin{equation}
	\label{p+}
	P_\pm'=-\frac{\partial\mathcal{C}_\text{BIX}^{\text{ADM-poly}}}{\partial \beta_\pm}=-\frac{3b_\alpha(4\pi)^4e^{4\alpha}}{\sin(2b_\alpha\mathcal{C}_\text{BIX}^{\text{ADM-poly}})}\frac{\partial U(\beta_\pm)}{\partial\beta_\pm},
\end{equation}
\begin{equation}
	\label{Halpha}
	\big(\mathcal{C}_\text{BIX}^{\text{ADM-poly}}\big)'=4e^{4\alpha-8\beta_+}\frac{3b_\alpha(4\pi)^4}{\sin(2b_\alpha\mathcal{C}_\text{BIX}^{\text{ADM-poly}})},
	\end{equation}
\end{subequations}
where the symbol $'$ denotes the derivative with respect to $\alpha$.

Due to the steepness of the potential walls, we initially study the dynamics in the regime $U\sim0$. Under this condition, it can be demonstrated that there is a logarithmic relation between the time variable $t$ and the isotropic one $\alpha$, so we have that $\alpha\sim\ln(t)\rightarrow-\infty$ for $t\rightarrow 0$, even if $\alpha$ is described in the polymer formulation. This result points out that the discrete nature of the isotropic variable $\alpha$ does not prevent the Universe volume from vanishing, so the cosmological singularity is still present. We remark that the hypothesis $U\sim0$ becomes more reliable near the singularity because for $\alpha\to-\infty$, the walls move outwards, and the region where the approximation is valid asymptotically covers the whole $(\beta_+,\beta_-)$ plane.

Regarding the study of the chaotic features in the polymer modified picture, we have to analyze the relative motion between the particle-Universe and the potential walls, whose velocity is still $|\beta'_{\text{wall}}|=\frac{1}{2}$ since the polymer representation leaves the potential unchanged. On the other hand, if we study the anisotropy velocity of the particle while varying the values of the polymer parameters, we can see that the following holds:
\begin{equation}
\label{caos}
\beta'(b_\alpha,b,P_\pm)\geq1\;\forall\;P_\pm\in\mathbb{R}\Leftrightarrow\frac{b_\alpha}{b}\geq1\,,
\end{equation}
where
\begin{equation}
\label{velocity}
\beta'\equiv\sqrt{\beta_+'^2+\beta_-'^2}=\sqrt{\frac{\sin^2(b P_+)\cos^2(b P_+)+\sin^2(b P_-)\cos^2(b P_-)}{r^2(b P_+,b P_-)\big[1-\frac{b_\alpha^2}{b^2}r^2(b P_+,b P_-)\big]}}\,,
\end{equation}
and $r(x,y)=\sqrt{\sin^2(x)+\sin^2(y)}$. We notice that $(P_+,P_-)$ (and consequently $\mathcal{C}_\text{BIX}^{\text{ADM-poly}}$) are constants of motion in the free particle regime. In addition, it is easy to verify that the relation \eqref{beta} is recovered for $b,b_\alpha\to0$.

\begin{figure}[H]
	\centering
	\includegraphics[scale=0.2]{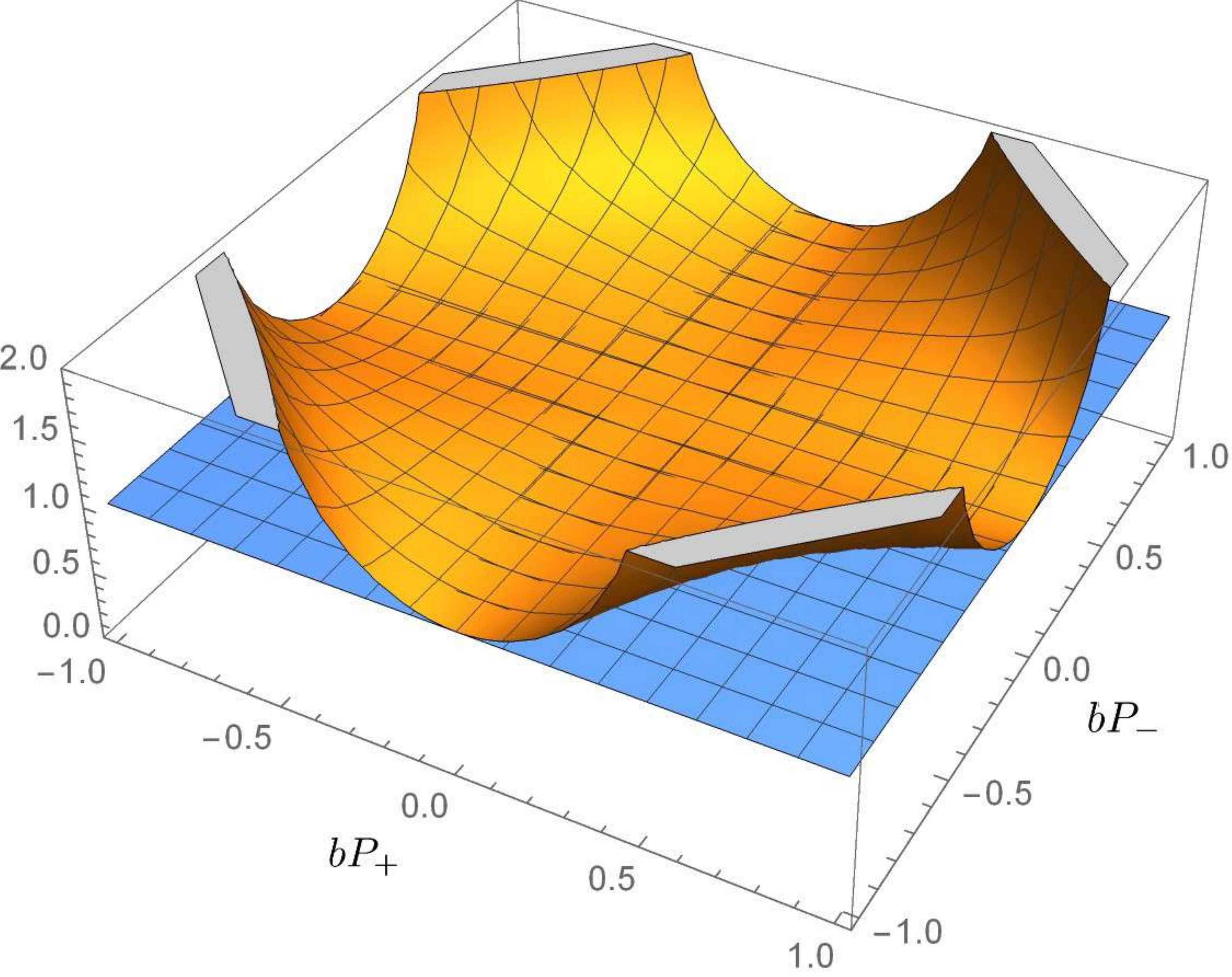}\quad\includegraphics[scale=0.2]{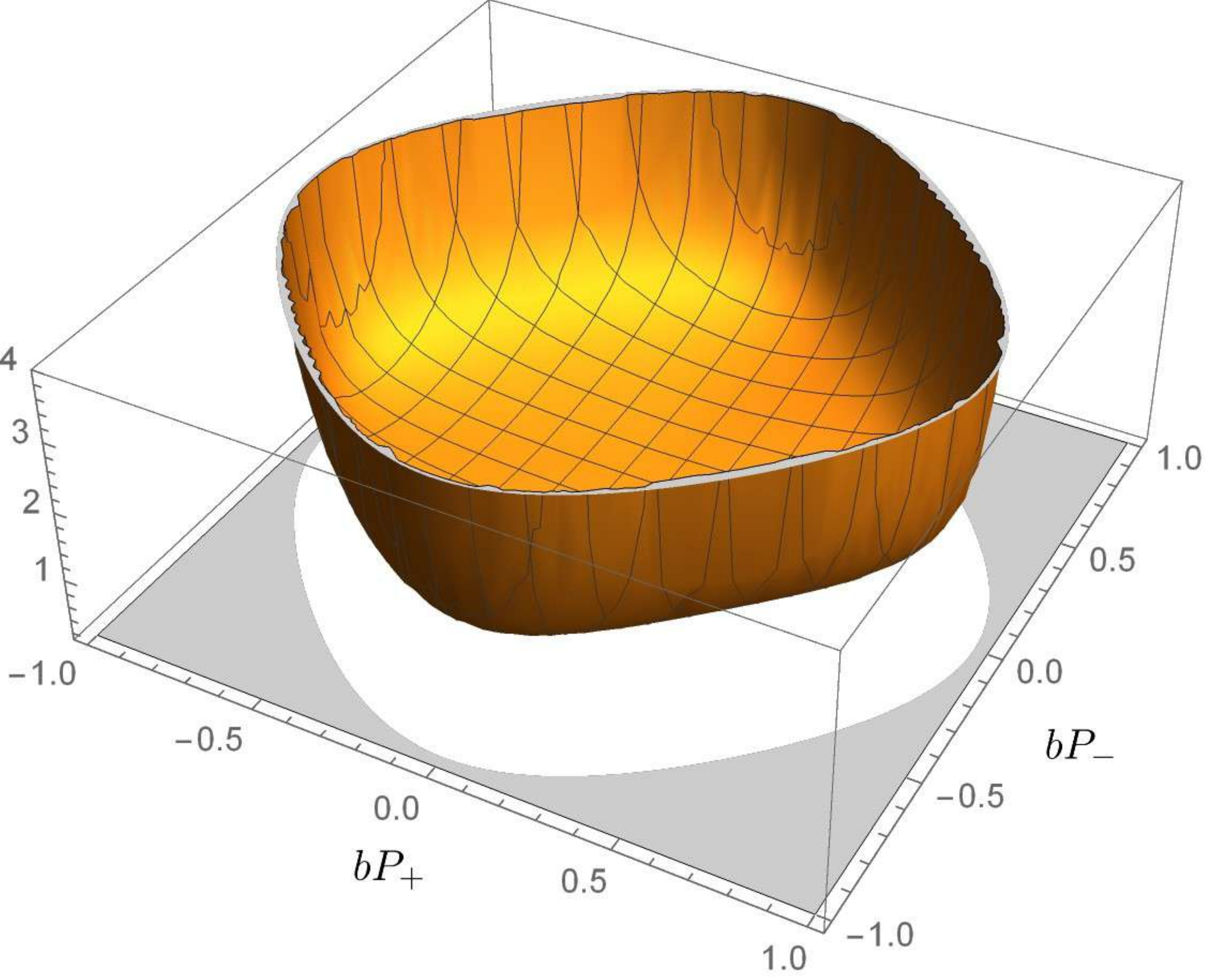}
	\caption{\label{fig:4} The 3D-profiles of the anisotropy velocity \eqref{velocity} with $b_\alpha/b=1$ (left panel) and $b_\alpha/b=1.5$ (right panel).}
\end{figure}

\begin{figure}[H]
\vspace{-6pt}
	\centering
	\includegraphics[width=0.7\linewidth]{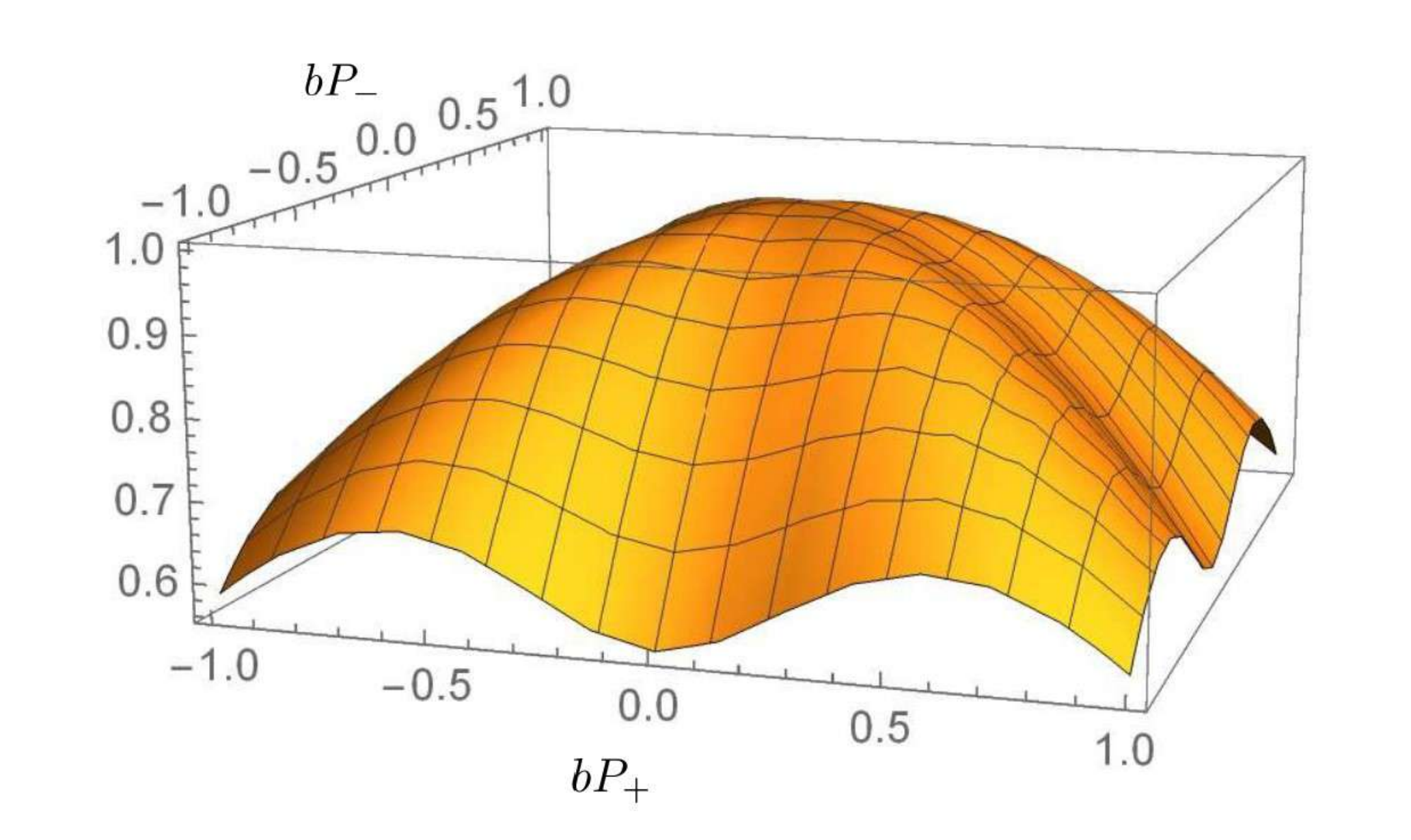}
	\caption{\label{fig:5} The 3D-profile of the anisotropy velocity \eqref{velocity} with $b_\alpha/b=0.1$.}
\end{figure}

The figures represented in Figure \ref{fig:4} show that the anisotropy velocity represented on the vertical axis is always greater than $1$ only if we choose the ratio $b_\alpha/b$ to be greater than or equal to one. Therefore, the dynamics of the Mixmaster model is expected to be still chaotic because of the existence of the singularity and the presence of a never-ending series of rebounds against the potential walls. Instead, if we choose the polymer parameters such that $b_\alpha/b<1$ (Figure \ref{fig:5}), the series of rebounds occurs until the particle velocity becomes smaller than the velocity of the potential walls; then, the point-Universe reaches the singularity with no other rebound.

Moreover, it can be demonstrated that for $b_\alpha/b<1$, we have $\theta_{\text{poly}}^{\text{max}}<\frac{\pi}{3}$, and this implies the absence of rebounds, even if the particle moves towards a specific wall. On the other hand, if $b_\alpha/b\geq1$, we have $\frac{\pi}{3}\leq\theta_{\text{poly}}^{\text{max}}<\frac{\pi}{2}$; this means that a rebound is always possible, given the triangular symmetry of the system. 

Regarding the polymer modified reflection law, by using analogous constants of motion with respect to the standard Misner case, we are able to derive the following law:
\begin{equation}
\begin{aligned}
	\frac{1}{4b}\bigg[\arcsin\bigg(&\cos\theta^f\frac{\sin\theta^i}{\sin\theta^f}\,r(2bP_+^i,2bP_-^i)\bigg)+\arcsin\bigg(\cos\theta^i\,r(2b P_+^i,2b P_-^i)\bigg)\bigg]=\\
	=\frac{1}{b_\alpha}\bigg[&\arcsin\bigg(\frac{b_\alpha}{b}\,r(b P_+^f,b P_-^f)\bigg)-\arcsin\bigg(\frac{b_\alpha}{b}\,r(b P_+^i,b P_-^i)\bigg)\bigg]\,,
\end{aligned}
\end{equation}
where $r(x,y)=\sqrt{\sin^2(x)+\sin^2(y)}$.

In order to make a comparison with the standard case \eqref{legge}, an expansion up to the second order for $b$ and $b_\alpha$ is required:
\begin{equation}
\label{reflection}
\frac{\sin(\theta^i+\theta^f)}{2}=\sin\theta^i(1+\Pi_f^2)(1+R_f)-\sin\theta^f(1+\Pi_i^2)(1+R_i)
\end{equation}
where
\begin{subequations}
\begin{equation}
\Pi^2=\frac{1}{6}b_\alpha^2(P_+^2+P_-^2),
\end{equation}
\begin{equation}
R=\frac{2}{3}b^2\frac{(P_+)^4+(P_-)^4}{(P_+)^2+(P_-)^2},
\end{equation}
\end{subequations}
and it is easy to show that in the limit $b,b_\alpha\rightarrow0$, we find the standard reflection law \eqref{legge} obtained by Misner.

Recovering this limit leads us to infer that the map above still admits stochastic properties; therefore, when $b_\alpha<b$, sooner or later the parameter region where no rebound takes place is reached, as in \cite{Lecian_2013}.

Given the results of this analysis and that of previous sections, we can infer that not only the physical nature of the Big Bounce, but also its very existence depends on the geometrical dimension of the variable chosen as discrete: we have seen here that using a logarithmic variable does not indeed avoid the singularity. Therefore, in the last model that we present in the next subsection, we choose to use a mixed representation involving the anisotropies as they are defined by Misner, together with an isotropic volume variable instead of $\alpha\propto\ln{v}$.

\subsubsection{The Inhomogeneous Mixmaster Model in the Volume Variable}
In \cite{stefano}, the semiclassical polymer dynamics of the inhomogeneous Mixmaster model is analyzed by choosing the cubed scale factor (i.e., the Universe volume) alone as the discretized configurational variable. In addition, a massless scalar field and a cosmological constant are included in the dynamics, accounting respectively for a quasi-isotropization and inflationary-like mechanisms. The resulting dynamics is a singularity-free Kasner-like phase that is linked with a homogeneous and isotropic de Sitter evolution. Moreover, the~chaotic character of the Mixmaster dynamics is absent in this framework. Thus, the~study presented in \cite{stefano} demonstrates that the generic cosmological solution is singularity-free and  non-chaotic once the polymer discretization of the Universe volume variable is performed at a semiclassical level. This result is alike to that achieved also in LQC, as~presented in Section \ref{anisotropicLQC}.

In order to summarize more in detail all the main results presented in \cite{stefano}, let us start by characterizing the generic cosmological problem and its relationship with the homogeneous Mixmaster model. The most general line element can be written as follows:
\begin{equation}
    ds^2=\text{N}^2dt^2-h_{ab}(dx^a+\text{N}^adt)(dx^b+\text{N}^bdt)\,,
\end{equation}
where $\text{N}^{a,b}(t,\vec{x})$ is the shift vector and the following holds:
\begin{equation}
h_{ab}(t,\vec{x})=e^{m_1(t,\vec{x})}l^1_al^1_b+e^{m_2(t,\vec{x})}l^2_al^2_b+e^{m_3(t,\vec{x})}l^3_al^3_b\,,
\end{equation}
with $a,b=1,2,3$. Rigorously, this is not the most general picture since we are not considering the rotation of the Kasner vectors $l^{1,2,3}_{a,b}(t,\vec{x})$. Differently from the homogeneous Bianchi line element \eqref{gioba12}, the lapse function $\text{N}(t,\vec{x})$ depends explicitly on the spatial coordinates, as well as the Misner-like variables, that in their generalized version are defined as follows:
\begin{equation}
    m_a(t,\vec{x})=\frac{2}{3}\ln[v(t,\vec{x})]+2\beta_a(t,\vec{x})\,,
\end{equation}
where $a=1,2,3$ and $\beta_a=(\beta_++\sqrt{3\,}\,\beta_-,\beta_+-\sqrt{3\,}\,\beta_-,-2\beta_+)$. Here, we are considering $v$ as the isotropic variable proportional to the spatial volume of the Universe (instead of $\alpha$ as in the proper Misner variables), while $\beta_+,\beta_-$ are the anisotropies. Let us include a massless, self-interacting scalar field $\phi$ in the dynamics, taking into account the slow-roll condition $\dot{\phi}^2\ll U(\phi)$ as well as the following hypothesis:
\begin{equation}
    |\nabla\phi|^2\ll U[\phi(t,\vec{x})]\sim\Lambda(\vec{x})\,,
\end{equation}
both typical of the inflationary paradigm. 

The action in the inhomogeneous case reads as follows:
\begin{equation}
S_\text{B}^{\text{inhom}}=\int d^3x\,dt\bigg(P_v\dot{v}+\sum_jP_j\dot{\beta_j}-\text{N}\mathcal{C}_\text{B}^{\text{inhom}}-\text{N}^i\mathcal{C}_i\bigg)\,,
\end{equation}
where $j=+,-,\phi$ and $\beta_\phi\equiv\phi$. The super Hamiltonian constraint is as follows:
\begin{equation}
\label{Hinho}
\mathcal{C}_\text{B}^{\text{inhom}}=\frac{3}{4}\bigg[-vP_v^2+\sum_j\frac{P_j^2}{9v^2}+\frac{v^{1/3}}{3}U^\text{inhom}_\text{B}+v\Lambda(\vec{x})\bigg]\,,  
\end{equation}
while $\mathcal{C}_i$ represents the supermomentum one. We remark how the self-interacting scalar field, under the slow-roll condition, is equivalent to a free scalar field plus a cosmological constant $\Lambda(\vec{x})$. 

The potential term is due to the three-dimensional scalar curvature and can be split as $U^\text{inhom}=W+U^\text{inhom}_{\text{B}}$, where the contribution due to the spatial gradients of the configurational variables is encoded in $W$, and the term $U_\text{B}^{\text{inhom}}$ represents the inhomogeneous generalization of the Bianchi model's potential (the Bianchi VIII and Bianchi IX models are the most general choices). By using the Belinskii--Khalatnikov--Lifshitz (BKL) conjecture~\cite{BKL82}, the term $W$ is considered to be negligible and so the super Hamiltonian \eqref{Hinho} reduces to that one of the Bianchi IX model, where the spatial coordinates appear only as parameters. The reliability of this assumption can be verified by explicitly evaluating the term $W$ by means of the supermomentum constraint. Moreover, by means of the gauge choice $\text{N}^i=0$, each point of space evolves independently and can be described, using a homogeneous Bianchi IX model. This way, point by point, the inhomogeneous cosmological evolution is approximated by a Mixmaster-like one. We notice that the corresponding solutions of the dynamics must satisfy also the supermomentum constraint, which identically vanishes for a homogeneous spacetime. In addition, with the hypothesis that the physical scale of the inhomogeneities is much bigger than the average Hubble horizon, each causal connected region verifies a Mixmaster-like evolution, instead of each point of space \cite{Cianfrani_2010Isotropization}.

Let us now analyze the semiclassical polymer dynamics of the model. Using the substitution \eqref{sin2} for the momentum conjugate to the variable $v$, i.e., $P_v\to\frac{\sin(b_0P_v)}{b_0}$, the~Hamiltonian constraint becomes the  following:
\begin{equation}
\label{HS}
\mathcal{C}_\text{BIX}^\text{poly}=\frac{3}{4}\bigg[-v\frac{\sin^2(b_0P_v)}{b_0^2}+\sum_j\frac{P_j^2}{9v^2}+\frac{v^{1/3}}{3}U_\text{BIX}+v\Lambda\bigg]\,,  
\end{equation}
where the spatial gradients are neglected and the potential term is that one of the Bianchi IX model. Firstly, we can notice that the following condition is imposed by the form of the Hamiltonian when the potential term is negligible:
\begin{equation}
\label{bounce}
\sin^2(b_0P_v)=b_0^2\bigg(\frac{\sum_jP_j^2}{9v^2}+\Lambda\bigg).
\end{equation}

As a consequence, the right-hand side of \eqref{bounce} must be smaller than $1$, and therefore, the Universe volume acquires a lower bound:
\begin{equation}
\label{VB}
v>v_\text{B}\equiv\frac{b_0}{3}\sqrt{\frac{\sum_jP_j^2}{\-b_0^2\Lambda}}\,,
\end{equation}
with the condition $b_0^2\Lambda<1$, which is physically reasonable in the Planck units. So, in~the polymer semiclassical dynamics, the initial singularity is replaced by a Big Bounce. We~notice that the minimum value of the Universe volume depends not only on the polymer scale $b_0$, but also on the cosmological constant $\Lambda$ and on the conjugate momenta $P_j$, which~are constants of motion when the potential term is negligible. This feature is in accordance with the results obtained in LQC \cite{Ashtekar2_2006}. The absence of singularity thanks to the semiclassical polymer formalism can be shown more in general by solving the Hamilton equation for the Universe volume, from which we obtain the following:
\begin{equation}
v(t)=\sqrt{\frac{\sum_jP_j^2\big[\cosh(3\sqrt{\Lambda}\sqrt{1-b_0^2\Lambda}t)-1+2b_0^2\Lambda\big]}{18\Lambda(1-b_0^2\Lambda)}}\,.
\end{equation}

\begin{figure}[H]
	\centering
	\includegraphics[width=0.8\linewidth]{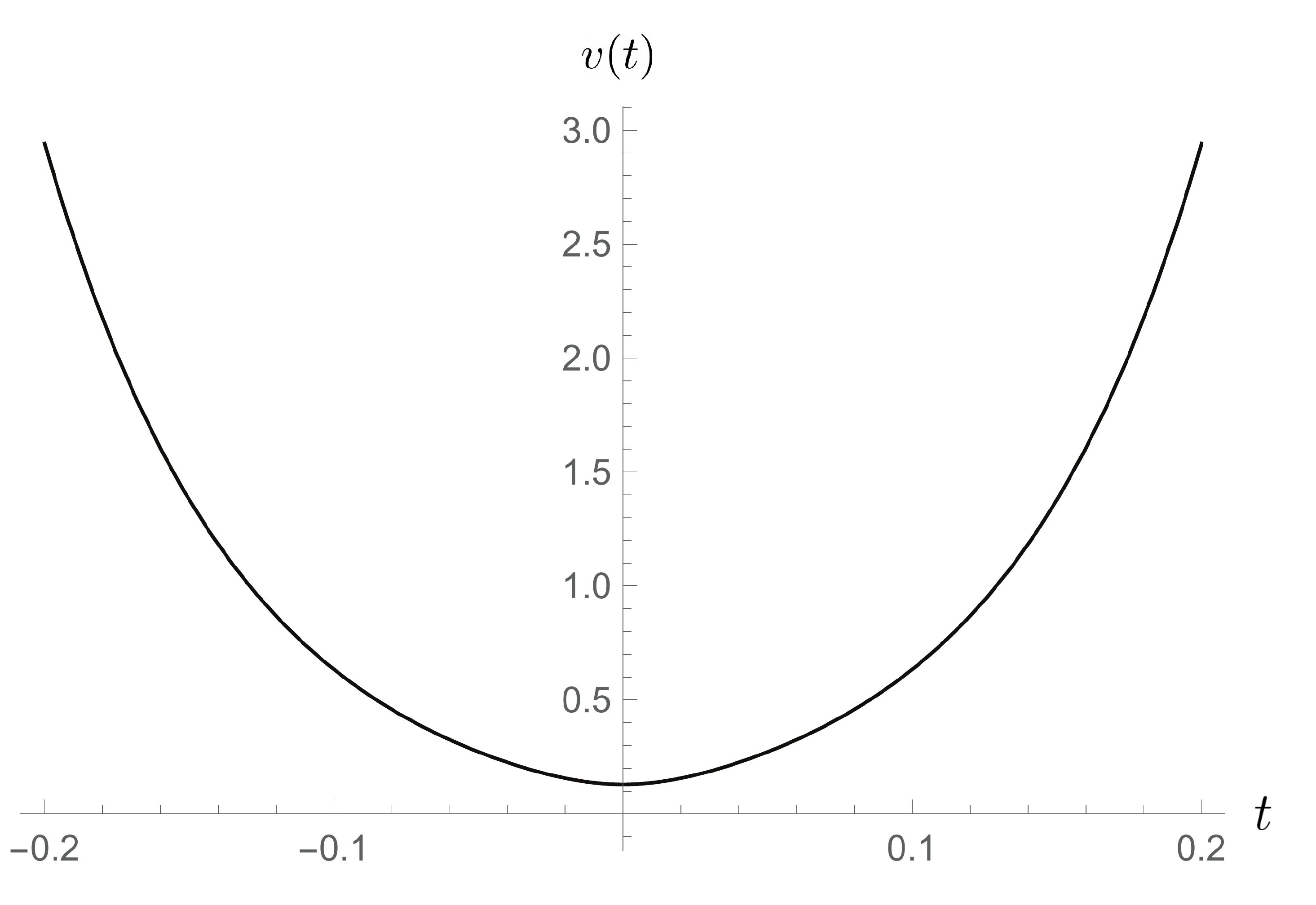}
	\caption{Semiclassical polymer trajectory of the Universe volume $v$ in the synchronous time $t$. Image from \cite{stefano}.}
	\label{Vs}
\end{figure}

In particular, as shown in Figure \ref{Vs} the Bounce occurs when the volume reaches its minimum, which is given by Equation \eqref{VB}.

Moreover, by performing the ADM reduction and choosing $v$ as the relational time, the Hamilton equations for the anisotropies and the scalar field can be solved, showing that the former do not diverge approaching the Big Bounce, as it happens in the standard case. Additionally, far from the Bounce, the anisotropies tend to a constant value that can be absorbed by an appropriate and local rescaling of the coordinate system. In this sense, the~quasi-isotropization mechanism described in \cite{Cianfrani_2010Isotropization,Kirillov_2002} results to be preserved under the polymer modification.

In conclusion, the semiclassical polymer dynamics is well described by a bridge solution connecting an initial non-singular, inhomogeneous and anisotropic Kasner-like regime with a later homogeneous and isotropic de Sitter phase. We notice that this result is local, i.e., valid in a nearly homogeneous region. Nonetheless, with the hypothesis of having an inhomogeneity scale much larger than the average Hubble horizon, each~causal connected region undergoes such an isotropization phase and can be expanded by the inflationary mechanism on a scale much larger than the Hubble radius. Thus, if the inflationary phase is long enough, it is possible to obtain a homogeneous and isotropic region larger than the Hubble horizon today.

Finally, in \cite{stefano}, it is also demonstrated that the chaotic properties of the dynamics are suppressed when the polymer representation is implemented at a semiclassical level. This property is a consequence of the following considerations. As we have seen in \mbox{Section \ref{giobaBIX}}, the chaoticity of the Bianchi IX model is derived by analyzing the relative velocity between the point-Universe and the potential walls. It is possible to verify that, similar to the analysis in Section \ref{Semiclassical}, in this model, the velocity of the walls is finite at the Bounce, while that of the point-Universe diverges in the polymer framework. So, rebounds always occur, even in the presence of a scalar field, even though in the classical framework, its presence is able to remove the chaos \cite{BelinskiScalarFieldChaos}. However, due to the polymer modifications, the following condition holds:
\begin{equation}
\label{chaos}
\sum_jP_j^2+9\Lambda v^2>\text{max}\bigg(3v^{4/3}U_\text{BIX}\bigg)\,.
\end{equation}

Therefore, there are two possible scenarios for the removal of chaos, similar to the case of the Loop quantized homogeneous Bianchi IX model described in Section \ref{anisotropicLQC}: either the condition \eqref{chaos} is satisfied before the Bounce, resulting in a final stable Kasner-like epoch (similarly to the AVTD regime reached classically when a scalar field is present), or chaos is removed by the fact that the presence of a Big Bounce makes the number of rebounds on the potential walls finite.

\subsection{The Link between Polymer Quantum Mechanics and Loop Quantum Cosmology: Canonical~Equivalence \label{Discussione}} 
In Sections \ref{A}--\ref{M}, we have presented the main results gained thanks to the application of the polymer representation at the level of the semiclassical dynamics, with some extensions to the full quantum regime in Sections \ref{A} and \ref{VL}. Along the presentation, we~have proceeded by considering more and more general cosmological models, starting from the FLRW one and then extending the analysis to the Bianchi I and IX models afterwards. For each model, we have compared the obtained results in different sets of variables, always performing the canonical transformation between the phase space variables before implementing the polymer approximation for the variables conjugate to those represented as discrete on the polymer lattices. If, on one hand, the polymer formulation permits to successfully overcome the GR shortcoming of the initial singularity, on~the other hand, it reproduces different physical pictures for the same cosmological model when represented in different sets of variables canonically linked in the classical regime. In~this respect, the~polymer behavior of the Bianchi IX cosmological model in the standard Misner variables is of particular relevance since it shows a singular dynamics in contrast with the bouncing cosmology and the chaos removal ensured by the use of a volume-like variable (note that in the Misner variables the presence of chaos depends on the ratio of the different discretization parameters associated to the isotropic variable and the anisotropy coordinates $\beta_{\pm}$ \cite{Giovannetti_2019}).

This fact raises an issue about what it really means to do a canonical transformation in the polymer framework. For a proposal about recovering the equivalence between different sets of variables after the polymer is implemented on the Hamiltonian constraint, see \cite{Federico}. In this work, it is demonstrated how, by fixing the preferred system of generalized coordinates and assigning a constant lattice step (actually in the considered case, there is only the scale factor, but a generalization of the analysis can be easily inferred), the dynamical features of the model remain the same in any other set of coordinates, at~the price of considering the discretization step as a function of the adopted variables. For example, starting from the volume coordinates and then changing to the Ashtekar connections maintains the canonical equivalence between the dynamics thanks to the dependence of the polymer parameter on $p$. Therefore, the dynamics is ruled by the set of variables in which the polymer spacing results in being constant. We note that this study was performed only at a semiclassical level since the implementation of a translational operator depending on the coordinates is not trivial, so the equivalence in the full quantum theory is still an open question.

Nevertheless, this result achieved in \cite{Federico} about the recovery of an equivalence class of configurational variables could have relevant implications for the $\bar{\mu}$ scheme of LQC. In particular, in the $\bar{\mu}$ scheme of LQC the translational operator acting on the physical states acquires a parameter depending on the momentum variable (due to the physical rescaling of the area element) that, in the polymer framework, plays exactly the role of $p$. This analogy suggests that a change of coordinates is at the basis of the improved LQC approach \cite{Ashtekar2_2006,Ashtekar_2008,Ashtekar_2011Review} in a way that a translational operator of constant step is restored and the theory is made technically viable. 

However, such a change of variables opens a possible criticism about the quantization procedure, which is no longer referred to the natural $SU(2)$ connection so that LQC would be departing from full LQG. While on a quantum level, this question is somewhat an open issue, on the level of an effective LQC theory, the result mentioned above can somewhat support the validity of the change in variables in view of the link between PQM and LQC. Indeed, since the physical picture of the model is dictated by the scheme with a constant polymerization parameter, we can conclude that the semiclassical dynamics of the $\bar{\mu}$ formulation is the same of that obtained in the Ashtekar variables if the lattice parameter is canonically transformed so that no ambiguity arises about the universal character of the Big Bounce, i.e., about its independence on the initial conditions.

\section{Concluding Remarks}\label{secConcl}
In this review, we analyzed basic aspects of LQC and presented polymer quantum cosmology in depth in order to trace some solid implications on the early Universe evolution precisely regarding the regularization of the singularity. 

In this respect, we can firmly say that a bouncing cosmology always emerges in LQC and in polymer quantum cosmology, when the Universe volume is considered one of the configurational variables. Furthermore, the Big Bounce has also the features of a universal cut-off since the critical energy density of the Universe takes a maximum value depending only on fundamental constants and the Immirzi parameter. This result is obtained in the improved scheme of LQC as shown in Section \ref{improvedLQC} and also demonstrated in the context of PQM for the isotropic model in \cite{Mantero}. Moreover, similar results to those obtained in the $\bar{\mu}$ scheme are outlined in \cite{Federico,Silvia} when PQM is implemented on volume-like~variables. 

The situation is slightly different when we polymerize different sets of generalized variables in PQM. In particular, when we consider the polymerization of the Ashtekar--Barbero--Immirzi connections, the Big Bounce still emerges in the semiclassical dynamics, but the critical energy density now depends on the initial conditions on the motion, or equivalently on the wave packets when the full quantum approach is implemented (see~\cite{Federico,Silvia}). In these analyses, the polymerization of the natural connection seems to reproduce a physical picture very similar to 
the $\mu_0$ scheme of LQC presented in Section \ref{standardLQC}.

The importance of the choice of the adopted variable for the polymer quantization of the Universe is outlined by observing that the polymerization of the Bianchi IX model (both in the homogeneous and inhomogeneous cases) expressed in the standard Misner variables yields a model that is still singular \cite{CrinoPintaudi,Giovannetti_2019}, while discretizing the Universe volume in the metric formulation does avoid the singularity \cite{stefano}. This result highlights the link between the dimensionality of the discretized variable and the regularization of the singularity in PQM; also, it opens the non-trivial question about which equivalence class among all possible choices of canonically related coordinates leads to the same physical picture.

An interesting result toward the solution of this puzzling question is elucidated from the analysis in \cite{Federico}. We have discussed how the equivalence is ensured for the isotropic model on a semiclassical level by the possibility to restate the problem in terms of a new polymerization step dependent on the configurational coordinates. This result can have positive implications in stating the equivalence of the effective dynamics in the $\mu_0$ and $\bar{\mu}$ schemes of LQC \cite{Ashtekar_2008,AshtekarEffectiveLQC}, as discussed in Section \ref{Discussione}, at least on a semiclassical level.

On the base of the results illustrated here, we can conclude that, if on one hand, some basic features of cut-off quantum cosmology are well traced, many other detailed points must be addressed. For instance, it is necessary to better characterize the physical properties of the Big Bounce, both by providing a pure quantum description instead of just a semiclassical one, and by further investigating the thermodynamical nature of the quantum cosmological fluid. Moreover, it is crucial to understand how a collapsing Universe before our expanding branch is generated and how it can influence the morphology of the present Universe (see some pioneering works \cite{Mantero,DiAntonio,Grain_2021,Lyth_2002,Khoury_2002,Khoury2_2002,Steinhardt_2003,Brandenberger2013,Alexander}). A~typical example of this issue affecting bouncing cosmologies is provided in \cite{Mantero}, where it is argued that the flatness paradox is still present, even though the horizon paradox is naturally solved, thanks to the pre-existing collapsing Universe. These kinds of physical questions become even more meaningful if we postulate that the Universe dynamics is associated to a cyclical evolution between a Big Bounce and a later classical turning point~\cite{BarrowCyclic,BarrowMixmaster}: in this context, the behavior of the Universe entropy becomes a delicate feature to be addressed. 

\end{document}